\documentclass[12pt]{article}
\usepackage{natbib}
\usepackage{url} 
\usepackage[inline]{enumitem}
\usepackage{tcolorbox}

\RequirePackage{amsmath}
\usepackage{amssymb}
\RequirePackage[colorlinks,citecolor=blue,urlcolor=blue]{hyperref}
\RequirePackage{amsfonts}
\RequirePackage{graphicx}
\RequirePackage{mathrsfs}
\RequirePackage{verbatim}
\RequirePackage{float}
\RequirePackage{amsthm,amsfonts,amssymb}
\RequirePackage{multirow}
\RequirePackage{makecell}
\RequirePackage{subfigure}
\usepackage{enumitem}
\usepackage{algorithm, algorithmic}
\allowdisplaybreaks

\usepackage{hyperref}

\newcommand{\blind}{1}

\newtheorem{proposition}{Proposition}
\newtheorem{theorem}{Theorem}

\newtheorem{remark}{Remark}
\newtheorem{corollary}{Corollary}
\newtheorem{assumption}{Assumption}
\newtheorem{example}{Example}

\newtheoremstyle{named}{}{}{\itshape}{}{\bfseries}{.}{.5em}{\thmnote{#3 }}
\theoremstyle{named}

\def\diag{\mathrm {diag}}
\newcommand{\bm}{\boldsymbol}

\def\0{{\bf 0}}

\def\boxit#1{\vbox{\hrule\hbox{\vrule\kern6pt  \vbox{\kern6pt#1\kern6pt}\kern6pt\vrule}\hrule}}

\begin{document}
\nocite{Noroozi2021Estimation}
\nocite{Majid2021Sparse}

\def\spacingset#1{\renewcommand{\baselinestretch}
{#1}\small\normalsize} \spacingset{1}
\if1\blind
{
  \title{\bf Strongly consistent community detection in popularity adjusted block models}
  \author{Quan Yuan$^\dag$,  
  Binghui Liu$^\dag$, 
  Danning Li$^\dag$ and Lingzhou Xue$^\ddag$ 
\hspace{.2cm}\\
   $^\dag$ Northeast Normal University\\
    $^\ddag$The Pennsylvania State University}
    \date{}
  \maketitle
} \fi

\if0\blind
{
  \bigskip
  \bigskip
  \bigskip
  \begin{center}
    {\Large\bf Strongly Consistent Community Detection in Popularity Adjusted Block Models}
\end{center}
  \medskip
} \fi
\begin{abstract}
{
The Popularity Adjusted Block Model (PABM) provides a flexible framework for community detection in network data by allowing heterogeneous node popularity across communities. However, this flexibility increases model complexity and raises key unresolved challenges, particularly in effectively adapting spectral clustering techniques and efficiently achieving strong consistency in label recovery. To address these challenges, we first propose the Thresholded Cosine Spectral Clustering (\textsf{TCSC}) algorithm and establish its weak consistency under the PABM. We then introduce the one-step Refined \textsf{TCSC} algorithm and prove that it achieves strong consistency under the PABM, correctly recovering all community labels with high probability.  {We further show that the two-step Refined \textsf{TCSC} markedly accelerates clustering error convergence, especially with small sample sizes.} Additionally, we propose a data-driven approach for selecting the number of communities, which outperforms existing methods under the PABM. The effectiveness and robustness of our methods are validated through extensive simulations and real-world applications. 
}
\end{abstract}

\spacingset{1.6} 
\section{Introduction}
\label{intro}

Networks provide a powerful tool for representing and analyzing the relationships between interacting units in complex systems \citep{ wasserman1994,Anna2010}. A central problem in network data analysis is community detection, which seeks to partition the nodes of the network into distinct groups or communities. Identifying such communities can yield valuable insights into the network's underlying organization and structure. A substantial body of research on community detection spans multiple disciplines, including business\citep{chen2024model}, social science \citep{moody2003structural}, environmental studies \citep{agarwal2020model}, and genetics \citep{spirin2003protein,Zhang2017Finding}. For comprehensive reviews of this topic, see \citet{fortunato2010community, fortunato2016community, kim2018review}.

In recent decades, there has been substantial progress in statistical models for community detection with network data. The Erdős–Rényi (ER) random graph model \citep{Erdos1959} assumes that all node pairs in the network are connected independently with a common edge probability. To accommodate community structure, \citet{Holland1983stochastic} proposed the stochastic block model (SBM), in which the $n$ nodes are partitioned into $K$ communities, and the probability of an edge between two nodes depends solely on their community memberships. Specifically, nodes $i$ and $j$ are connected with probability $P_{kl}$ if they belong to communities $k$ and $l$, respectively, for $i, j \in \{1, \dots, n\}$ and $k, l \in \{1, \dots, K\}$. To account for degree heterogeneity often observed in real-world networks,  \citet{Karrer2011Stochastic} introduced the degree-corrected stochastic block model (DCSBM), which extends the SBM by assigning each node a degree parameter that uniformly scales its connectivity across all communities, thereby modeling varying node popularity.

Nevertheless, neither the SBM nor the DCSBM adequately captures variations in node popularity across communities, a phenomenon frequently observed in real-world networks. For instance, even if nodes $i$ and $j$ belong to the same community, there may exist communities $k$ and $l$ such that community $k$ exhibits stronger connectivity to node $i$ than to node $j$, while community $l$ connects more strongly to node $j$ than to node $i$. To model such asymmetries, \citet{Sengupta2017aBlock} proposed the Popularity Adjusted Block Model (PABM) with parameter matrix $\bm \Lambda = (\lambda_{i'k' }) \in \mathbb{R}^{n \times K}$, where $\lambda_{i' k'}$ encodes the popularity of node $i'$ with respect to community $k'$, in which the probability of an edge between nodes $i$ and $j$—belonging to communities $k$ and $l$, respectively—is given by $\lambda_{i l} \lambda_{j k}$.

Several methods have been proposed for community detection under the PABM. \citet{Sengupta2017aBlock} optimized a modified modularity function and established the weak consistency of its proposed estimator, namely, as $n\rightarrow \infty$, for any $\varepsilon >0$, 
\begin{eqnarray*}
 \mathbb{P}
 \left\{ \min_{\pi : [K]\rightarrow [K]} \frac{1}{n} \sum_{i=1}^{n}  \mathbb{I} \{\bm c^{\ast}(i) \neq  \pi(\hat{\bm c}(i))\}
 < \varepsilon
  \right\}
\rightarrow 1
 ,
\end{eqnarray*}
where $\hat{\bm c} \in [K]^{n}$ is an estimate of the true community label vector $\bm c^{\ast}$. Subsequently, \citet{Majid2021Sparse,Noroozi2021Estimation} noted that the structure of the edge probability matrix in the PABM poses challenges for the direct application of spectral clustering. To address this, they proposed an alternative community detection method based on subspace clustering in practice
and also proved the weak consistency of their estimator. More recently, \citet{John2023Popularity} analyzed the PABM within the framework of generalized random dot product graphs and showed that community labels can be recovered with high accuracy, {provided that} the rows of matrix $\bm \Lambda$ for nodes within the same community are assumed to be independently and identically distributed (\emph{i.i.d.}). This assumption imposes a stronger constraint than those in \citet{Sengupta2017aBlock} and \citet{Noroozi2021Estimation}, and as shown in Section~\ref{S_simu_heterogeneity}, when this \emph{i.i.d.} assumption is violated, the performance of its algorithm deteriorates significantly.

Despite the important progress made in developing and analyzing methods for community detection under the PABM, two key open questions remain:
\begin{itemize}
    \item [(1)] Can spectral clustering be effectively implemented under the PABM? Spectral methods are widely studied and known for their practical efficiency in community detection.  Classical spectral clustering techniques—such as those in \citet{Rohe2010SpectralCA, Sussman2011ACA, Jing2015Consistency, Abbe2015Community, Su2020Strong}—and their numerous variants—e.g., \citet{amini2013pseudo, Jin2015Fast, Sihan2023PCABM, Jin2024Mixed}—have demonstrated strong empirical performance and theoretical guarantees under various network models. Owing to their computational efficiency,  these methods are often used as initialization steps in more complex procedures \citep{amini2013pseudo, Gao2015, NIPS2016Yun, Wang2020Fast, Zhao2024Variational}. However, it remains unclear whether such strategies can be effectively adapted to the PABM, given its more flexible and heterogeneous structure.
   \item [(2)] {Is it possible to efficiently achieve strong consistency for community label estimation under the PABM?
}
  As discussed in \cite{Zhao2011Consistency}, consistency in community detection can be classified as weak or strong. Strong consistency refers to the exact recovery of all community labels (up to a permutation) with high probability: 
     \begin{equation*}
 \mathbb{P}
 \left\{
 \underset{\pi : [K]\rightarrow [K]}{\cup}
 \left(\hat{\bm c}=\pi[\bm c^{\ast}]\right)
 \right\}
\rightarrow 1,
\end{equation*}
where $\pi[\bm c^{\ast}]=\big(\pi(\bm c^{\ast}(1)),\cdots,\pi(\bm c^{\ast}(n))\big)^{\top}$.
A strongly consistent estimation of community labels is particularly valuable in downstream inference tasks. For example, in hypothesis testing for network models \citep{Zhang2017AHT,yuan2022testing} or in estimating the number of communities \citep{saldana2017many,Chen2018Network,li2020network, Hu2021Using, Jin2023Optimal}, strongly consistent community estimates can be directly incorporated into test statistics, thereby enhancing the reliability and validity of the resulting inference. When the initialization is reasonable, a refinement step often improves estimation accuracy, as shown for SBM \citep{Gao2015} and DCSBM \cite{Gao2018}. However, efficiently refining estimates under the PABM poses significantly greater challenges and remains an open problem.
\end{itemize}

In this paper, we provide affirmative answers to these two open questions. To address the first, we derive the explicit form of the eigenvector matrix of the PABM's edge probability matrix (see Proposition~\ref{cPABMlemma:orthogonal}). This characterization reveals that, unlike in the SBM where the rows of the eigenvector matrix corresponding to nodes within the same community are identical, or in the DCSBM where they are proportional, the PABM exhibits a more intricate structure: rows within the same community are neither identical nor proportional. As a result, standard spectral clustering methods and their variants, such as spherical spectral clustering, are ineffective under the PABM. However, our result suggests that the angle between the representations of nodes in the eigenspace can serve as a meaningful similarity measure. With a simple transformation, we construct a similarity matrix from the eigenspace and apply K-means clustering, following the general principle of standard spectral clustering.

To address the second question, we propose an one-step refinement algorithm based on the inter-cluster angular similarity of the adjacency matrix, specifically designed to enhance the accuracy of community label estimation.
Importantly, our approach—consistent with the modeling assumptions in \citet{Sengupta2017aBlock} and \citet{Noroozi2021Estimation}—does not impose any distributional assumptions on the matrix $\bm \Lambda$. As demonstrated by the simulation results in Figure~\ref{cPABMiid_differ_plot}, this flexibility enables our method to perform well even when the rows of $\bm \Lambda$ are not \emph{i.i.d.}. 

Our contributions in this paper are threefold. First, we provide a comprehensive analysis of the eigenspace structure of the edge probability matrix under the PABM. Building on this structural analysis, we develop the Thresholded Cosine Spectral Clustering (\textsf{TCSC}) algorithm, which serves as an effective initialization procedure and yields a weakly consistent estimator of the community labels. {Second, we show that one-step Refined \textsf{TCSC} is sufficient to upgrade the weakly consistent estimate to a strongly consistent one, that is, as $n \to \infty$, all nodes are correctly clustered with high probability. Moreover, we theoretically demonstrate that for finite samples, two-step Refined \textsf{TCSC} can substantially accelerate the convergence rate of the clustering error.}
Third, we propose a data-driven method for selecting the number of communities $K$ under the PABM, which outperforms the loss-plus-penalty approach proposed by \citet{Noroozi2021Estimation} in simulations. 

The rest of this paper is organized as follows. Section~\ref{cPABMmodel} introduces the PABM framework. Section~\ref{cPABMStructural_PABM} studies the eigenspace properties and presents our proposed algorithms \textsf{TCSC} and \textsf{R-TCSC}. In Section~\ref{cPABMtheorem}, we establish upper bounds on the clustering error rates for both algorithms. Section~\ref{cPABMsec:simulation} presents simulation studies demonstrating the numerical properties of \textsf{TCSC} and \textsf{R-TCSC}. Section~\ref{cPABMsec:SelectionK} addresses the selection of the number of communities when it is unknown. Section~\ref{cPABMRealdata} shows the practical utility of our methods through applications to two real-world datasets. Section \ref{conclusion} includes a few concluding remarks.

\section{Preliminaries}\label{cPABMmodel}
We begin by introducing useful notations. Let $[n]\doteq \{1,\cdots,n\}$ for any $n\in\mathbb{N}_+$, and $|\mathcal{S}|$ be the cardinality of a set $\mathcal{S}$. Throughout the paper, we use $C$ to denote generic constants whose values may vary from line to line. For positive sequences $\{x_{n}\}_{n=1}^\infty$ and $\{y_{n}\}_{n=1}^\infty$, $x_{n}\gtrsim y_{n}$ means that
$x_{n}\geq C y_{n}$ for some $C>0$; $x_{n} \lesssim y_{n}$ means that
$x_{n}\leq C y_{n}$ for some $C>0$; $x_{n}\asymp y_{n}$ means that
$\frac{1}{C} y_{n} \leq x_{n}\leq C y_{n}$ for some $C\geq 1$;
$x_{n} \gg y_{n}$ means that $y_{n}= o( x_{n})$; $x_{n} \ll y_{n}$ means $x_{n}= o( y_{n})$.
For $\bm x=(x_{1},\cdots,x_{n})^{\top} \in \mathbb{R}^{n},$
$\|\bm x\|=\sqrt{ \sum_{i=1}^{n} x_{i}^{2} }.$
For $\bm M=(M_{ij})_{m\times n} \in \mathbb{R}^{m\times n}, $ let
$\|\bm M\|_{\textrm{F}}=\sqrt{ \sum_{i=1}^{m}\sum_{j=1}^{n} M_{ij}^{2}}$, $ \sigma_{r}(\bm{M}) $ denote the $ r $-th largest singular value of $ \bm{M}$ and
$\|\bm M\|_{\textrm{op}}= \sigma_{1}(\bm M)$, where $ 1 \leq r \leq \min\{m, n\} $.
 For a square matrix $\bm M$, $\diag(\bm M)$ denotes the diagonal matrix whose diagonal entries are those of $\bm M$. 

We consider the undirected network $G=(V,E)$ with the node set $V=[n]$, the edge set $E \subseteq \{(i,j):i,j \in V\}$, and the adjacency matrix $\bm A=(A_{ij})_{n \times n}\in \{0,1\}^{n\times n}$.
Here, $A_{ij}=1$ if $(i,j)\in E$, otherwise $A_{ij}=0$.
Suppose that there is no self-loop in network $G$, i.e., $A_{ii}=0$ for each node $i \in V$. As in \cite{Sengupta2017aBlock} and \cite{Noroozi2021Estimation},
the network $G$ can be characterized by the Popularity Adjusted Block Model (PABM):
\begin{eqnarray}\label{cPABMPABM}
\begin{aligned}
A_{i j}=A_{j i} \stackrel{i n d}{\sim} \operatorname{Bern}\left(
\theta_{ij}
\right) \text { for all } i < j \in[n],\quad \text{and} \quad
A_{i i}\equiv 0 \text { for all } i \in [n],
\end{aligned}
\end{eqnarray}
where $\theta_{ij}=\lambda_{i \bm c^{\ast}(j)}\lambda_{j \bm c^{\ast}(i)}$, $\bm \Lambda = (\lambda_{ik})_{n \times K} \in [0,1]^{n \times K}$ and
$\bm c^{\ast} =(\bm c^{\ast}(1), \cdots, \bm c^{\ast}(n))^{\top} \in [K]^{n}$ is the community label. Let $\bm \Theta=(\theta_{ij})_{n \times n}$. And we have that $\mathbb{E} \bm A=\bm \Theta-\diag(\bm \Theta)$.

Our objective is to recover the true community labels $\bm c^{\ast}$. Following \cite{Sengupta2017aBlock} and \cite{Noroozi2021Estimation}, we are concerned only with the partition structure and not the specific label assignments. Accordingly, for any true label vector $\bm c^{\ast} \in [K]^{n},$ we evaluate an estimator $\bm c^{\ast} \in [K]^{n}$ using the following loss function:
\begin{eqnarray}
\ell(\bm c^{\ast},  \hat{\bm c} )= \min_{\pi : [K]\rightarrow [K]} \frac{1}{n} \sum_{i=1}^{n}  \mathbb{I} \{\bm c^{\ast}(i) \neq  \pi(\hat{\bm c}(i))\}.
\label{cPABMLoss}
\end{eqnarray}
Let $\pi^{\ast}(\cdot)$ be the optimal permutation achieving the minimum in \eqref{cPABMLoss}, which means $n\operatorname{\ell}(\bm c^{\ast}, \hat{\bm c})=\sum_{i=1}^{n}\operatorname{1}\{\bm c^{\ast}(i)\neq \pi^{\ast}(\hat{\bm c}(i))\}$. 
For the sake of notational simplicity, we replace $\pi^{\ast}[\hat{\bm c}]$ by $\hat{\bm c}$ later, where $\pi^{\ast}[\hat{\bm c}]=\big(\pi^{\ast}(\hat{\bm c}(1)),\cdots,\pi^{\ast}(\hat{\bm c}(n))\big)^{\top}$.

\citet{Sengupta2017aBlock} introduced the PABM likelihood modularity approach to estimate community labels by maximizing the likelihood function. \citet{Majid2021Sparse,Noroozi2021Estimation} leveraged a key structural property of the PABM that $\bm \theta_{i\cdot}$ can and only can be expressed by $\{\bm \theta_{j\cdot}\}_{j:\bm c^{\ast}(j)=\bm c^{\ast}(i)}$ to apply the sparse subspace clustering (SSC) directly to the adjacency matrix $\bm A$ to estimate the community label vector $\bm c^{\ast}$, where $\bm \theta_{i\cdot}$ is the $i$-th row of $\bm \Theta$. In contrast, \cite{John2023Popularity} recovered the community label vector $\bm c^{\ast}$ using the relationship between the eigenspace of $\bm \Theta$ and community labels $\bm c^{\ast}$. Instead of applying SSC to $\bm A$ as in \cite{Noroozi2021Estimation}, they applied the SSC to the spectral embedding of $\bm A$.

\section{Methodology}\label{cPABMStructural_PABM}

This section aims to implement spectral clustering under the PABM. Subsection \ref{cPABMsubsecEigenspace} examines the eigenspace properties of the edge probability matrix and explores its relationship with the underlying community structure. Building on these insights, we present the proposed \textsf{TCSC} algorithm in Subsection \ref{cPABMsubsecInitial_Al} and \textsf{R-TCSC} algorithm in Subsection \ref{cPABMsubsecInitial_Al2}.

\subsection{Structural Properties of the PABM}
\label{cPABMsubsecEigenspace}
{
The foundation of spectral clustering algorithms lies in analyzing the relationship between the eigenvectors of the adjacency matrix $\bm A$ with those of the edge probability matrix $\bm \Theta$, 
and in understanding how the eigenstructure of $\bm \Theta$ relates to the true community labels $\bm c^{\ast}$. For example, under mild conditions, \cite{Jing2015Consistency} showed that, in the SBM, the eigenvector matrix $\bm \Xi_{\text{SBM}}$ of the edge probability matrix $\bm \Theta_{\text{SBM}}$ satisfies that two nodes belong to the same community if and only if their corresponding rows in $\bm \Xi_{\text{SBM}}$ are identical. Similarly, in the DCSBM, the eigenvector matrix $\bm \Xi_{\text{DCSBM}}$ of the edge probability matrix $\bm \Theta_{\text{DCSBM}}$ satisfies that two nodes belong to the same community if and only if their corresponding rows in $\bm \Xi_{\text{DCSBM}}$ are proportional. }

To implement spectral clustering under the PABM, we start with a spectral analysis of the edge probability matrix $\bm \Theta$. As noted in \cite{Noroozi2021Estimation}, by permuting the nodes, $\bm \Theta$ can be rearranged into a block matrix consisting of $K^2$ rank-one submatrices:
\begin{eqnarray}
\label{cPABMTheta:rank1}
\bm \Theta
=
\begin{bmatrix}
\bm \lambda^{(1,1)} {\bm \lambda^{(1,1)}}^{\top} & \cdots & \bm \lambda^{(1,K)} {\bm \lambda^{(K,1)}}^{\top}
\\
\vdots &  & \vdots
\\
\bm \lambda^{(K,1)} {\bm \lambda^{(1,K)}}^{\top} & \cdots & \bm \lambda^{(K,K)} {\bm \lambda^{(K,K)}}^{\top}
\end{bmatrix},
\end{eqnarray}
where  $\bm \lambda^{(k,l)}=(\lambda_{il})_{i: \bm c^{\ast}(i)=k} \in [0,1]^{n_{k}(\bm c^{\ast}) \times 1}$ for each $k,l \in [K]$, with $n_{k}(\bm c)=\sum_{i=1}^{n} \mathbb{I} \{\bm c(i) = k\}$ for any general $\bm c \in [K]^{n}$. To avoid degeneracy in the model parameters, we 
suppose that $\bm \Theta$  has rank $K^2$. This is equivalent to Assumption $A 1^{\ast}$ in \cite{Noroozi2021Estimation}, which requires that $\bm \lambda^{(k,1)}, \cdots , \bm \lambda^{(k,K)}$ are linearly independent for each $k \in [K]$.

In the PABM, let $\bm \Xi \bm D \bm \Xi^{\top}$ be an eigenvalue decomposition of $\bm \Theta$, where $\bm \Xi
=[\bm \xi_{1\cdot}, \cdots,\bm \xi_{n\cdot}]^{\top}$. Among existing works on the PABM, only \cite{John2023Popularity} further analyzed the structure of $\bm \Xi$. Specifically, their Theorem 2 states that, when $\lambda_{i \bm c^{\ast}(i)} > 0$ for all $i\in[n]$, then $\bm \xi_{i \cdot}^{\top} \bm \xi_{j \cdot}=0$ if and only if $\bm c^{\ast}(i)\neq \bm c^{\ast}(j)$. The logic of their algorithms and theoretical developments relies entirely on this argument. However, we provide Example \ref{cPABMlemma:example:1} to highlight issues in the statement and proof of their Theorem $2$. Notably, their proof only establishes that $\bm \xi_{i \cdot}^{\top} \bm \xi_{j \cdot}=0$ if $\bm c^{\ast}(i)\neq \bm c^{\ast}(j)$, but does not prove the converse that orthogonality necessarily implies different community memberships (i.e., $\bm \xi_{i \cdot}^{\top} \bm \xi_{j \cdot}=0$ leads to $\bm c^{\ast}(i) \neq \bm c^{\ast}(j)$). In fact, when $\lambda_{i \bm c^{\ast}(i)} > 0$ for each $i\in[n]$, we will provide a simple example in Example \ref{cPABMlemma:example:1} to show that there can be many instances where $\bm \xi_{i \cdot}^{\top} \bm \xi_{j \cdot}=0$ despite $\bm c^{\ast}(i)=\bm c^{\ast}(j)$; that is, the eigenvectors corresponding to nodes within the same community can still be orthogonal.

\begin{example}
\label{cPABMlemma:example:1}
Set $n=8$, $K=2$,  
\scalebox{0.88}{$\bm \Lambda = \frac{1}{4} {
\begin{bmatrix}
2 & 2 & 2 & 2 & 1 & 1 & 2 & 2 \\
2 & 2 & 1 & 1 & 4 & 4 & 4 & 4
\end{bmatrix}
}^{\top}$} and $\bm c^{\ast}=(1,1,1,1,2,2,2,2)^{\top}$. Let $\bm{1}_{2}$ be an $2$-dimensional vector whose elements are all $1$. Then, we have 
\begin{eqnarray*}
\label{cPABMlemma:orthogonal:example}
(\bm \xi_{i \cdot}^{\top} \bm \xi_{j \cdot})_{i,j \in [8]}=\frac{1}{2}
\begin{bmatrix}
 \bm 1_{2}\bm 1_{2}^{\top}  & \bm 0_{2 \times 2} & \bm 0_{2 \times 2} & \bm 0_{2 \times 2} \\
 \bm 0_{2 \times 2}  & \bm 1_{2}\bm 1_{2}^{\top} & \bm 0_{2 \times 2} & \bm 0_{2 \times 2} \\
  \bm 0_{2 \times 2} & \bm 0_{2 \times 2}   & \bm 1_{2}\bm 1_{2}^{\top} & \bm 0_{2 \times 2} \\
   \bm 0_{2 \times 2}   & \bm 0_{2 \times 2} & \bm 0_{2 \times 2} & \bm 1_{2}\bm 1_{2}^{\top}\\
\end{bmatrix}.
\end{eqnarray*} 
{Note that 
$\bm \xi_{1 \cdot}^{\top} \bm \xi_{3 \cdot}=\bm \xi_{1 \cdot}^{\top} \bm \xi_{4 \cdot}=\bm \xi_{2 \cdot}^{\top} \bm \xi_{3 \cdot}=\bm \xi_{2 \cdot}^{\top} \bm \xi_{4 \cdot}=0$ even though $\bm c^{\ast}(1)=\bm c^{\ast}(2)=\bm c^{\ast}(3)=\bm c^{\ast}(4)=1$, which implies that $\bm \xi_{i \cdot}^{\top} \bm \xi_{j \cdot} >0$ may not hold even if $\bm c^{\ast}(i)= \bm c^{\ast}(j)$.}  
\end{example}

The absence of a comprehensive and rigorous spectral analysis of $\bm \Xi$ in the PABM has hindered the development of effective spectral clustering algorithms. To address this gap, we present the following proposition, which establishes explicit relationships between the eigenspace $\bm \Xi$ of the edge probability matrix $\bm \Theta$ and the true community labels $\bm c^{\ast}$.

\begin{proposition}
\label{cPABMlemma:orthogonal}
Suppose that the rank of $\bm \Theta$ is $K^2$, and let $\bm \Xi = (\xi_{iu})_{n \times K^{2}}$ be the eigenvector matrix of $\bm \Theta$ corresponding to the $K^{2}$ nonzero eigenvalues. Let $\bm \Xi^{(k)}= ( \xi_{iu})_{i: \bm c^{\ast}(i)=k, u \in [K^{2}]}$ and $\bm \Lambda^{(k,\cdot)}= ( \lambda_{il})_{i: \bm c^{\ast}(i)=k, l \in [K]}$.
Then, for each $k\in[K]$, we have
\begin{eqnarray}
\label{cPABMlemma:orthogonal:Zk}
\bm \Xi^{(k)}= \bm \Lambda^{(k,\cdot)} \bm Z_{k},
\end{eqnarray}
where $\bm Z_{k} \in \mathbf{R}^{K \times K^{2}}$ satisfies that $\bm Z_{k}\bm Z_{l}^{\top}=\bm 0_{K \times K}$ for each $l \neq k$ and $\bm Z_{k}\bm Z_{k}^{\top}=\big( {\bm \Lambda^{(k,\cdot)} }^{\top}\bm \Lambda^{(k,\cdot)}  \big)^{-1}$.
\end{proposition}
Proposition \ref{cPABMlemma:orthogonal} shows that each row of the eigenvector matrix $\bm \Xi$ satisfies $\bm \xi_{i\cdot}^{\top}=\bm \lambda_{i\cdot}^{\top}\bm Z_{\bm c^{\ast}(i)}$, where $\bm \lambda_{i\cdot}$ is the $i$-th row of $\bm \Lambda\in[0,1]^{n \times K}$. This structure underscores the inadequacy of directly applying K-means clustering to the rows of $\bm \Xi$, as also noted in \citet{Noroozi2021Estimation}. Specifically, the Euclidean distance $\|\bm \xi_{i\cdot}-\bm \xi_{j\cdot}\|$ depends not only on the community labels $\bm c^{\ast}(i)$ and $\bm c^{\ast}(j)$, but also on the latent node-specific vectors $\bm \lambda_{i\cdot}$ and $\bm \lambda_{j\cdot}$.

More importantly, Proposition \ref{cPABMlemma:orthogonal} also implies that $\bm \xi_{i \cdot}^{\top} \bm \xi_{j \cdot}=0$ as long as $\bm c^{\ast}(i)\neq \bm c^{\ast}(j)$. This motivates a different perspective based on angular similarity. Specifically, for each $i,j\in[n]$ we define the angle-based similarity between nodes $i$ and $j$ as $$
\tau_{ij} = |\cos\left(\bm \xi_{i \cdot},\bm \xi_{j \cdot}\right)|.$$ 
Although $\tau_{ij}$ is defined in terms of the eigenvector matrix $\bm \Xi$, our analysis will reveal an explicit relationship between $\bm \Xi$,  the parameter matrix $\bm \Lambda$, and the true labels $\bm c^{\ast}$.

For each $k\in[K]$, define $\bm P_{k}=\bm Z_{k} {\bm Z_{k}}^{\top}=\big( {\bm \Lambda^{(k,\cdot)}}^{\top}\bm \Lambda^{(k,\cdot)} \big)^{-1}$, and $D_{\bm P_{k}}(\bm \lambda_{i\cdot},\bm \lambda_{j\cdot})=\bm \lambda_{i\cdot}^{\top} \bm P_{k} {\bm \lambda_{j\cdot}}.$ 
Using Proposition \ref{cPABMlemma:orthogonal}, we directly obtain Corollary \ref{cPABMlemma:tau_ij}, which characterizes the relationship between the angle similarity $\tau_{ij}$ and  the underlying parameters $\bm \Lambda$, $\bm c^{\ast}(i)$, and $\bm c^{\ast}(j)$. This connection clarifies the assumptions for our theoretical guarantees and guides the development of subsequent results.

\begin{corollary}
\label{cPABMlemma:tau_ij}
Under the same conditions of Proposition \ref{cPABMlemma:orthogonal}, for $i,j \in [n]$, we have  
$$
\tau_{ij}
 =
\begin{cases}
\left| \frac{D_{\bm P_{\bm c^{\ast}(i)}}(\bm \lambda_{i\cdot},\bm \lambda_{j\cdot})}{D_{\bm P_{\bm c^{\ast}(i)}}^{1/2}(\bm \lambda_{i\cdot},\bm \lambda_{i\cdot})D_{\bm P_{\bm c^{\ast}(i)}}^{1/2}(\bm \lambda_{j\cdot},\bm \lambda_{j\cdot})} \right|
, & \bm c^{\ast}(i)= \bm c^{\ast}(j); \\
0, & \bm c^{\ast}(i)\neq \bm c^{\ast}(j).
\end{cases}
$$
\end{corollary}

\subsection{Thresholded Cosine Spectral Clustering (\textsf{TCSC})}
\label{cPABMsubsecInitial_Al}

To implement spectral clustering under the PABM, we begin by estimating the angle-based similarity measures $(\tau_{ij})_{i,j \in [n]}$. Specifically, we compute the eigenvectors of the adjacency matrix $\bm A$, and construct the matrix  $\hat{\bm \Xi}$ by selecting the eigenvectors corresponding to the $K^{2}$ largest eigenvalues in absolute value. 
For each $i, j\in[n]$, we estimate $\tau_{ij}$ as $$\hat{\tau}_{ij}=\left|\cos\left(\hat{\bm \xi}_{i \cdot},\hat{\bm \xi}_{j \cdot}\right)\right|,$$
where $\hat{\bm \xi}_{i}$ denotes the $i$-th row of $\hat{\bm \Xi}$. We then define
$\hat{\bm \tau}_{i}=(\hat{\tau}_{i1},\cdots,\hat{\tau}_{in})^{\top}$ for node $i$.

However, applying K-means clustering to $\{\hat{\bm \tau}_{i}\}_{i\in[n]}$ cannot effectively identify the underlying communities. To illustrate this issue, we provide Example \ref{cPABMlemma:example:2} below to show that, although node $8$ belongs to the second community, $\bm \tau_{8}$ is closer to the center of the first community, even using the true similarity measures $\{\bm \tau_{i}\}_{i \in [n]}$, as shown in Table \ref{cPABMTab_example2}.

\begin{example}
\label{cPABMlemma:example:2}
Set $n=8$, $K=2$, \scalebox{0.99}{$\bm \Lambda= \frac{1}{10}
\begin{bmatrix}
1 & 10 & 10 & 8 & 1 & 8 & 6 & 10 \\
2 & 8 & 8 & 10 & 2 & 10 & 8 & 8
\end{bmatrix}^{\top}
$}, and $\bm c^{\ast}=(1,1,1,1,2,2,2,2)^{\top}$.
Let $\bar{\bm \tau}_{k}=\frac{1}{4}\sum_{i:\bm c^{\ast}(i)=k} \bm \tau_{i}$ for $k\in[2]$.
Then, we can compute the distance from $\bm \tau_{i}$, for each $i\in [8]$, to community centers $\bar{\bm \tau}_{1}$ and $\bar{\bm \tau}_{2}$ in the following table.
\begin{table}[H]
\centering
\caption{The distance to two community centers $\bar{\bm \tau}_{1}$ and $\bar{\bm \tau}_{2}$.}
\label{cPABMTab_example2}
\begin{tabular}{c | c c c c c c c c}
  \hline	
$i$	&	$1$	&	$2$	&	$3$	&	$4$	&	$5$&	$6$	&	$7$	& $8$\\
  \hline	
$\|\bm \tau_{i}-\bar{\bm \tau}_{1}\|^{2}$ 	&  $\bm{1.76}$ 	&	$\bm{1.28}$ 	&	$\bm{1.28}$  &	$\bm{0.93}$ &	$3.43$ 	&	$3.37$  &	$3.51$  &	$\bm{1.99}$	\\
$\|\bm \tau_{i}-\bar{\bm \tau}_{2}\|^{2}$	&	$3.43$ 	&	$3.45$ 	&	$3.45$  &	$3.23$ &	$\bm{0.65}$ 	&	$\bm{0.22}$  &	$\bm{0.31}$  &	$2.81$	\\
	\hline										
\end{tabular}
\end{table} 
\end{example}

To address this issue, we define the thresholded cosine measures $\tilde{\bm \tau}_{i}=(\tilde{\tau}_{ij},\cdots,\tilde{\tau}_{in})^{\top}$ with $\tilde{\tau}_{ij}=\mathbb{I} \{\hat{\tau}_{ij} \geq d_{n}\}$  for each $i \in[n]$. This thresholding step is motivated by the observation that, for most node pairs $(i,j)$ within the same community, $\bm \xi_{i \cdot}$ and $\bm \xi_{j \cdot}$ are not orthogonal, and their true similarity $\tau_{ij}$ does not decay rapidly toward zero. To formalize this idea, we define
$\mathcal{G}_i = \left\{ j \in [n]: \bm{c}^{\ast}(j) = \bm{c}^{\ast}(i), \ |\tau_{ij}| \geq \phi_{1,n} \right\}$ for each $i \in [n]$, where $\bm{c}^{\ast}$ denotes the true community assignment, and $\phi_{1,n} \in (0,1]$ is a sequence bounded away from zero.
We assume that there exists a sequence $\phi_{1,n}$ sufficiently large to be robust to noise, such that $
|\mathcal{G}_i| = n_{\bm{c}^{\ast}(i)}(\bm{c}^{\ast})(1 - o(1)),
$
where $n_{\bm{c}^{\ast}(i)}(\bm{c}^{\ast})$ denotes the size of community $\bm{c}^{\ast}(i)$. Given this, we set the threshold as $
d_n = {\phi_{1,n}}/{2},
$
which ensures the following desirable properties: for all $i \in [n]$ and all $j$ such that $\bm{c}^{\ast}(j) \neq \bm{c}^{\ast}(i)$, we have $\tilde{\tau}_{ij} = 0$; and for all $j \in \mathcal{G}_i$, we have $\tilde{\tau}_{ij} = 1$. Consequently, $\tilde{\bm \tau}_{i}$ closely approximates the vector $\bm a_{\bm c^{\ast}(i)}$, where for each $k\in[K]$, $\bm a_{k}=(a_{k}(1),\cdots,a_{k}(n))^{\top}$ with $a_{k}(i)=1$ if $\bm c^{\ast}(i)=k$ and  $a_{k}(i)=0$ otherwise.

After obtaining  $(\tilde{\bm \tau}_{i})_{i\in[n]}$, we apply K-means clustering to recover the community labels $\bm c^{\ast}$. It is known that finding the global optimum for the $K$-means clustering problem is NP-hard. Thus, we employ an approximation algorithm that computes a  $(1+\varepsilon)$-approximate solution in polynomial time, for any fixed constant $\varepsilon>0$. Such algorithms have been studied and theoretically justified in \cite{Kumar2004} and \cite{Jing2015Consistency}. Specifically, we solve the following $(1+\varepsilon)$-approximate $K$-means optimization problem to find some $\hat{\bm c}^{(0)}=\left(\hat{\bm c}^{(0)}(1), \cdots, \hat{\bm c}^{(0)}(n)\right)^{\top} \in[K]^n$, such that
\begin{eqnarray}
\label{cPABMAlgorithmInitialKmeansEq}
\sum_{k=1}^K \min _{\tilde{\bm \nu}_{k} \in \mathbb{R}^n} \sum_{i: \hat{\bm c}^{(0)}(i)=k}\left\|\tilde{\bm{\tau}}_{i}-\tilde{\boldsymbol{\nu}}_k\right\|^2 \leqslant(1+\varepsilon) \min _{\tilde{\boldsymbol{c}} \in [K]^n} \sum_{k=1}^K \min _{\tilde{\bm \nu}_{k} \in \mathbb{R}^n} \sum_{i: \tilde{\bm c}(i)=k}\left\|\tilde{\bm{\tau}}_{i}-\tilde{\boldsymbol{\nu}}_k\right\|^2.
\end{eqnarray}

These three steps are integrated into the proposed Thresholded Cosine Spectral Clustering (\textsf{TCSC}), which is summarized in the following Algorithm \ref{cPABMAlgorithmInitialization}. 

\begin{algorithm}[ht]
\caption{{Thresholded Cosine Spectral Clustering (\textsf{TCSC})}}
\renewcommand{\algorithmicrequire}{\textbf{Input}:
}
\renewcommand{\algorithmicensure}{\textbf{Output}:
}
\begin{algorithmic}[1]
  \REQUIRE Adjacency matrix $\bm A $, and the number of communities $K \geq 2$.
  \ENSURE An estimator $\hat{\bm c}^{(0)} \in[K]^{n}$ of the community label vector $\bm c^{\ast} \in[K]^{n}$.
    \STATE  Estimate the angle-based similarity measures via $(\hat{\tau}_{ij})_{i,j \in[n]}$ with $\hat{\tau}_{ij}=\left|\cos\left(\hat{\bm \xi}_{i \cdot},\hat{\bm \xi}_{j \cdot}\right)\right|$.

    \STATE
    \label{Al:thresholding}
  For $i \in[n]$, solve the thresholded measure $\tilde{\bm \tau}_{i}=(\tilde{\tau}_{ij},\cdots,\tilde{\tau}_{in})^{\top}$ with $\tilde{\tau}_{ij}=\mathbb{I} \{\hat{\tau}_{ij} \geq d_{n}\}$.
    \STATE For $\varepsilon >0$, solve $\hat{\bm c}^{(0)}$ from the $(1+\varepsilon)$-approximation $K$-means problem \eqref{cPABMAlgorithmInitialKmeansEq}. 
\end{algorithmic}
\label{cPABMAlgorithmInitialization}
\end{algorithm}

\begin{figure}[ht!]
  \centering
  \subfigure[$(\hat{\bm \xi}_{i\cdot})_{i \in [n]}$]{
    \label{Al:a} 
    \includegraphics[width=1.98in]{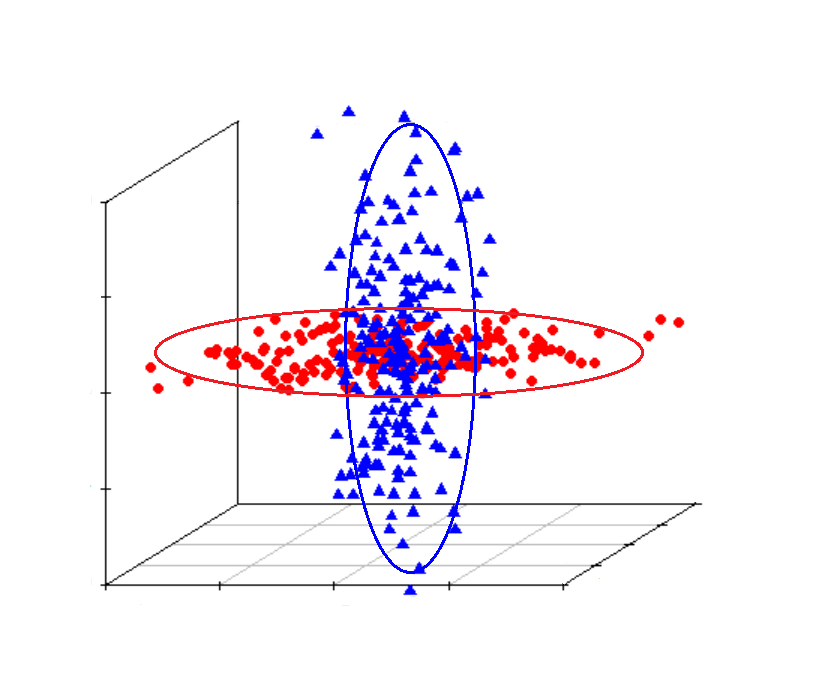}
  }
  \subfigure[$(\hat{\bm{\tau}}_{i})_{i \in [n]}$]{
    \label{Al:b} 
    \includegraphics[width=1.98in]{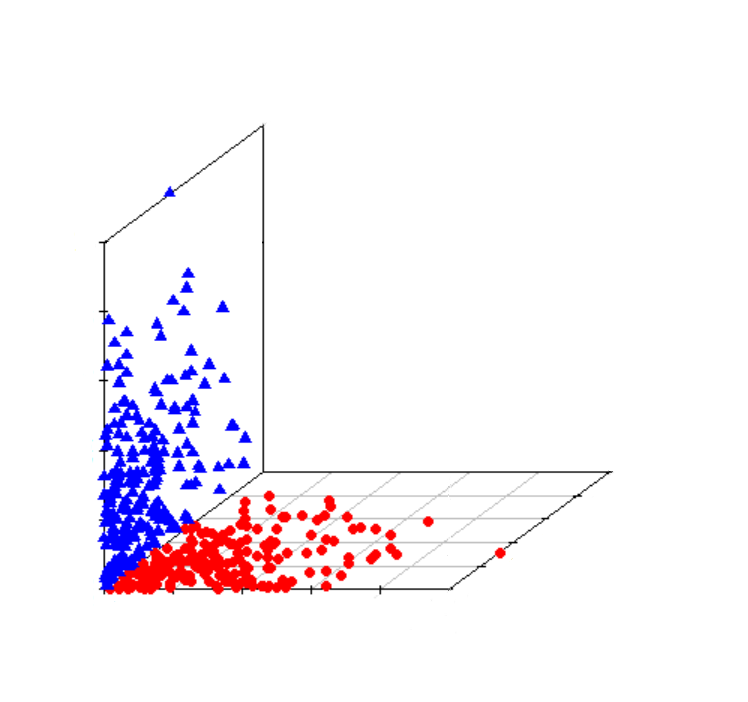}
  }
  \subfigure[$(\tilde{\bm{\tau}}_{i})_{i \in [n]}$]{
    \label{Al:c} 
    \includegraphics[width=1.98in]{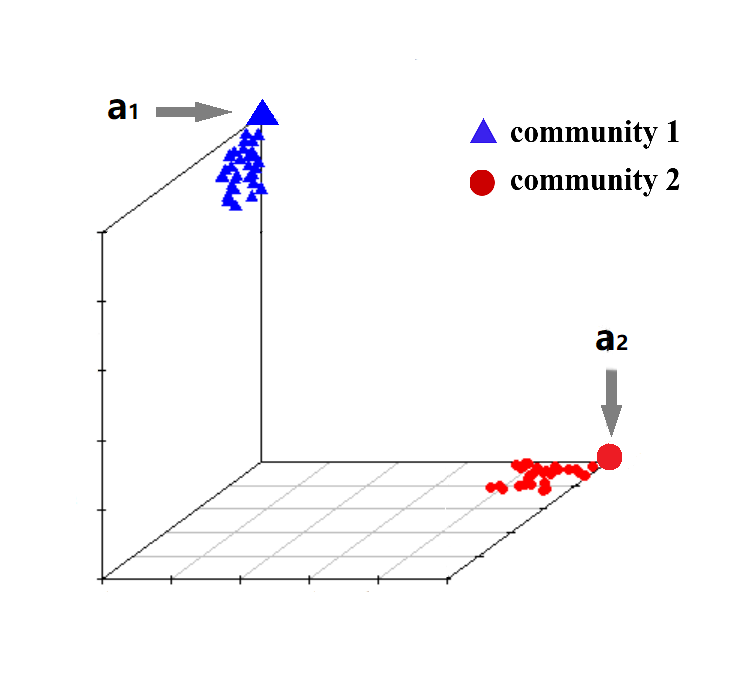}
  } 
  \caption{Conceptual Figure. An illustration  of Algorithm \ref{cPABMAlgorithmInitialization} for $K = 2$. The red circles and blue triangles represent nodes from two different communities. Although $\hat{\bm \xi}_{i\cdot}$, $\hat{\bm{\tau}}_{i}$ and $\tilde{\bm{\tau}}_{i}$ lie in high-dimensional spaces, this figure provides a low-dimensional visualization for intuitive understanding only.
  }
  \label{Al_plot}
\end{figure}

Figure~\ref{Al_plot} provides an intuitive overview of Algorithm \ref{cPABMAlgorithmInitialization}. When the signal-to-noise ratio is sufficiently high, the vectors $\hat{\bm{\xi}}_i$ and $\hat{\bm{\xi}}_j$ corresponding to nodes from different communities are nearly orthogonal, as illustrated in Figure~\ref{Al:a}. We then compute the estimated angular similarity $\hat{\tau}_{ij}$ to capture the similarity between node pairs $(i,j)$. However, as shown in Example~\ref{cPABMlemma:example:2}, even in the noiseless population case, it is generally impossible to directly perform $K$-means clustering on the vectors $\{{\bm{\tau}}_{i}\}_{i\in[n]}$. To overcome this, we apply a thresholding step to $\hat{\bm{\tau}}_i$, resulting in the vectors $\tilde{\bm{\tau}}_i$, which concentrate around the idealized community centers $\bm{a}_{\bm{c}^{\ast}(i)}$, as shown in Figure~\ref{Al:c}. This concentration effect justifies applying $K$-means clustering to the thresholded vectors $\tilde{\bm{\tau}}_i$.

\subsection{Refined Thresholded Cosine Spectral Clustering (\textsf{R-TCSC})}
\label{cPABMsubsecInitial_Al2}

The \textsf{TCSC} provides an effective implementation under the PABM and yields a weakly consistent estimator, as shown in Section \ref{cPABMtheorem}. However, under mild conditions, it does not guarantee exact recovery.  To improve estimation accuracy and achieve strong consistency, we incorporate a refinement step to enhance the accuracy of community label estimation.

Refinements are commonly used to improve initial estimators and have proven effective across various network models. \citet{Gao2015, NIPS2016Yun, Gao2018, Min2020Optimal} employed refinement techniques in the Stochastic Block Model (SBM), Labeled SBM, Degree-Corrected SBM, and Weighted SBM, respectively. The refinement step can be viewed as a node-wise classification task: for each node, its community label is updated by treating the labels of the remaining $ n-1 $ nodes as known. Different refinement strategies vary in how they measure similarity between the target node and the $ K $ communities. Likelihood-based similarity measures are commonly used in prior works such as \cite{Gao2015, NIPS2016Yun, Min2020Optimal}. Due to the complexity of DCSBM, which requires estimating  $ n + K^2 $ parameters in addition to the community labels, \cite{Gao2018}  instead adopted a simpler approach based on edge frequencies.

However, for the more flexible and complex PABM, neither likelihood-based nor edge frequency–based similarity measures are suitable. This is because the PABM involves \( nK \) parameters, exceeding those in the DCSBM, yet it does not impose higher within-community edge densities. Consequently, achieving exact recovery of community labels through refinement under reasonable conditions presents two key challenges:  
(1) developing a refinement strategy with a similarity measure tailored to the PABM; and (2) determining whether such a refinement can upgrade a weakly consistent initializer to strong consistency.

To address these challenges, we first propose an angle-based similarity measure tailored to the structure of the PABM. We then show in Theorem~\ref{cPABMtheorem:oneRefine} that this measure enables exact recovery of the true community labels \( \bm c^{\ast} \)  with high probability through a one-step refinement step. Specially,  suppose for each $i,j\in[n]$ and $k\in[K]$, we have $\lambda_{ik}, \lambda_{jk} \neq 0$, and the set of vectors $\{\bm \lambda^{(k,l)}\}_{l\in[K]}$ is linearly independent, where $\bm \lambda^{(k,l)} = (\lambda_{il})_{i: \bm{c}^{\ast}(i) = k}$. Under these conditions, it can be shown that $\cos(\bm{\theta}_{i \cdot}^{(l)}, \bm{\theta}_{j \cdot}^{(l)})=\cos(\lambda_{il}\bm \lambda^{(l,\bm c^{\ast}(i))}, \lambda_{jl}\bm \lambda^{(l,\bm c^{\ast}(j))})$ equals $1$ if $\bm c^{\ast}(i)=\bm c^{\ast}(j)$ and less than $1$ if $\bm c^{\ast}(i)\neq \bm c^{\ast}(j)$.
Let $\bm{\theta}_{i\cdot}^{(l)} = (\theta_{ij})_{j:\bm{c}^{\ast}(j)=l}$ be the edge probability vector between node $i$ and nodes in community $l$. Define $\bar{\bm{\theta}}^{(k,l)}=\sum_{v: \bm c^{\ast}(v)=k} \bm{\theta}_{v \cdot}^{(l)}/n_{k}(\bm c^{\ast})$ for each $k \in [K]$. We have that $\cos(\bm{\theta}_{i \cdot}^{(l)}, \bar{\bm{\theta}}^{(k,l)})=\cos(\lambda_{il}\bm \lambda^{(l,\bm c^{\ast}(i))}, \frac{1}{n_{k}(\bm c^{\ast})}\bm{1}_{n_{k}(\bm c^{\ast})}^{\top}\bm \lambda^{(k,l)} \bm \lambda^{(l, k)})$ equals $1$ if $\bm c^{\ast}(i)=k$ and less than $1$ if $\bm c^{\ast}(i)\neq k$,
where $\bm{1}_{m}$, $m\in\mathbb{N}_+$, denotes a vector whose elements are all $1$. If the community labels of all nodes except $i$ are known and denoted $\bm{c}_{-i}^{\ast}$, the label $\bm{c}^{\ast}(i)$ can be recovered by solving 
\begin{eqnarray}
\label{cPABMMutiRefinement:population}
\underset{k \in [K]}{\arg \max} \sum_{l=1}^{K} \cos\left(\bm{\theta}_{i \cdot}^{(l)}, \bar{\bm{\theta}}^{(k,l)}\right).
\end{eqnarray}
The population-level insights motivate refining community assignments by maximizing the aggregated inter-cluster cosine similarity between each node and the estimated community centers, using the initial label estimate $ \hat{\bm{c}}^{(0)} = ( \hat{\bm{c}}^{(0)}(1), \ldots, \hat{\bm{c}}^{(0)}(n) )^{\top} $ obtained from Algorithm \ref{cPABMAlgorithmInitialization}.

{Based on $\hat{\bm c}^{(0)}$, we estimate $\bm{\theta}_{i \cdot}^{(l)}= (\theta_{ij})_{j: \bm c^{\ast}(j) = l}$ by $\bm A_{i\cdot}^{(l)}(\hat{\bm c}^{(0)})=(A_{ij})_{j:\hat{\bm c}^{(0)}(j)=l}$, and estimate $\bar{\bm{\theta}}^{(k,l)}=\sum_{v: \bm c^{\ast}(v)=k} \bm{\theta}_{v \cdot}^{(l)}/n_{k}(\bm c^{\ast})$ by $\bar{\bm{A}}^{(k,l)}(\hat{\bm c}^{(0)})=\sum_{v: \hat{\bm c}^{(0)}(v)=k} \bm{A}_{v \cdot}^{(l)}(\hat{\bm c}^{(0)})/n_{k}(\hat{\bm c}^{(0)})$, replacing $\bm{\Theta}$ with $\bm{A}$ and $\bm c^{\ast}$ with $\hat{\bm c}^{(0)}$. 
} 
Each node $i$ updates its community label by solving:
\begin{equation}
\label{cPABMMutiRefinement_0}
\hat{\bm{c}}^{(1)}(i) = \underset{k \in [K]}{\operatorname{argmax}} 
\sum_{l=1}^{K}
\frac{
\bm{A}_{i\cdot}^{(l)}(\hat{\bm{c}}^{(0)})^\top \bar{\bm{A}}^{(k,l)}(\hat{\bm c}^{(0)})
}{
\|\bm{A}_{i\cdot}^{(l)}(\hat{\bm{c}}^{(0)})\| \|\bar{\bm{A}}^{(k,l)}(\hat{\bm c}^{(0)})\|
}.
\end{equation}
Applying this update to all nodes yields the one-step refined estimate $\hat{\bm{c}}^{(1)}$.

Before proceeding, we introduce the leave-one-out version $\bar{\bm{A}}_{-i}^{(k,l)}(\hat{\bm c}^{(0)})$, a modification of $\bar{\bm{A}}^{(k,l)}(\hat{\bm c}^{(0)})$ that excludes the contribution from node $i$. It is formally defined as
$\bar{\bm{A}}_{-i}^{(k,l)}(\hat{\bm c}^{(0)}) ={n_{k}^{-1} (\hat{\bm{c}}^{(0)})}\sum_{j\in[n]} \bm{A}_{j\cdot}^{(l)}(\hat{\bm c}^{(0)}) \, \mathbb{I}\{\hat{\bm{c}}^{(0)}(j)=k, \, j\neq i\}.
$ This construction ensures that, $\bar{\bm{A}}_{-i}^{(k,l)}(\tilde{\bm c})$ is independent of $\bm A_{i\cdot}^{(l)}(\tilde{\bm c})$ for any fixed $\tilde{\bm c} \in [K]^{n}$, thereby simplifying the theoretical analysis. Thus, the refinement update is modified as:
\begin{equation}
\hat{\bm c}^{(1)}(i) = \underset{k \in [K]}{\operatorname{argmax}}
\sum_{l=1}^{K}
\frac
{
\bm A_{i\cdot}^{(l)}(\hat{\bm c}^{(0)})^{\top}
\bar{\bm{A}}_{-i}^{(k,l)}(\hat{\bm c}^{(0)})
}
{
\|\bm A_{i\cdot}^{(l)}(\hat{\bm c}^{(0)})\|
\|\bar{\bm{A}}^{(k,l)}(\hat{\bm c}^{(0)})\|
}.
\end{equation}

Integrating both initialization and refinement steps, the proposed One-step Refined Thresholded Cosine Spectral Clustering (\textsf{R-TCSC}) Algorithm is presented as Algorithm \ref{cPABMAlgorithmOneRefine}. 

{
\begin{algorithm}[H]
\caption{The One-Step Refined Thresholded Cosine Spectral Clustering (\textsf{R-TCSC})}
\renewcommand{\algorithmicrequire}{\textbf{Input}:}
\renewcommand{\algorithmicensure}{\textbf{Output}:}
\begin{algorithmic}[1]
\REQUIRE Adjacency matrix $\bm{A}$, number of communities $K \geq 2$.
\ENSURE A refined estimator $\hat{\bm{c}}:= \hat{\bm{c}}^{(1)} \in [K]^n$ of the community label vector $\bm c^{\ast} \in[K]^{n}$.

\STATE Apply Algorithm \ref{cPABMAlgorithmInitialization} to obtain the initial label vector $\hat{\bm{c}}^{(0)}$. 
\STATE 
{ For each $i \in [n]$, update each $\hat{\bm{c}}^{(1)}(i)$ by maximizing the cosine similarity: 
\begin{equation}
\label{cPABMMutiRefinement}
\hat{\bm c}^{(1)}(i) = \underset{k \in [K]}{\operatorname{argmax}}
\sum_{l=1}^{K}
\frac
{
\bm A_{i\cdot}^{(l)}(\hat{\bm c}^{(0)})^{\top}
\bar{\bm{A}}_{-i}^{(k,l)}(\hat{\bm c}^{(0)})
}
{
\|\bm A_{i\cdot}^{(l)}(\hat{\bm c}^{(0)})\|
\|\bar{\bm{A}}^{(k,l)}(\hat{\bm c}^{(0)})\|
}.
\end{equation}
}
\end{algorithmic}
\label{cPABMAlgorithmOneRefine}
\end{algorithm}

\begin{figure}[!ht]
  \centering
  \subfigure[Before the $1$st refinement]{
    \label{Al:Refine:a} 
    \includegraphics[width=2.68in]{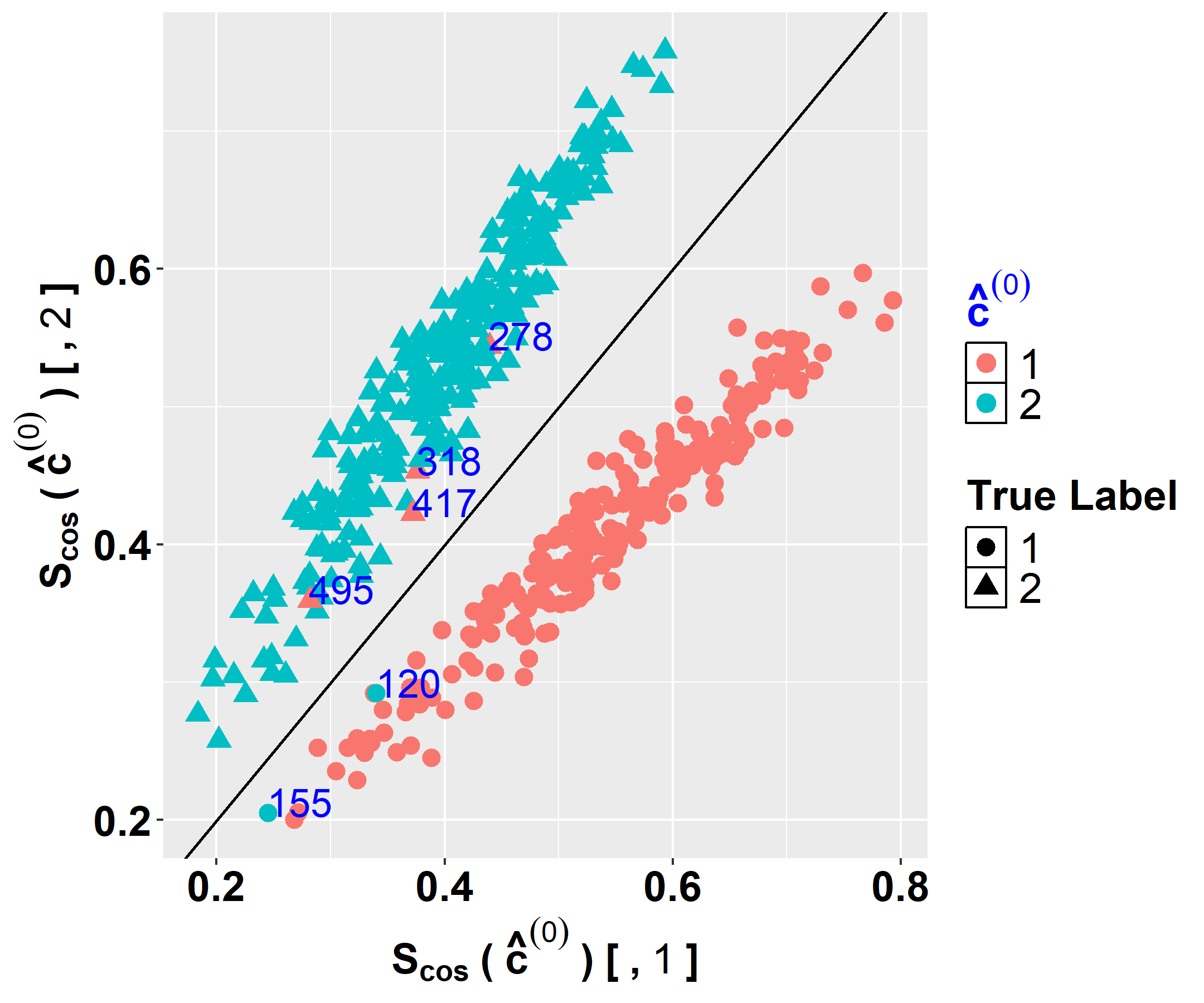}
  }
  \subfigure[After the $1$st refinement]{
    \label{Al:Refine:b} 
    \includegraphics[width=2.68in]{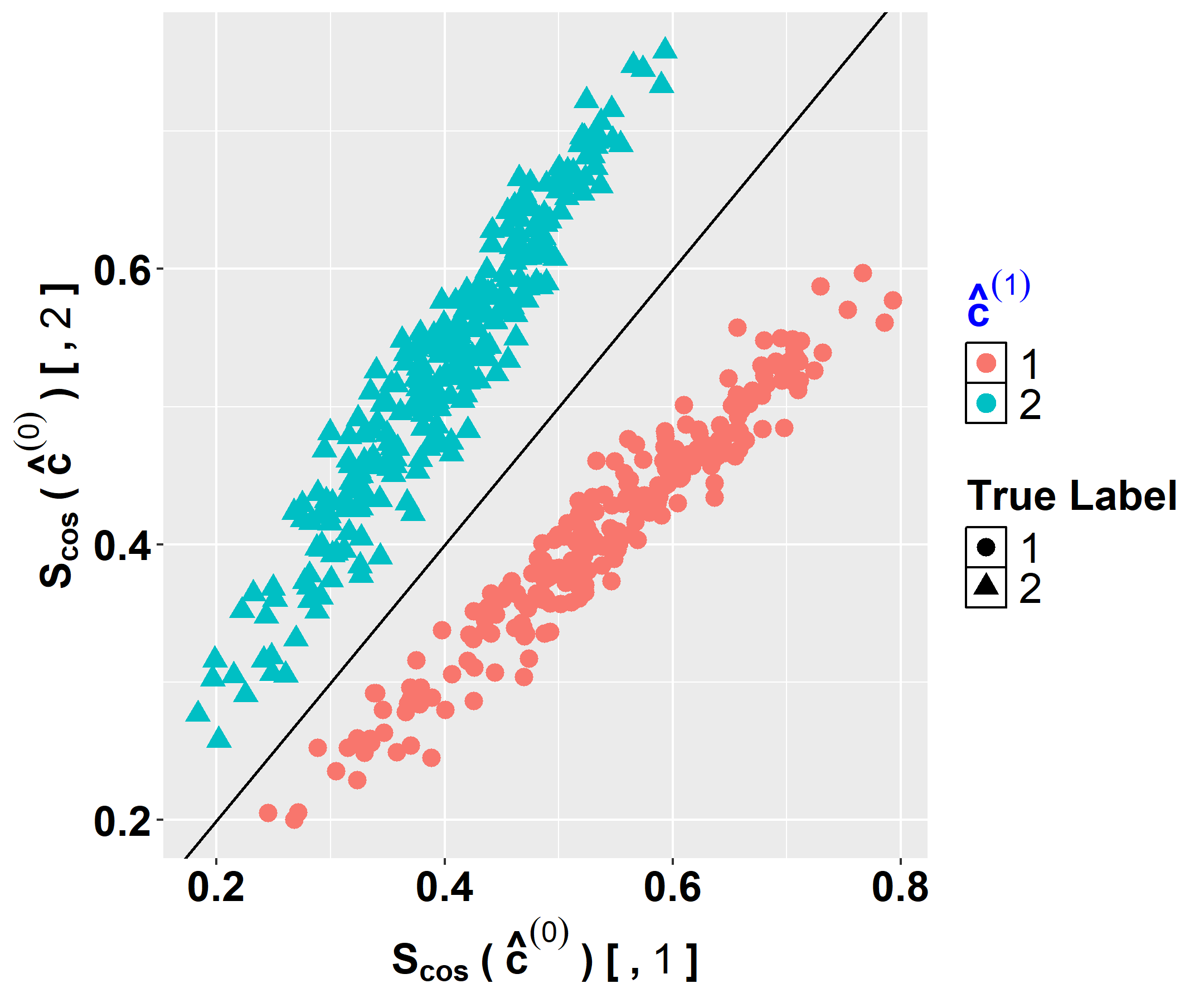}
  }
  \caption{An illustration of the one-step refinement for a network of $512$ nodes with $K=2$ and incorrectly estimated node numbers marked in each plot.}
  \label{Refine_plot}
\end{figure}

Figure~\ref{Refine_plot} illustrates the refinement process in Algorithm \ref{cPABMAlgorithmOneRefine} for a network with $512$ nodes. The one-step \textsf{R-TCSC}  improves the initial estimate and successfully corrects all $6$ initially misclustered nodes. We define the similarity score between node $i$ and community $k$ as $S_{\cos,ik}(\hat{\bm c}^{(0)})= \sum_{l \in [K]}\cos(\bm A_{i\cdot}^{(l)}(\hat{\bm c}^{(0)}), \bar{\bm{A}}^{(k,l)}(\hat{\bm c}^{(0)}))$, leading to the update $\hat{\bm c}^{(1)}(i)=\underset{k \in [2]}{\arg \max} \;\, S_{\cos,ik}(\hat{\bm c}^{(0)})$. 
Let $\bm S_{\cos}(\hat{\bm c}^{(1)})=(S_{\cos,ik}(\hat{\bm c}^{(1)}))_{i\in[n],k \in [K]}$, with columns $ \bm S_{\cos}(\hat{\bm c}^{(1)})[,1]$, $ \bm S_{\cos}(\hat{\bm c}^{(1)})[,2]$ representing similarity scores to each community. The diagonal line of Figure~\ref{Refine_plot} serves as the decision boundary between communities. Figures~\ref{Al:Refine:a} and \ref{Al:Refine:b} show that all misclustered nodes ($120$, $155$, $278$, $318$, $417$ and $495$) are correctly clustered after the one-step refinement.

Extensive numerical studies (see Section~\ref{S_onestep_multistep}) confirm that when $n$ is sufficiently large (e.g., $n = 2^{10}$), the one-step refinement $\hat{\bm{c}}^{(1)}$ reliably recovers the true community labels, aligned with the strong consistency result established for Algorithm~\ref{cPABMAlgorithmOneRefine} in Theorem~\ref{cPABMtheorem:oneRefine}.

\begin{algorithm}[H]
\caption{Two-Step Refined Thresholded Cosine Spectral Clustering (\textsf{R-TCSC})}
\renewcommand{\algorithmicrequire}{\textbf{Input}:}
\renewcommand{\algorithmicensure}{\textbf{Output}:}
\begin{algorithmic}[1]
\REQUIRE Adjacency matrix $\bm{A}$, number of communities $K \geq 2$.
\ENSURE A refined estimator $\hat{\bm{c}}:= \hat{\bm{c}}^{(2)} \in [K]^n$ of the community label vector $\bm c^{\ast} \in[K]^{n}$.

\STATE Apply Algorithm~\ref{cPABMAlgorithmOneRefine} to obtain the estimated label vector $\hat{\bm{c}}^{(0)}$ via one-step refinement.
\STATE 
For each $i \in [n]$, update each $\hat{\bm{c}}^{(2)}(i)$ by maximizing the cosine similarity: 
\begin{equation}
\label{cPABMMutiRefinement2}
\hat{\bm c}^{(2)}(i) = \underset{k \in [K]}{\operatorname{argmax}}
\sum_{l=1}^{K}
\frac
{
\bm A_{i\cdot}^{(l)}(\hat{\bm c}^{(1)})^{\top}
\bar{\bm{A}}_{-i}^{(k,l)}(\hat{\bm c}^{(1)})
}
{
\|\bm A_{i\cdot}^{(l)}(\hat{\bm c}^{(1)})\|
\|\bar{\bm{A}}^{(k,l)}(\hat{\bm c}^{(1)})\|
}.
\end{equation}
\end{algorithmic}
\label{cPABMAlgorithmMutiRefine}
\end{algorithm}

However, when the sample size is relatively small (e.g., $n = 2^{7}$ or $2^{8}$), applying an additional refinement step (i.e., two-step refinement) can further enhance the accuracy of community label recovery beyond that of the one-step procedure. This motivates the design of Algorithm~\ref{cPABMAlgorithmMutiRefine}, which builds on Algorithm~\ref{cPABMAlgorithmOneRefine} by incorporating an extra refinement step.

Our work shares similar philosophy with the one-step estimation in the maximum likelihood estimation \citep{bickel1975one} and the one-step and two-step local linear approximation solutions in folded concave penalized estimation \citep{fan2014strong}. Theorem~\ref{cPABMtheorem:MutiRefine} provides a theoretical justification for Algorithm~\ref{cPABMAlgorithmMutiRefine}: compared to the one-step refinement, the two-step refinement significantly accelerates the convergence rate of the clustering error. When the sample size $n$ is large, the effect is negligible; however, for smaller $n$, the improvement becomes substantial, which is consistent with the simulation results in Section~\ref{S_onestep_multistep}.

\section{Theoretical Properties}\label{cPABMtheorem}
This section gives upper bounds on the community labeling error rate for the initialization algorithm \textsf{TCSC} and the refinement algorithm \textsf{R-TCSC}, respectively.

The following assumptions ensure that the true labels $\bm c^{\ast} \in [K]^{n}$ are estimable. Recall that $n_{k}(\bm c^{\ast})=\sum_{i=1}^{n} \mathbb{I} \{\bm c^{\ast}(i) = k\}$, $\forall k \in [K]$. Assumption \ref{cPABMassump_2} ensures that the $K$ communities do not degenerate as $n$ grows, i.e., the ratio $n_{k}(\bm c^{\ast})/n_{l}(\bm c^{\ast})$ 
does not vanish for any $k$, $l\in[K]$. This condition is standard in the theoretical works on community detection (see, e.g., \cite{Harrison2016,Gao2018,NIPS2016Yun,ariu2023instanceoptimal}).

\begin{assumption}
\label{cPABMassump_2}
There exists a constant $C\geq 1$, such that $\frac{n}{CK} \leq n_{k}(\bm c^{\ast})\leq \frac{Cn}{K}$, $\forall k\in[K]$.
\end{assumption}

Recall that $\bm \Lambda = (\lambda_{ik})_{n \times K} \in [0,1]^{n \times K}$, 
where $\lambda_{ik}$ captures the popularity of node $i$ in community $k$. {Suppose \( \bm{\Lambda} = \rho_n \bm{\Lambda}_0 \), where \( \bm{\Lambda}_0 \in (0,1)^{n \times K} \) is a constant matrix and \( \rho_n \in (0,1] \) is the sparsity parameter, following \citet{Chen2018Network}. This implies the edge probability matrix \( \bm{\Theta} = \rho_n^{2} \bm{\Theta}_0 \) for a constant matrix \( \bm{\Theta}_0 \). As $\rho_n$ decreases, connectivity becomes sparser, reducing available information and making community detection more difficult.} We relax the conditions from \cite{Chen2018Network}, which required $n \rho_n^4 \gg (\log n)^2$ and a constant $K$, to the weaker condition in Assumption \ref{cPABMassump_initial_pho}.

\begin{assumption}
\label{cPABMassump_initial_pho}
The sparsity parameter $\rho_{n}$ satisfies
$n \rho_{n}^{2}/K^{2} \geq C (\log n)^{2}$,
for a large constant $C>0$.
\end{assumption}

Recall that $\tau_{ij}=0$ for each $i,j$ with $\bm c^{\ast}(i) \neq \bm c^{\ast}(j)$, and $
\tau_{ij}
 =
| \frac{D_{\bm P_{\bm c^{\ast}(i)}}(\bm \lambda_{i\cdot},\bm \lambda_{j\cdot})}{D_{\bm P_{\bm c^{\ast}(i)}}^{1/2}(\bm \lambda_{i\cdot},\bm \lambda_{i\cdot})D_{\bm P_{\bm c^{\ast}(i)}}^{1/2}(\bm \lambda_{j\cdot},\bm \lambda_{j\cdot})} |$
for each $i,j$ with $\bm c^{\ast}(i)= \bm c^{\ast}(j)$ as shown in Corollary \ref{cPABMlemma:tau_ij}.
{Accordingly, we impose the following assumptions on $\{\tau_{ij}\}_{i,j \in [n]}$ to ensure that nodes within the same community remain sufficiently similar to be reliably identified.
}

\begin{assumption}
\label{cPABMassump_initial_1}
For each $i \in [n]$, define
$\mathcal{G}_{i}=
\left\{j \in [n]: \bm c^{\ast}(j)=\bm c^{\ast}(i), 
|\tau_{ij}|
\geq \phi_{1,n}\right\},
$ for a sequence $\phi_{1,n} \in (0,1]$. Let $\mathcal{H}=\{i \in [n]: |\mathcal{G}_{i}|\geq n_{\bm c^{\ast}(i)}(\bm c^{\ast})(1-\tilde{\eta}_{n, 0}) \}$ and $\bar{\mathcal{H}} = [n]\setminus \mathcal{H}$. Assume that $|\bar{\mathcal{H}}|\leq \tilde{\eta}_{n, 1}n$, for two sequences $\tilde{\eta}_{n, 1}=o((n \rho_{n}^{2})^{-1}(\log n)^{2})$ 
and $\tilde{\eta}_{n, 0}=o((n \rho_{n}^{2})^{-1/2}\log n)$.
\end{assumption}

Note that $\mathcal{G}_{i}$ in Assumption \ref{cPABMassump_initial_1} can be written as $\{j \in [n]: \bm c^{\ast}(j)=\bm c^{\ast}(i), |\cos(\bm \xi_{i \cdot}, \bm \xi_{j \cdot})|\geq \phi_{1,n}\}$, meaning that a major portion of node pairs within the same community occupy non-orthogonal positions in the eigenspace (i.e., $\bm \xi_{i \cdot}$, $\bm \xi_{j \cdot}$). 
The set \( \mathcal{H} \) consists of nodes with relatively clear community signals, generally aligned with a major subset of their peers, while the small size of \( \bar{\mathcal{H}} \) allows for a few nodes to have weak or ambiguous affiliations.

\begin{assumption}
\label{cPABMassump_initial_2}
Let $\mathcal{I}=\{i \in [n]: D_{\bm P_{\bm c^{\ast}(i)}}^{1/2}(\bm \lambda_{i\cdot},\bm \lambda_{i\cdot}) \geq \phi_{2,n} \}$ for a sequence $\phi_{2,n}$. Assume that $|\mathcal{I}|>n(1-\tilde{\eta}_{n, 2})$
for some $\tilde{\eta}_{n, 2}=o((\log n)^{2}(n \rho_{n}^{2})^{-1})$, 
and $ \max_{i \in [n]}D_{\bm P_{\bm c^{\ast}(i)}}(\bm \lambda_{i\cdot},\bm \lambda_{i\cdot}) \lesssim K^{2}/n$.
\end{assumption}

Note that $D_{\bm P_{\bm c^{\ast}(i)}}^{1/2}(\bm \lambda_{i\cdot},\bm \lambda_{i\cdot})=\|\bm \xi_{i \cdot}\|$, so
$\mathcal{I}$ in Assumption \ref{cPABMassump_initial_2} can be written as $\{i \in [n]: \|\bm \xi_{i \cdot}\|\geq \phi_{2,n} \}$. {This ensures that the eigenvector rows retain sufficient information to capture the spatial positions of the nodes.
}  
Moreover, the condition $ \max_{i \in [n]}D_{\bm P_{\bm c^{\ast}(i)}}(\bm \lambda_{i\cdot},\bm \lambda_{i\cdot}) \lesssim K^{2}/n=\|\bm \Xi\|_{\textrm{F}}^{2}/n$ is well justified: in SBM, \cite{Jing2015Consistency} showed that $\|\bm U_{i \cdot}\|^{2}\asymp K/n=\|\bm U\|_{\textrm{F}}^{2}/n$ under Assumption \ref{cPABMassump_2}, where $\bm U \bm \Gamma \bm U^{\top}$ is the eigendecomposition of its $ \mathbb{E}\bm A$, and $\bm U_{i\cdot}$ denotes its $i$-th row.

We now present an upper bound on the error rate of the \textsf{TCSC} in Theorem \ref{cPABMtheorem:Initial}.

\begin{theorem} [Error Rate of the \textsf{TCSC}]
\label{cPABMtheorem:Initial}
Assume that $\bm \Theta$ has $K^2$ nonzero singular values $\sigma_{1,n} \geq \cdots \geq \sigma_{K^2,n}$, with ${\sigma_{1,n}}/{\sigma_{K^2,n}} \leq C_1 \sqrt{\log n}$ for some constant $C_1 > 0$. {Suppose Assumptions \ref{cPABMassump_2}-\ref{cPABMassump_initial_pho} hold, and Assumptions \ref{cPABMassump_initial_1}-\ref{cPABMassump_initial_2} are satisfied for some $\phi_{1,n}$ and $\phi_{2,n}$ such that,} for a large constant $C_2 > 0$,
\begin{eqnarray}
\label{cPABMassump_initial_eq1}
\frac{K}{\sqrt{n}} \gtrsim \phi_{1,n} \phi_{2,n} \geq C_2 \frac{\rho_n K \sqrt{\log n}}{\sigma_{K^2,n}}.
\end{eqnarray}
If we set $d_n = {\phi_{1,n}}/{2}$ in Algorithm \ref{cPABMAlgorithmInitialization}, there exists a constant $C > 0$ such that
\begin{equation}
\label{cPABMassump_initial_eq2}
 \mathbb{P}
\left\{
\operatorname{\ell}(\bm c^\ast, \hat{\bm c}^{(0)}) \leq \tilde{\eta}_n
\right\} \geq 1 - C n^{-4},
\end{equation}
for a sequence $\tilde{\eta}_n = o(1)$, 
and $\hat{\bm c}^{(0)}$ is the output of Algorithm \ref{cPABMAlgorithmInitialization}.
\end{theorem}

{
Condition~\eqref{cPABMassump_initial_eq1} holds under mild conditions. Specifically, by the assumptions, we have $\sigma_{1,n}\geq \|\bm \Theta\|_{\textrm{F}}/K \geq Cn\rho_{n}^{2}/K$ and $\sigma_{K^{2},n} \geq C \sigma_{K^{2},1}/\sqrt{\log n} \geq Cn\rho_{n}^{2}/(K\sqrt{\log n})$. Together with Assumption~\ref{cPABMassump_initial_pho}, we can obtain ${\rho_n K \sqrt{\log n}}/{\sigma_{K^2,n}} \lesssim (K/\sqrt{n})\{K \log n/(n \rho_{n}^{2})^{1/2}\} \leq CK/\sqrt{n}$ for some small constant $C>0$.
}

Theorem~\ref{cPABMtheorem:Initial} establishes that Algorithm~\ref{cPABMAlgorithmInitialization} yields a weakly consistent estimator of $\bm c^\ast$ under mild conditions. Next, we will show that applying the refinement via Algorithm \ref{cPABMAlgorithmOneRefine}  upgrades this weakly consistent estimator to a strongly consistent one.

To establish strong consistency, we introduce two additional assumptions. 

\begin{assumption}
\label{cPABMassump_delta}
For each $l \in [K]$ and all distinct $k, k' \in [K]$ with $k \neq k^{\prime}$, the popularity vectors satisfy that $|\cos(\bm \lambda^{(l,k)}, \bm \lambda^{(l,k^{\prime})})|\leq 1- \delta$ for some constant $\delta>0$.
\end{assumption}

Assumption \ref{cPABMassump_delta} ensures the detectability of the true community labels $\bm c^{\ast}$. Similar assumptions are standard in the PABM literature—for instance, \citet{Noroozi2021Estimation} require the set $\{\bm \lambda^{(k,l)}\}_{l \in [K]}$ to be linearly independent for each $k \in [K]$. Notably, we can allow $\delta$ to approach zero at a certain rate, which just requires strengthening other assumptions.

Intuitively, the success of Algorithm~\ref{cPABMAlgorithmOneRefine} depends on the quality of the initial estimates. The initialization must provide a reasonably accurate starting point. We show that a weakly consistent initial estimator satisfying a mild error condition is sufficient, provided Assumptions~\ref{cPABMassump_initial_pho}–\ref{cPABMassump_initial_2} are moderately strengthened. Under this condition, the refinement procedure in Algorithm~\ref{cPABMAlgorithmOneRefine} substantially reduces the error rate. We formalize these strengthened conditions in Assumption \ref{cPABMCondition1}.
{
\begin{assumption}
\label{cPABMCondition1}
(Strengthened Assumption~\ref{cPABMassump_initial_pho}).
\begin{eqnarray}
\label{cPABMgammaMulti_1}
  \frac{K\log n}{ n\rho_{n}^{4}} = o(1) \; \text{and} \; K\leq {n}^{1/6}.
\end{eqnarray}
(Strengthened Assumptions~\ref{cPABMassump_initial_1} and \ref{cPABMassump_initial_2}). $\tilde{\eta}_{n,1},\tilde{\eta}_{n,2} = o(1)\min\{(n \rho_{n}^{2})^{-1}(\log n)^{2}, {\rho_n^4} / K^2\}$, and $\tilde{\eta}_{n,0} = o(1)\min\{(n \rho_{n}^{2})^{-1/2} \log n, \rho_n^4 / K\}$.
\end{assumption} 
The strengthened assumptions remain quite mild. For instance, under the PABM, \cite{Chen2018Network} required $n\rho_n^4 \gg (\log n)^2$ with fixed $K$, while \cite{Noroozi2021Estimation} showed in their Corollary ~1 that for $K = 2$, weak consistency of their community label estimator requires $n\rho_n^6 \to \infty$.

We then show that a strongly consistent estimate of the community label vector $\bm c^{\ast}$ can be obtained via one-step \textsf{R-TCSC} in Theorem~\ref{cPABMtheorem:oneRefine}.

\begin{theorem} [Error Rate of the One-Step \textsf{R-TCSC}]
\label{cPABMtheorem:oneRefine}
If Assumptions \ref{cPABMassump_2}, \ref{cPABMassump_delta} and \ref{cPABMCondition1} hold, 
then there exist positive constants $C_1$ and $C_2$ such that
\begin{equation}
\label{cPABMThRACEineq_1}
 \mathbb{P}
 \left\{
 \underset{\pi : [K]\rightarrow [K]}{\cup}
 \left(\hat{\bm c}^{(1)} = \pi[\bm c^{\ast}]\right)
 \right\}
 > 1 - C_1 n^{-(1 + C_2)},
\end{equation}
where $\hat{\bm c}^{(1)}$ is the output of Algorithm \ref{cPABMAlgorithmOneRefine}, and $\pi[\bm c^{\ast}] = \left(\pi(\bm c^{\ast}(1)), \cdots, \pi(\bm c^{\ast}(n))\right)^{\top}$.
\end{theorem}

Theorem~\ref{cPABMtheorem:oneRefine} establishes that the one-step \textsf{R-TCSC} refines the weakly consistent estimate of the community label vector $\bm{c}^\ast$ into a strongly consistent one, achieving exact recovery with high probability. And all the simulation results in Section~\ref{cPABMsec:simulation_1} is consistent it.

However, as shown in Section \ref{cPABMsubsecInitial_Al2}, two-step refinement provides additional benefits when the number of nodes $n$ is small, motivating a theoretical investigation of Algorithm~\ref{cPABMAlgorithmMutiRefine}. 
In fact, it follows from the proof of Theorem~\ref{cPABMtheorem:oneRefine} that $$
\mathbb{P}\left\{
n \ell(\bm c^{\ast},  \hat{\bm c}^{(1)} ) =o(1)
 \right\}
\geq
1-C_{1}n^{-(1+C_{2})}
,
$$
for two constants $C_{1}, C_{2} >0$, as $n \to \infty$.
We then show the two-step \textsf{R-TCSC} achieves an additional gain in convergence rate over the one-step \textsf{R-TCSC} in Theorem~\ref{cPABMtheorem:MutiRefine}.

\begin{theorem} [Additional gain in convergence rate of the Two-Step \textsf{R-TCSC}]
\label{cPABMtheorem:MutiRefine}
Under the conditions of Theorem \ref{cPABMtheorem:oneRefine}, 
there exist constants $ C$ and $C'> 0$ such that
\begin{equation}
\label{cPABMThRACEineq}
 \mathbb{P}
 \left\{
{\ell(\bm c^{\ast}, \hat{\bm c}^{(2)} )} =o\left(\frac{1}{n \rho_n^2}\right)
 \right\}
 > 1 - C n^{-(1 + C')},
\end{equation}
where $\hat{\bm c}^{(2)}$ denote the outputs of Algorithm~\ref{cPABMAlgorithmMutiRefine},
{
and there exist a constant $C_0>0$ such that
\begin{equation}
\label{cPABMThRACEineq00}
 \mathbb{P}
 \left\{
 \underset{t=1,2}{\bigcap}
\left(\frac{\ell(\bm c^{\ast}, \hat{\bm c}^{(t)} )}{\ell(\bm c^{\ast}, \hat{\bm c}^{(t-1)} )} < \frac{C_0}{n \rho_n^2}\right)
 \right\}
 > 1 - C n^{-(1 + C')},
\end{equation}
where $\hat{\bm c}^{(0)}$, $\hat{\bm c}^{(1)}$ and $\hat{\bm c}^{(2)}$ denote the outputs of Algorithms~\ref{cPABMAlgorithmInitialization}, \ref{cPABMAlgorithmOneRefine}, and \ref{cPABMAlgorithmMutiRefine}, respectively.
}
\end{theorem}
As $n \to \infty$, $n\ell(\bm c^{\ast}, \hat{\bm c}^{(1)} )=o(1) < 1$ guarantees the strong consistency of the one-step refinement. However, when $n$ is finite, the convergence rate implied by $o(1)$ may be insufficiently fast, whereas $\ell(\bm c^{\ast}, \hat{\bm c}^{(2)} )=o(1/({n \rho_n^2}))$ represents a much faster rate. In particular, when $\rho_n$ is fixed, we have $\ell(\bm c^{\ast}, \hat{\bm c}^{(2)} )=o(1/n)$. This explains why two-step refinement can lead to additional gains when $n$ is small, while the improvement over one-step refinement becomes negligible as $n$ increases, as corroborated by the empirical results presented in Section~\ref{S_onestep_multistep}. Thus, to accommodate small-sample settings, we employ the two-step \textsf{R-TCSC} algorithm in all simulation studies and real-data analyses. {Moreover, inequality~\eqref{cPABMThRACEineq00} indicates that the clustering error rate decreases monotonically and geometrically with each refinement step ($t = 1, 2$), which is also consistent with the simulation results in Section~\ref{S_onestep_multistep}, where the first refinement step yields a larger error reduction than the second refinement step when $n$ is not so large.
}
}}

\section{Simulation Study}\label{cPABMsec:simulation}
In this section, we compare the performance of the proposed $\textsf{TCSC}$ and $\textsf{R-TCSC}$ against several competitors: 
{
$\textsf{EP}$ that optimizes the likelihood modularity under the PABM \citep{Chen2018Network},  
$\textsf{SSC-A}$ that uses using the rows of $\bm A$ \citep{Noroozi2021Estimation}, and $\textsf{OSC}$, $\textsf{SSC-ASE}$ \citep{John2023Popularity} that use the spectral embedding of $\bm A$. For $\textsf{EP}$, since only the algorithm for $K=2$ is provided by \citet{Chen2018Network}, we compare with it only in case $K=2$. 
For the other competing methods, we use the publicly available source code from \citet{John2023Popularity}, following their recommended configurations and default settings. 
}
To adaptively select the threshold $d_{n}$ in Algorithm \ref{cPABMAlgorithmInitialization}, we compute all values
$(\hat{\tau}_{ij})_{1\leq i < j \leq n}$ and choose the point of the steepest drop in their frequency histogram as the threshold. We use the two-step $\textsf{R-TCSC}$, Algorithm \ref{cPABMAlgorithmMutiRefine}. 
{
Furthermore, in practice, we replace $\bar{\bm{A}}_{-i}^{(k,l)}(\hat{\bm c}^{(t)})$ with $\bar{\bm{A}}^{(k,l)}(\hat{\bm c}^{(t)})$ in ~\eqref{cPABMMutiRefinement} and~\eqref{cPABMMutiRefinement2} for efficiency, as the difference is negligible.}

\subsection{Comparison of Basic Settings}
\label{cPABMsec:simulation_1}
To compare the effectiveness of different methods, we use the community detection error defined in \eqref{cPABMLoss}. 
We consider the settings with $n=128,256,512,1024$ nodes and $K=2,3,4$ communities, and each experiment is repeated $100$ times. In each simulation, community labels $\bm c^{\ast}(1), \ldots, \bm c^{\ast}(n)$ are drawn 
\emph{i.i.d.} from a multinomial distribution with parameters $\left\{1,(\pi_1, \ldots, \pi_K)^{\top}\right\}$. The popularity vectors $\{\bm \lambda^{(k, l)}\}_{k,l \in [K]}$ are sampled from two different joint distributions, depending on whether $k=l$ or $k \neq l$. The edge probability matrix $\bm \Theta =(\theta_{ij})_{n \times n}$ is constructed as $\theta_{ij}=\lambda_{i \bm c^{\ast}(j)}\lambda_{j \bm c^{\ast}(i)}$, the upper triangular entries of the adjacency matrix \( (A_{ij})_{1 \leq i < j \leq n} \) are independently drawn from \( \operatorname{Bernoulli}(\theta_{ij}) \), and the full adjacency matrix is obtained by symmetrization and $A_{ii}=0$.

\begin{figure}[ht!]
  \centering
  \subfigure[$K=2$]{
    \label{cPABMfig1:subfig:a} 
    \includegraphics[width=1.98in]{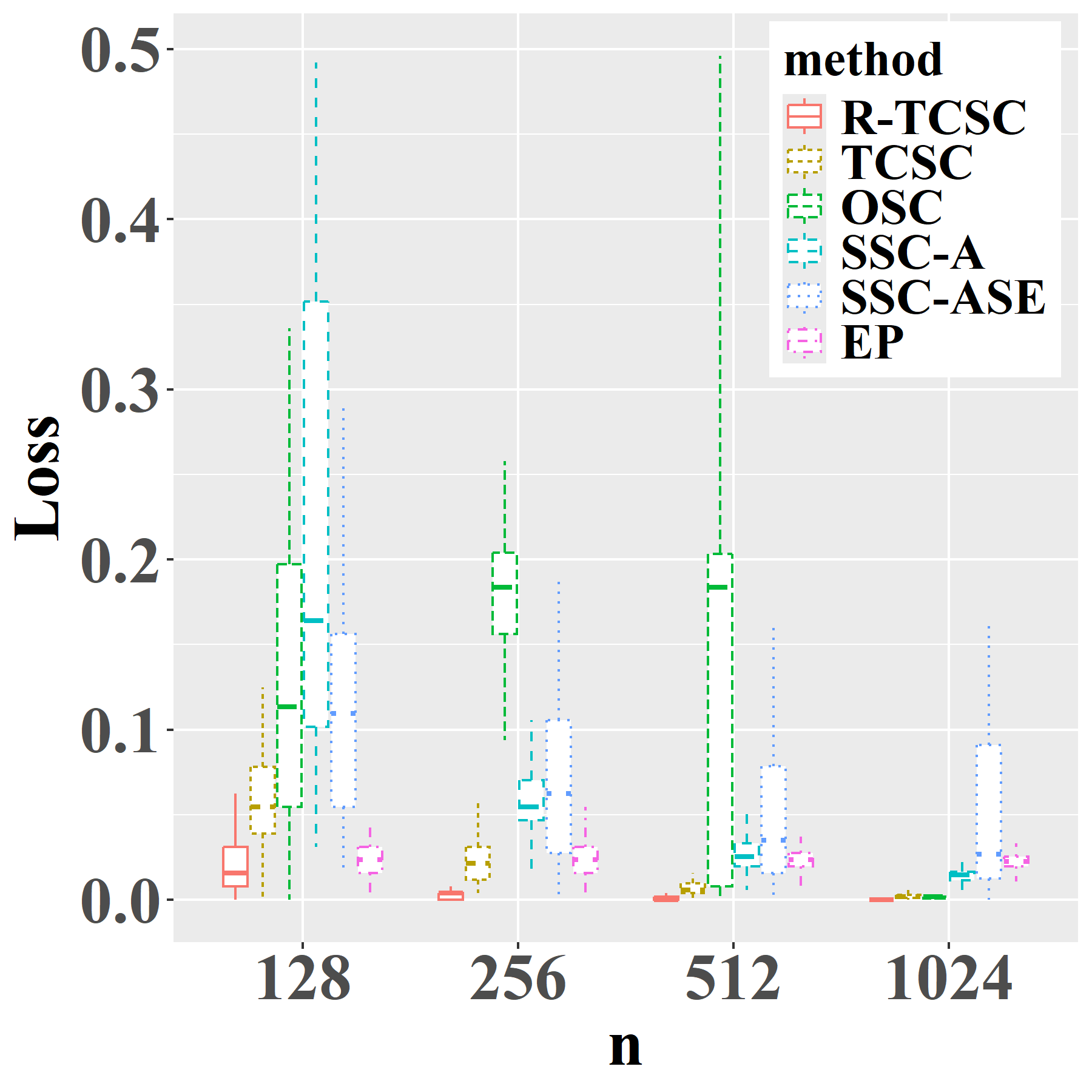}
  }
  \subfigure[$K=3$]{
    \label{cPABMfig1:subfig:b} 
    \includegraphics[width=1.98in]{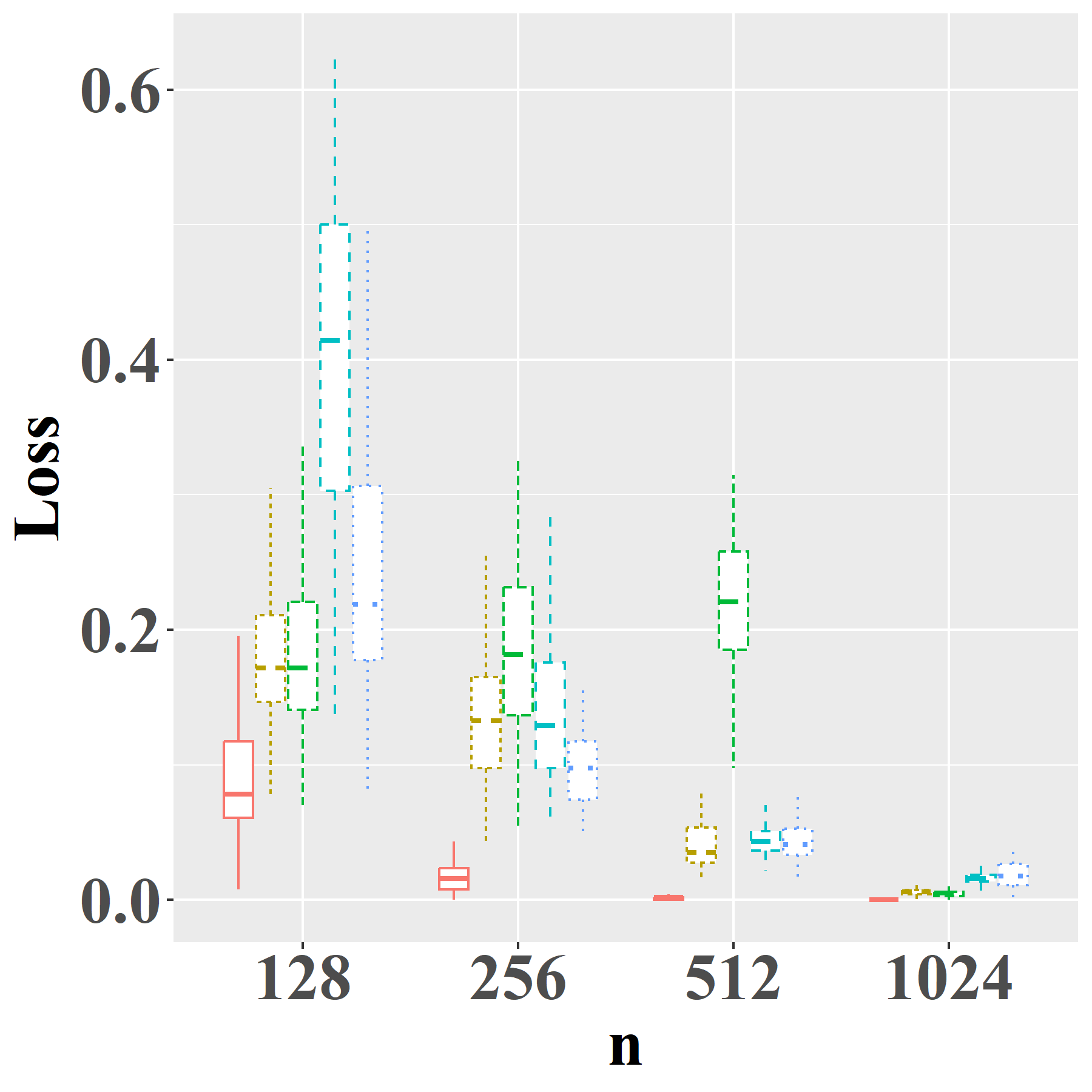}
  }
  \subfigure[$K=4$]{
    \label{cPABMfig1:subfig:c} 
    \includegraphics[width=1.98in]{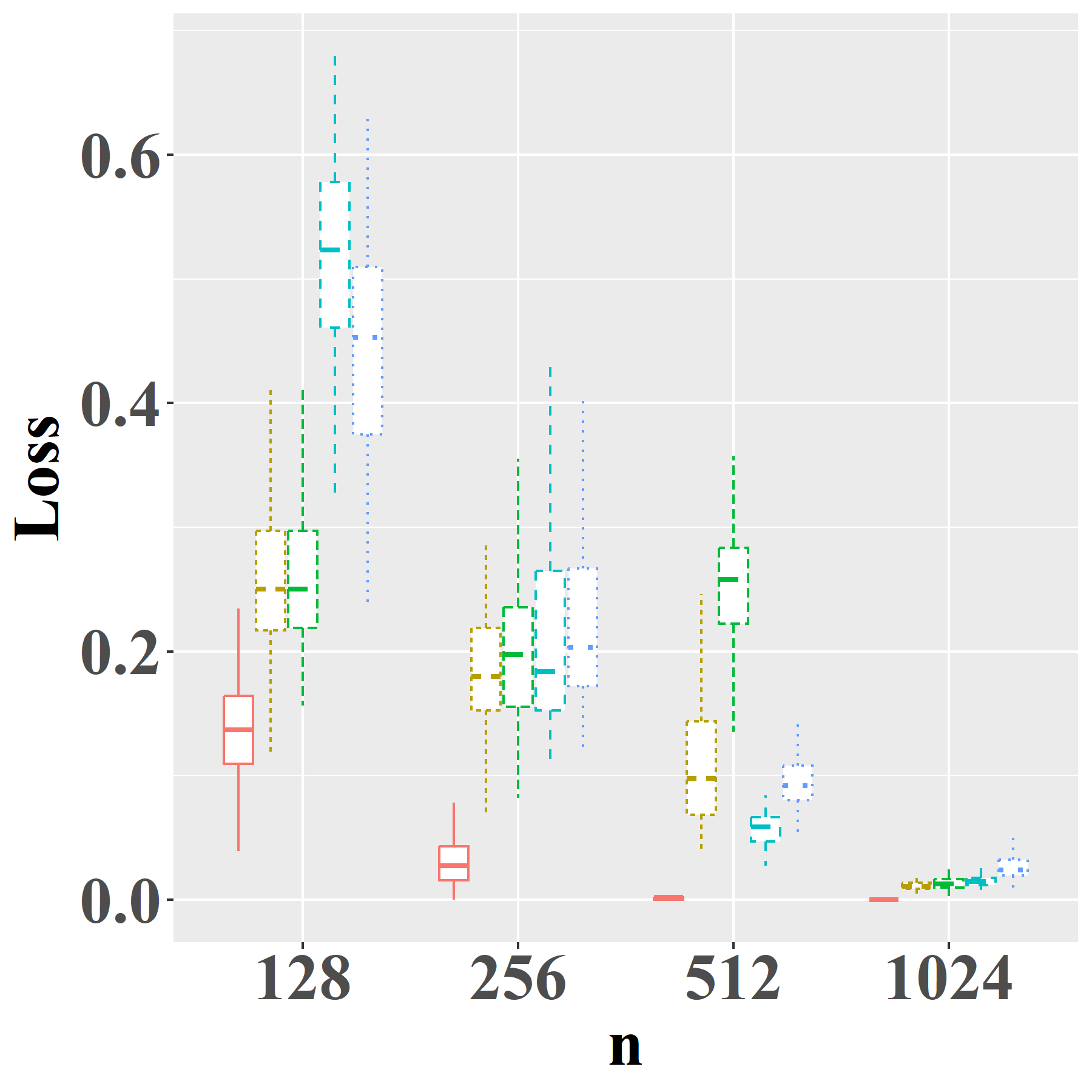}
  }
  \caption{Comparison of community detection performance under balanced communities.}
  \label{cPABMn_change_ab2112}
\end{figure}

First, we consider the case where community sizes are balanced: $\pi_k=1 / K$ for $k \in [K]$. The popularity parameters are generated as follows: intra-community parameters $\bm \lambda^{(k, k)} \stackrel{\text{ iid }}{\sim} \operatorname{Beta}(2,1)$ and inter-community parameters $\bm \lambda^{(k, l)} \stackrel{\text{ iid }}{\sim} \operatorname{Beta}(1,2)$ for $k \neq l \in [K]$. Figure~\ref{cPABMn_change_ab2112} presents the results for community detection under balanced community sizes, where the vertical axis shows the proportion of misclustered nodes as defined in \eqref{cPABMLoss}. The figure illustrates two key findings: (i) the error rates of both \textsf{TCSC} and \textsf{R-TCSC} decrease as $n$ increases, consistent with Theorems~\ref{cPABMtheorem:Initial} and \ref{cPABMtheorem:MutiRefine}; (ii) across all settings, \textsf{TCSC} and \textsf{R-TCSC} outperform competing methods, with \textsf{R-TCSC} rapidly converging to zero error.

\begin{figure}[ht!]
  \centering
  \subfigure[$K=2$]{
    \label{cPABMfig2:subfig:a} 
    \includegraphics[width=1.98in]{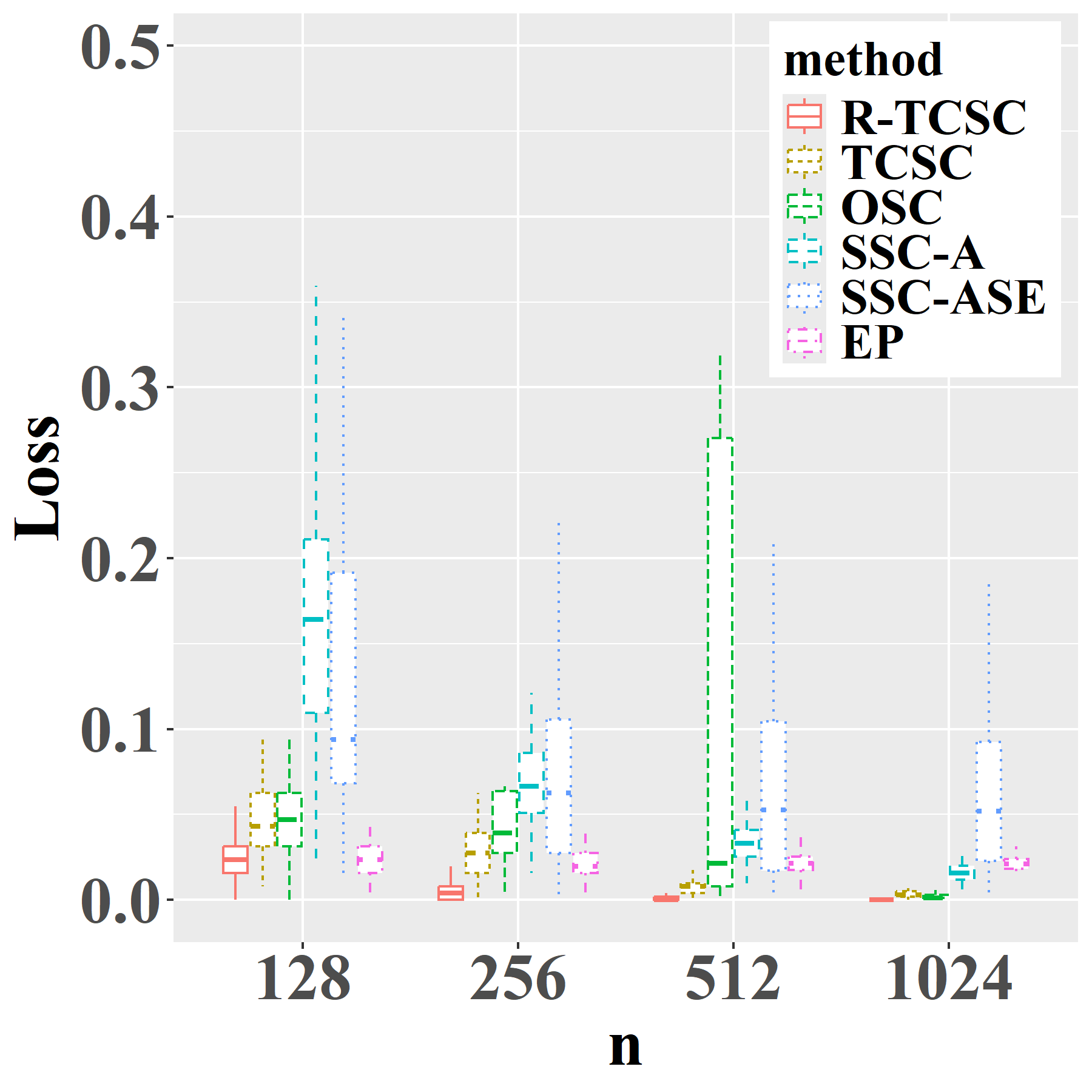}
  }
  \subfigure[$K=3$]{
    \label{cPABMfig2:subfig:b} 
    \includegraphics[width=1.98in]{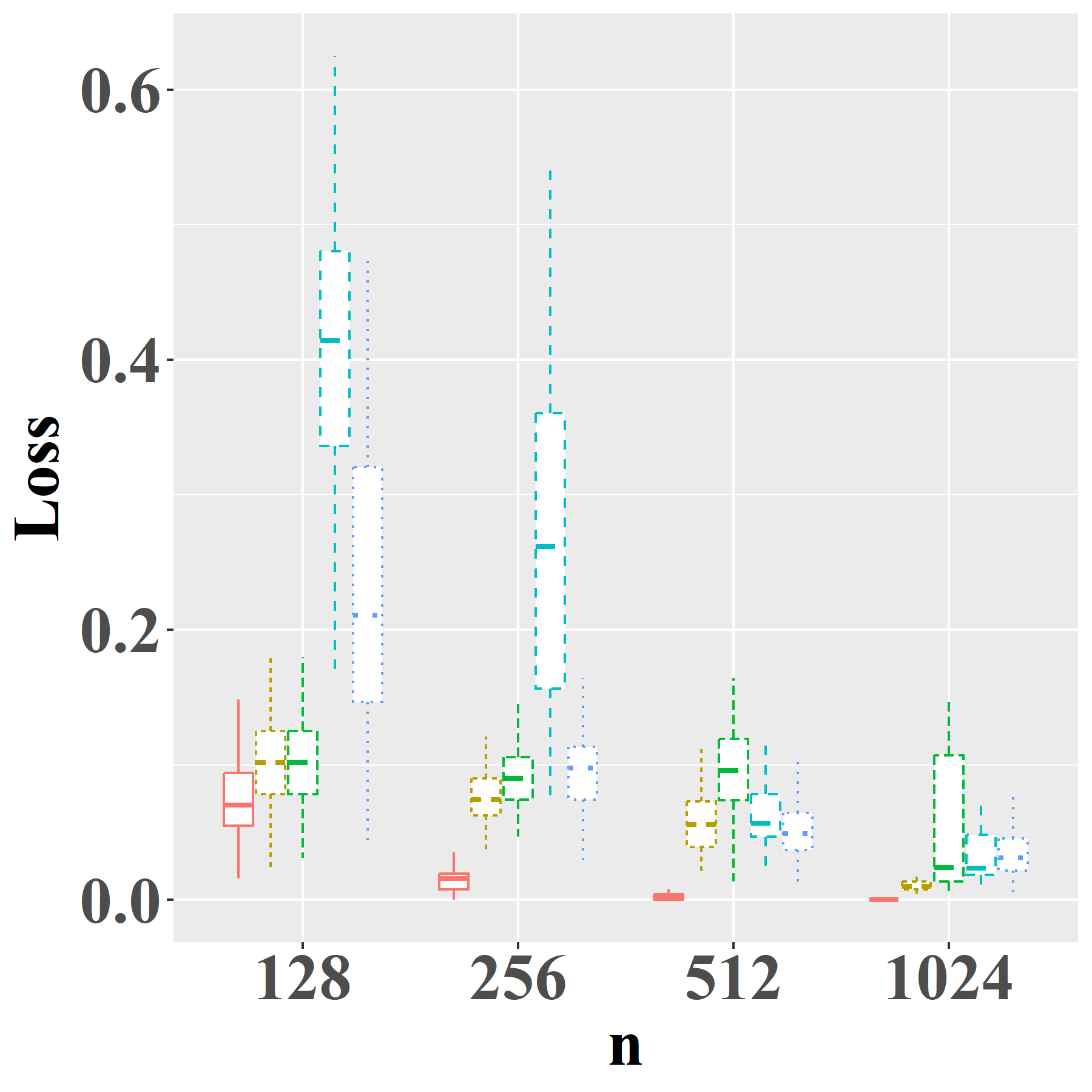}
  }
  \subfigure[$K=4$]{
    \label{cPABMfig2:subfig:c} 
    \includegraphics[width=1.98in]{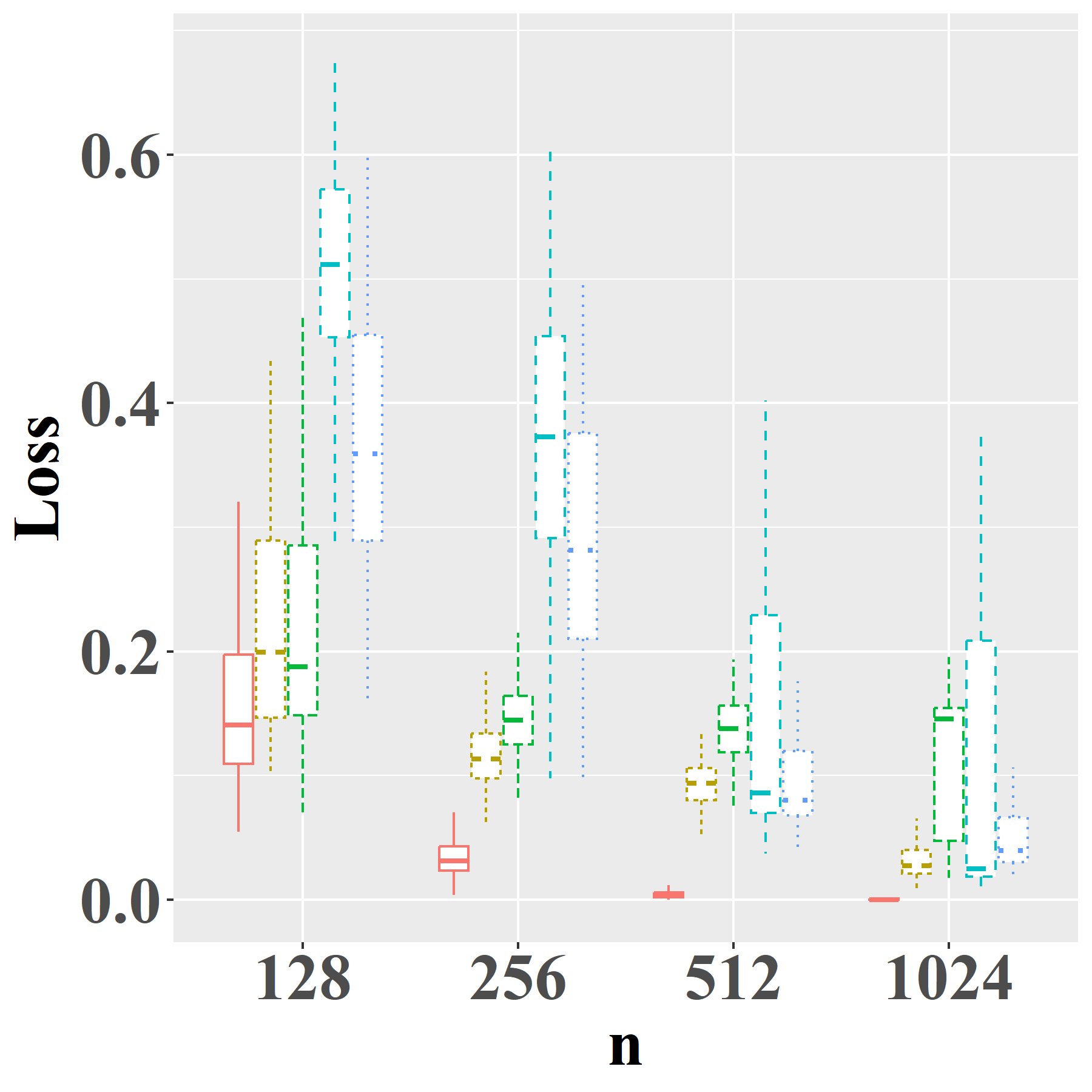}
  }
  \caption{Comparison of community detection performance under imbalanced communities.}
  \label{cPABMn_change_ab2112_unbalanced}
\end{figure}

Second, we consider the case of imbalanced community sizes: $\pi_k={k^{-1}}/({\sum_{\ell=1}^K \ell^{-1}})$ for $k \in [K]$. The popularity parameters are generated as in the balanced setting. The community detection results, shown in Figure \ref{cPABMn_change_ab2112_unbalanced}, lead to two main observations: (i), similar to the balanced case (Figure~\ref{cPABMn_change_ab2112}), both \textsf{TCSC} and \textsf{R-TCSC} outperform competing methods under imbalanced settings; (ii) by comparing the results to $K = 3, 4$, and $n = 1024$ across the two figures (Figures~\ref{cPABMn_change_ab2112} and \ref{cPABMn_change_ab2112_unbalanced}), we observe that \textsf{R-TCSC} exhibits more stable performance than the other methods as community sizes become more imbalanced.

\begin{figure}[ht!]
  \centering
  \subfigure[$K=2$]{
    \label{cPABMfig3:subfig:a} 
    \includegraphics[width=1.98in]{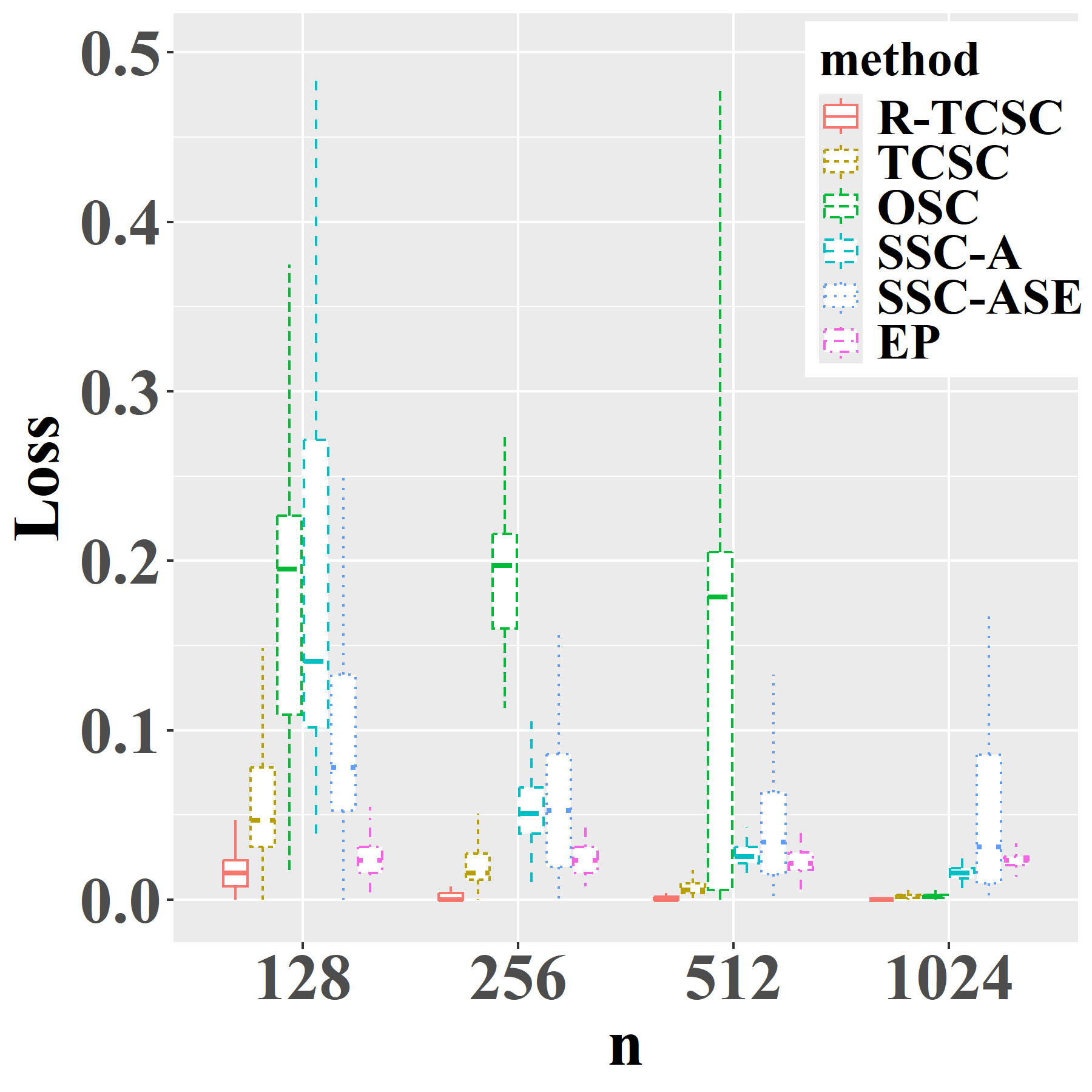}
  }
  \subfigure[$K=3$]{
    \label{cPABMfig3:subfig:b} 
    \includegraphics[width=1.98in]{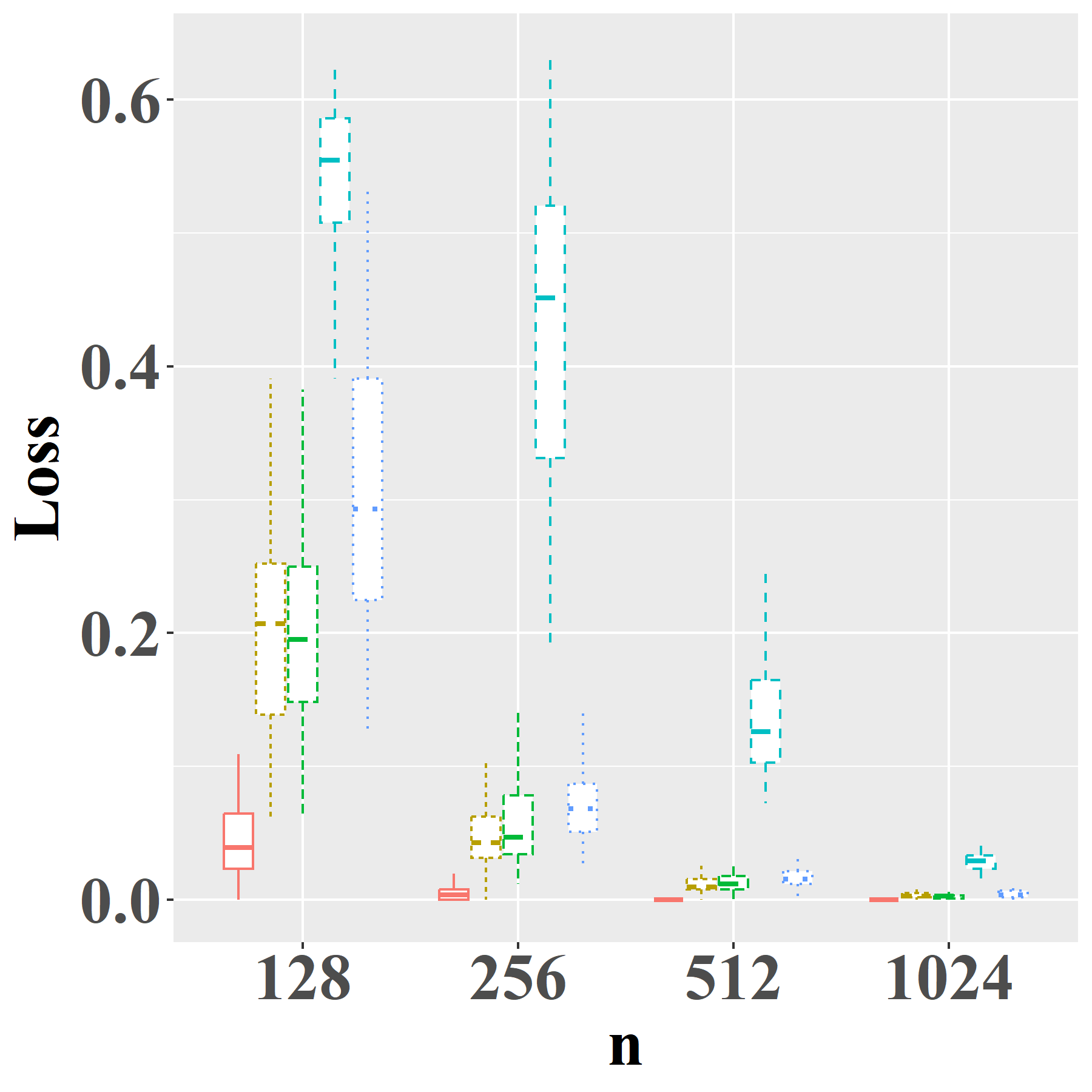}
  }
  \subfigure[$K=4$]{
    \label{cPABMfig3:subfig:c} 
    \includegraphics[width=1.98in]{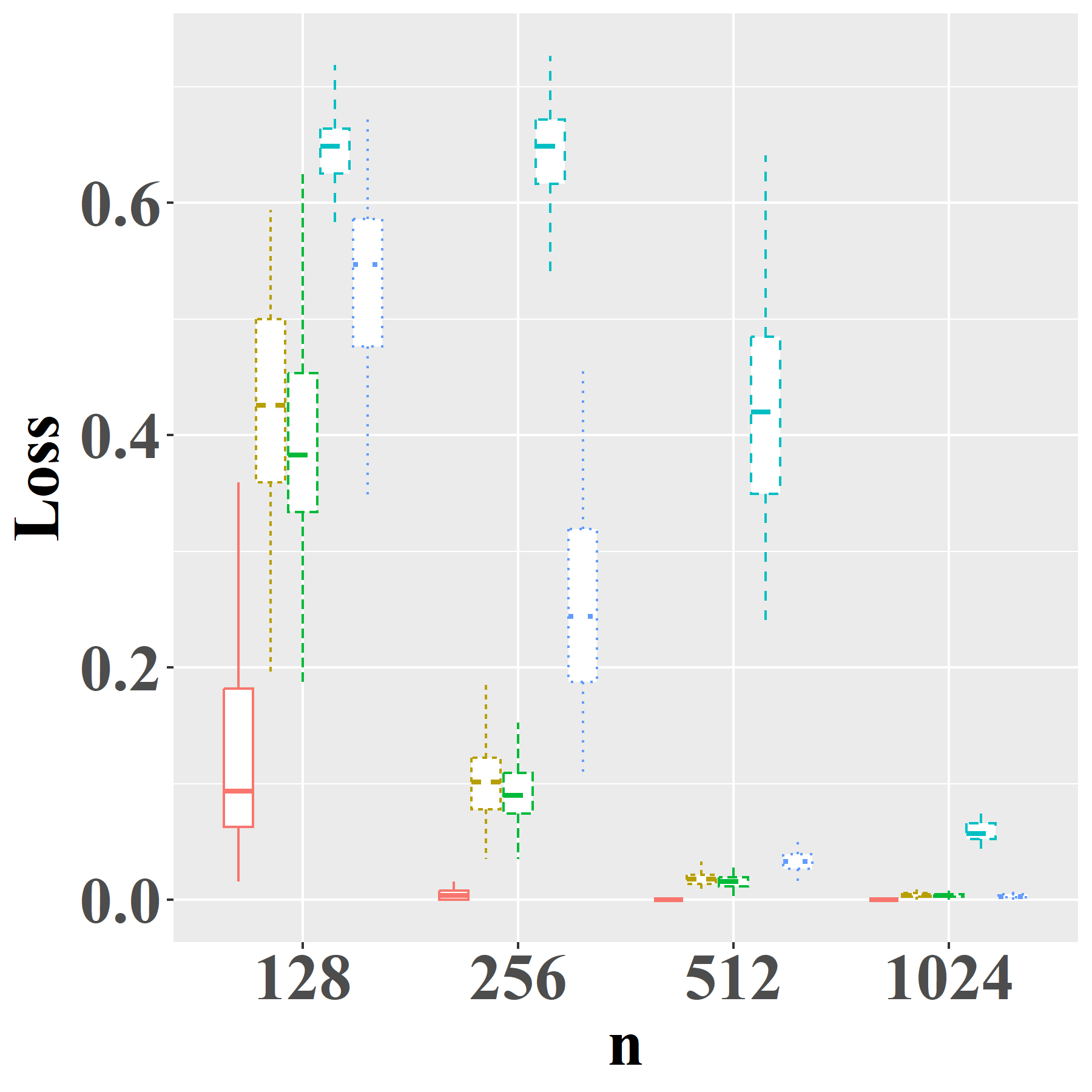}
  }
  \caption{Comparison of community detection performance under disassortative models.}
  \label{cPABMn_change_ab1221}
\end{figure}

Third, we consider the disassortative case, where intra-community popularity parameters are weaker than inter-community popularity parameters. Specifically, the popularity parameters are generated as follows: $\bm \lambda^{(k, k)}$ are drawn independently from $\operatorname{Beta}(1,2)$ and $\bm \lambda^{(k, l)}$ for $k \neq l \in [K]$ are drawn independently from  $\operatorname{Beta}(2,1)$. As shown in Figure \ref{cPABMn_change_ab1221},
$\textsf{R-TCSC}$ maintains strong and stable performance even in the disassortative case. 

\begin{figure}[ht!]
  \centering
  \subfigure[$K=2$]{
    \label{cPABMfig4:subfig:a} 
    \includegraphics[width=1.98in]{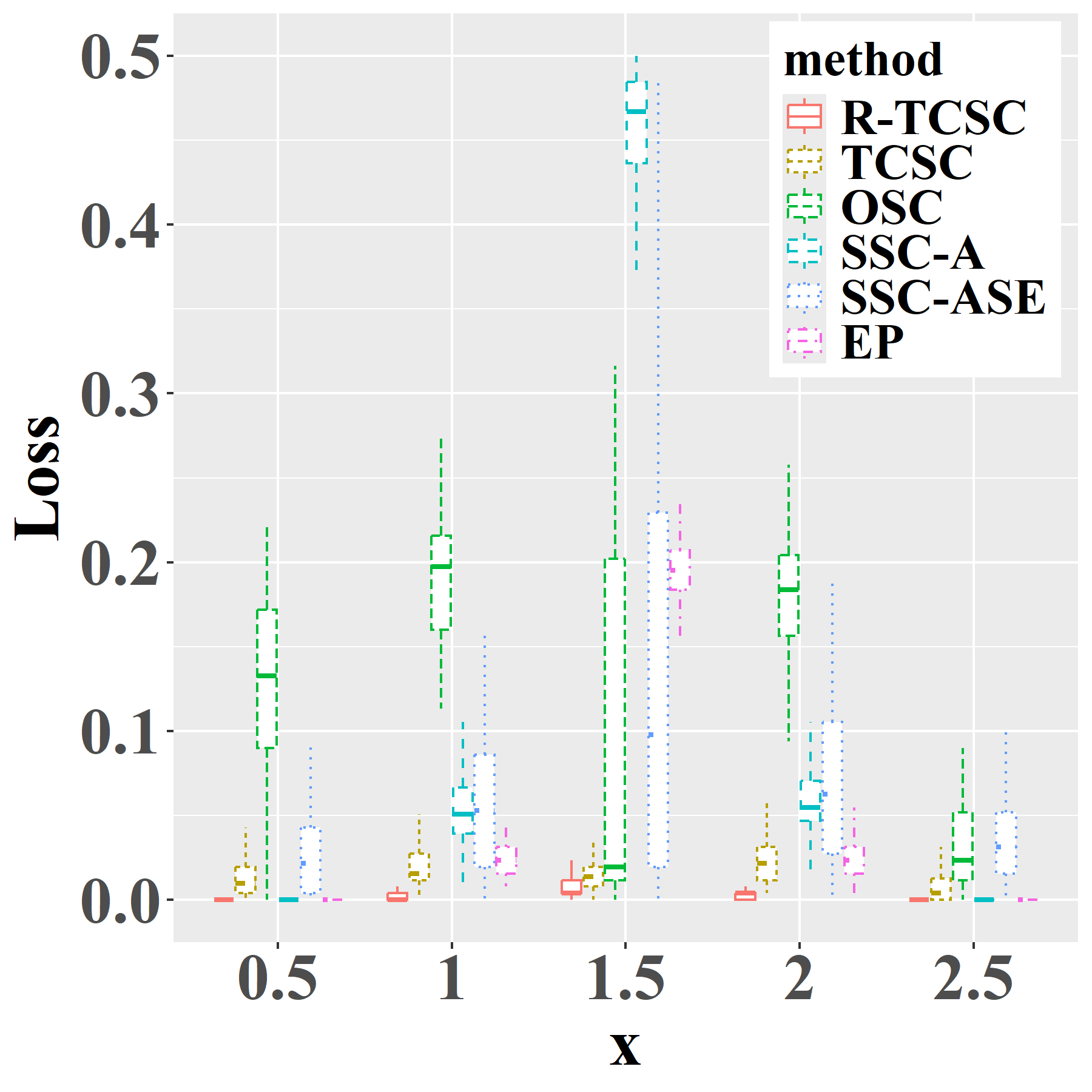}
  }
  \subfigure[$K=3$]{
    \label{cPABMfig4:subfig:b} 
    \includegraphics[width=1.98in]{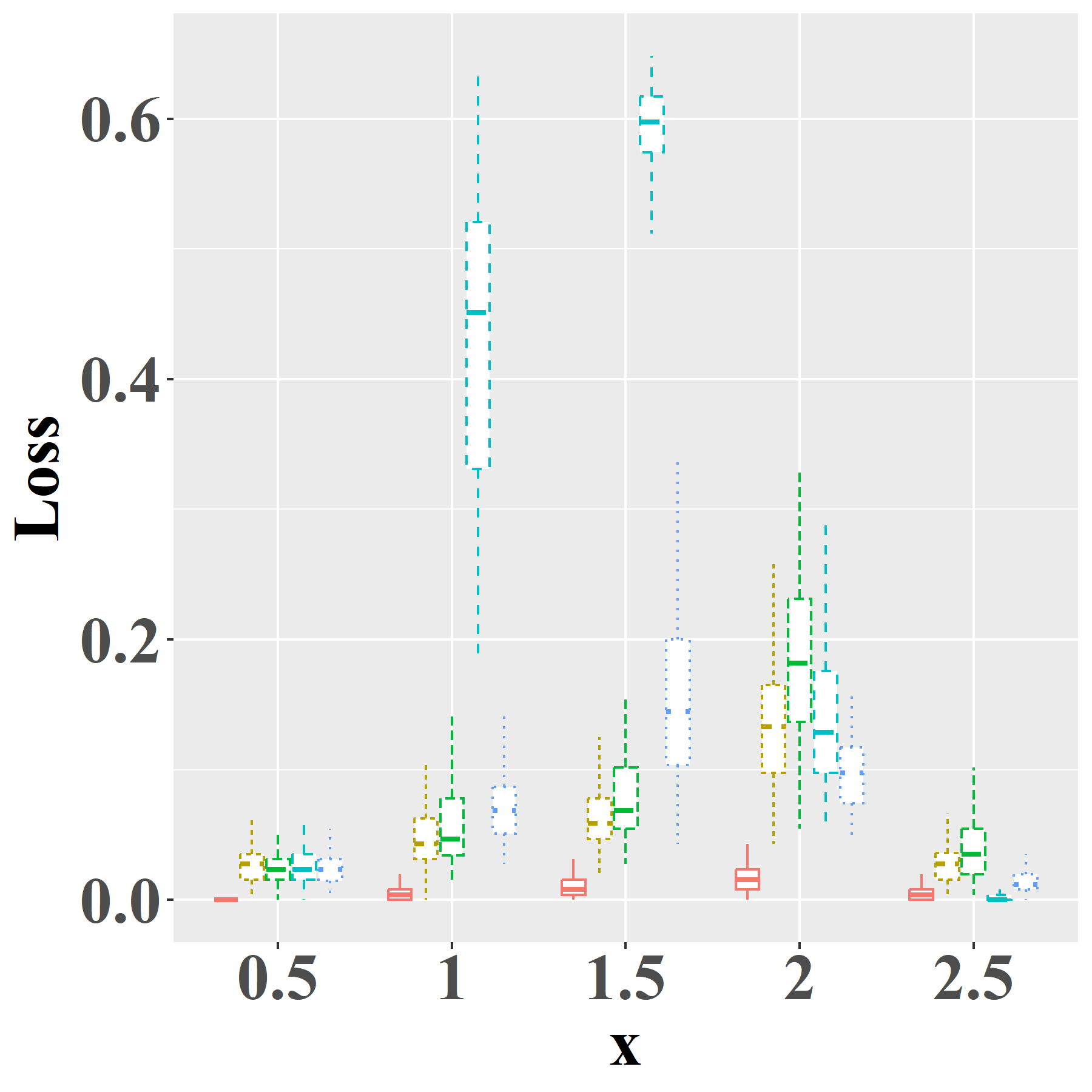}
  }
  \subfigure[$K=4$]{
    \label{cPABMfig4:subfig:c} 
    \includegraphics[width=1.98in]{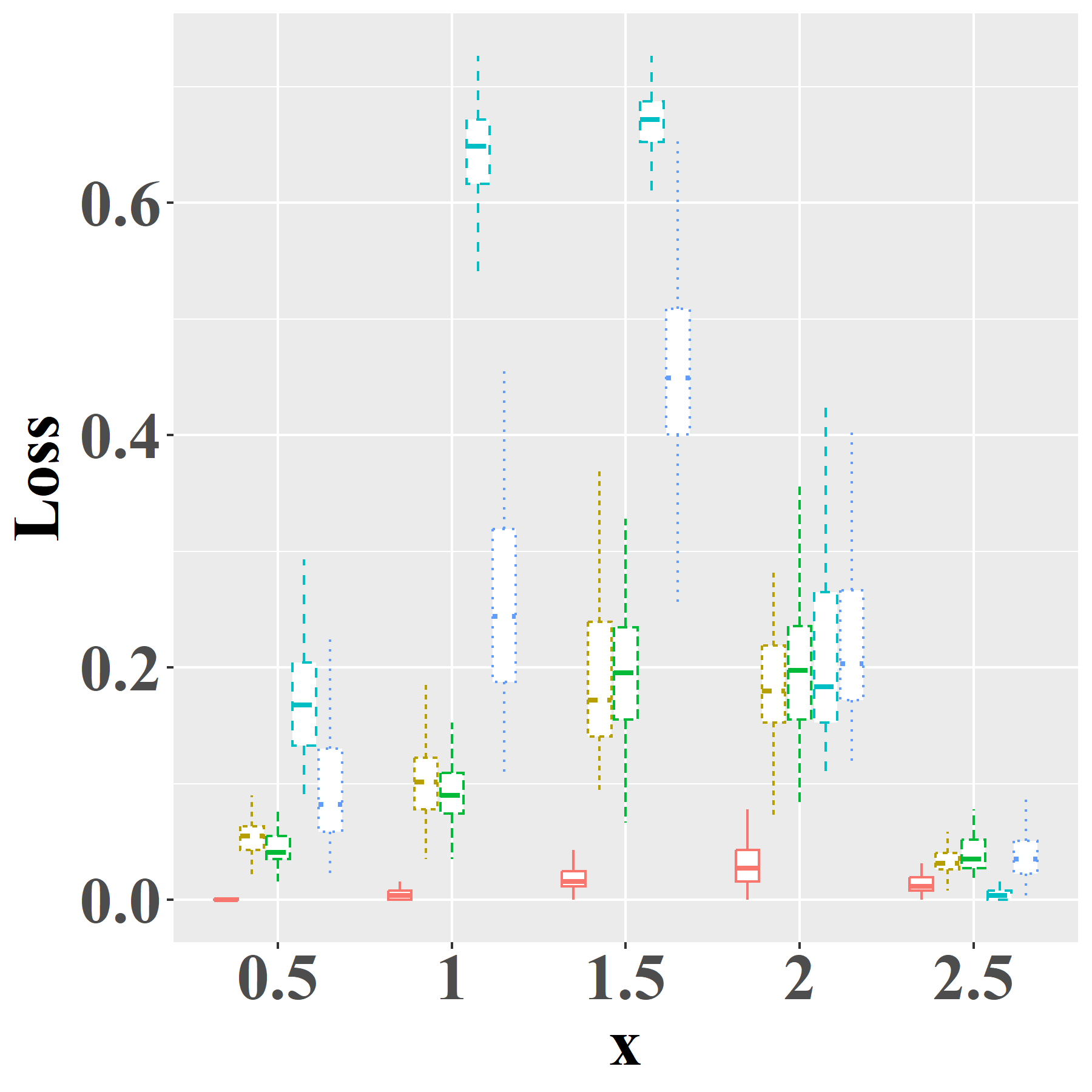}
  }
  \caption{The community detection performance from disassortative to assortative.}
  \label{cPABMdig_differ_plot}
\end{figure}

To further examine how various methods respond to assortative versus disassortative structures, we fix $n=512$ and systematically vary the degree of assortativity. Specifically, within-community popularity parameters $\bm \lambda^{(k, k)}$, are drawn \emph{i.i.d.} from $\operatorname{Beta}(x,3-x)$, while between-community parameters $\bm \lambda^{(k, l)}$ are drawn \emph{i.i.d.} from $\operatorname{Beta}(3-x,x)$ for $k \neq l \in [K]$, and let $x=0.5,1,1.5,2,2.5$. When $x=0.5$ or $1$, the model exhibits expected disassortative behavior. When $x=2$ or $2.5$, it becomes assortative. The case $x=1.5$ corresponds to a setting where, in expectation, within- and between-community popularity are equal, making label recovery particularly challenging. The results for community detection are presented in Figure \ref{cPABMdig_differ_plot}.
When $x = 1.5$ (the theoretically most difficult case, where within- and between-community popularity are equal in expectation), \(\textsf{SSC-A}\) performs poorly and nearly fails, {and although \(\textsf{EP}\) had been consistently stable under \( K=2 \), it also shows signs of deterioration.} In contrast, \textsf{R-TCSC} remains consistently strong across all settings. Although \textsf{TCSC} trails \textsf{R-TCSC}, it consistently ranks as one of the top-performing methods.

\subsection{Comparison of one-step and two-step refinements}
\label{S_onestep_multistep}
Following the numerical simulation setup outlined in Section \ref{cPABMsec:simulation_1}, we compared the performance of Algorithm~\ref{cPABMAlgorithmOneRefine}, the one-step \textsf{R-TCSC} (abbreviated as \textsf{R-TCSC-1}), and Algorithm~\ref{cPABMAlgorithmMutiRefine}, the two-step \textsf{R-TCSC} (abbreviated as \textsf{R-TCSC-2}), relative to the initialization algorithm \textsf{TCSC}. Simulation results demonstrate that \textsf{R-TCSC-1} achieves substantial improvements over the initialization. When the number of nodes is large (e.g., $n = 1024$), \textsf{R-TCSC-1} and \textsf{R-TCSC-2} perform similarly, with the second refinement step providing negligible additional gains. In contrast, when $n$ is small (e.g., $n = 128$ or $256$), the second refinement step further enhances performance. Consequently, we employ \textsf{R-TCSC-2} throughout the main text and refer to it as \textsf{R-TCSC} for simplicity.

\begin{figure}[!htp]
  \centering
  \subfigure[$K=2$]{
    \label{cPABMfig1:subfig:a_multistep} 
    \includegraphics[width=1.98in]{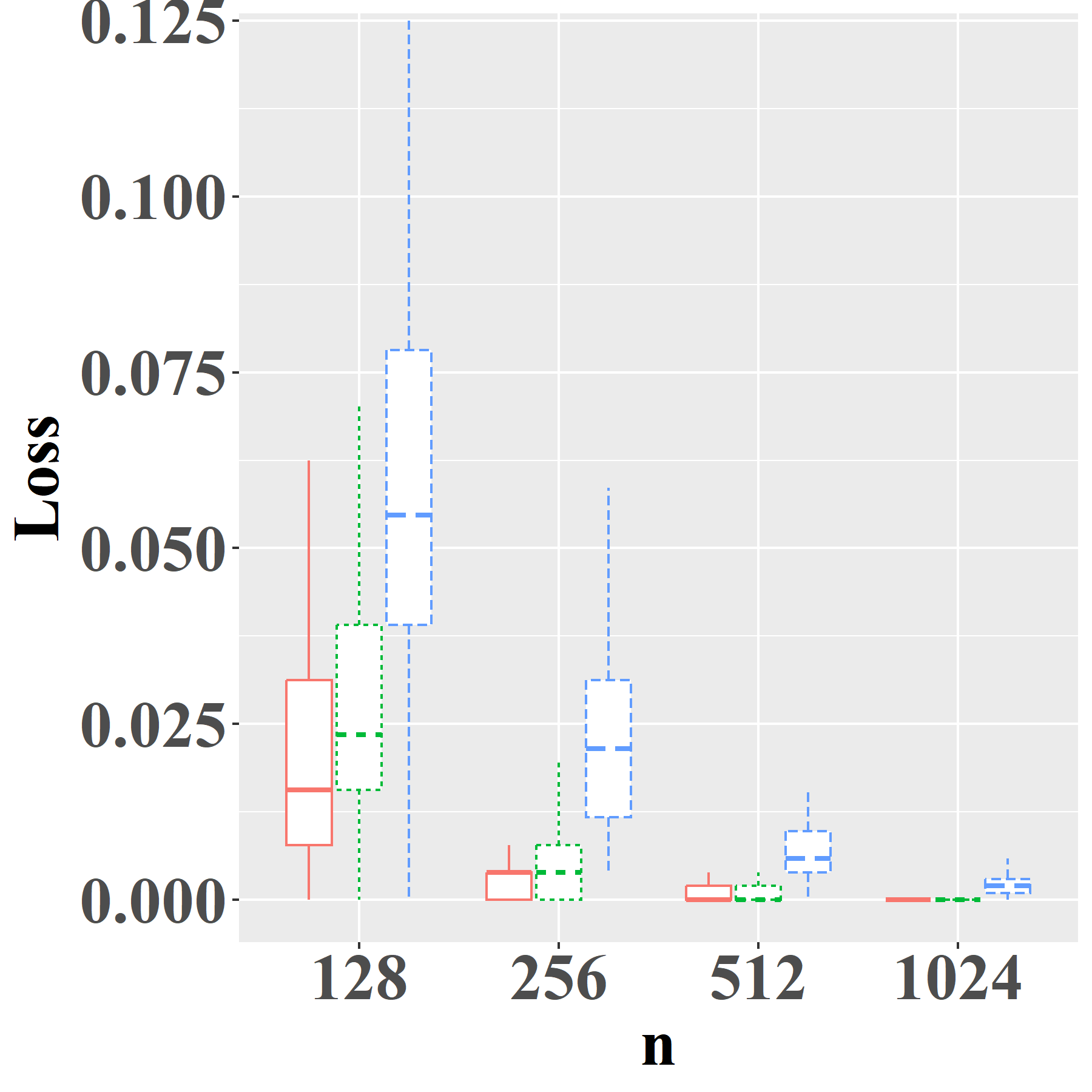}
  }
  \subfigure[$K=3$]{
    \label{cPABMfig1:subfig:b_multistep} 
    \includegraphics[width=1.98in]{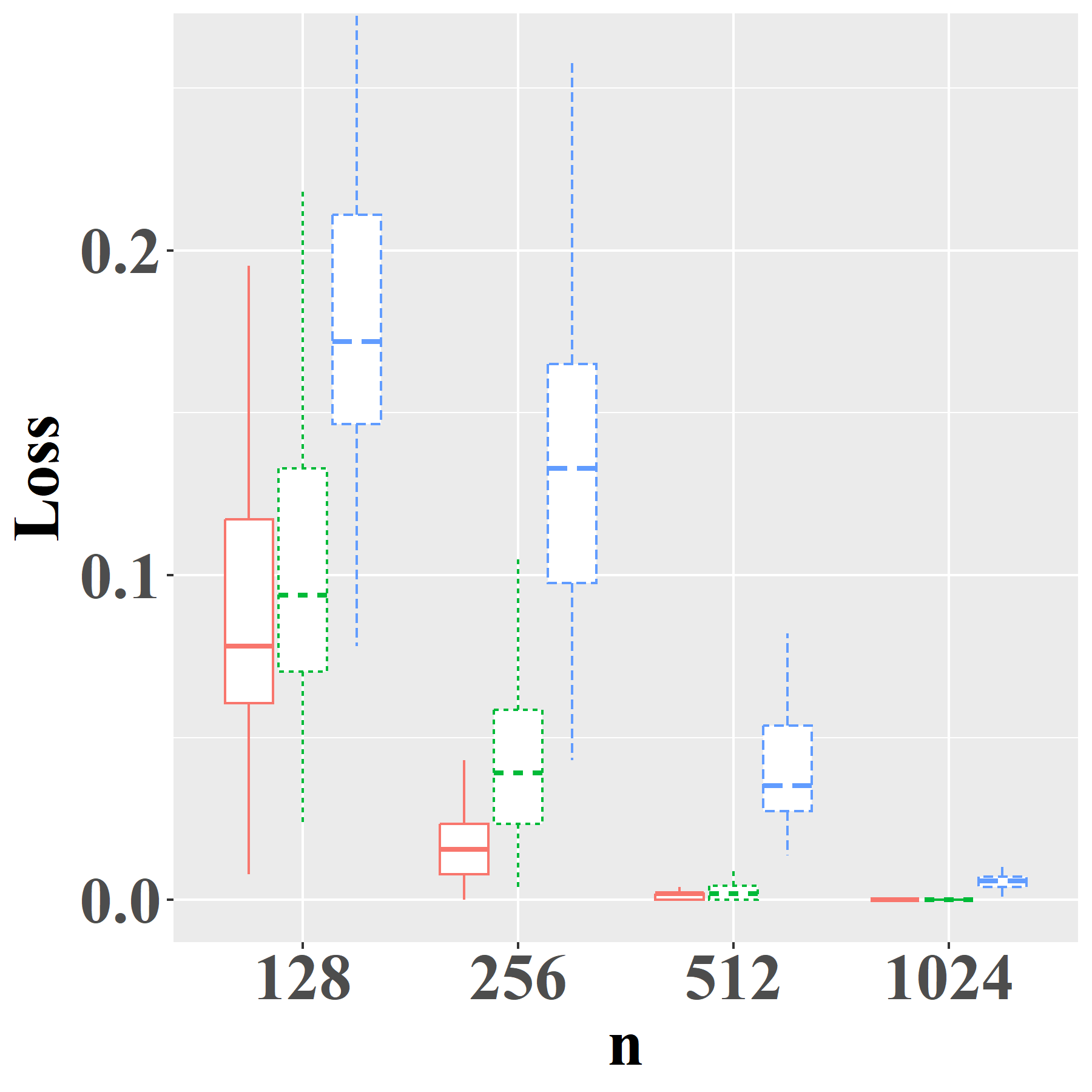}
  }
  \subfigure[$K=4$]{
    \label{cPABMfig1:subfig:c_multistep} 
    \includegraphics[width=1.98in]{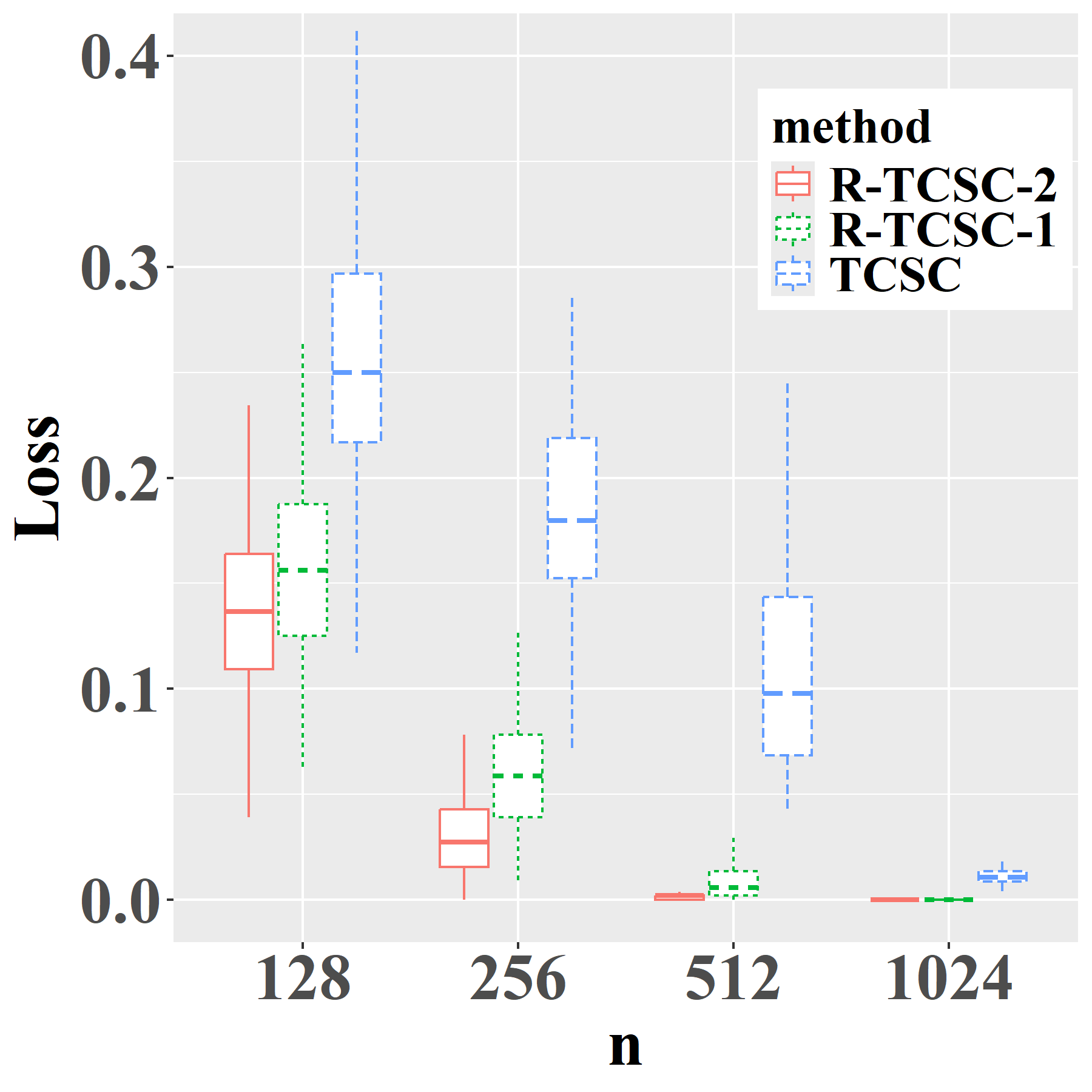}
  }
  \caption{\textsf{R-TCSC-1} vs \textsf{R-TCSC-2} in case of balanced communities.}
  \label{cPABMn_change_ab2112_multistep}
\end{figure}

\begin{figure}[t!]
  \centering
  \subfigure[$K=2$]{
    \label{cPABMfig2:subfig:a_multistep} 
    \includegraphics[width=1.98in]{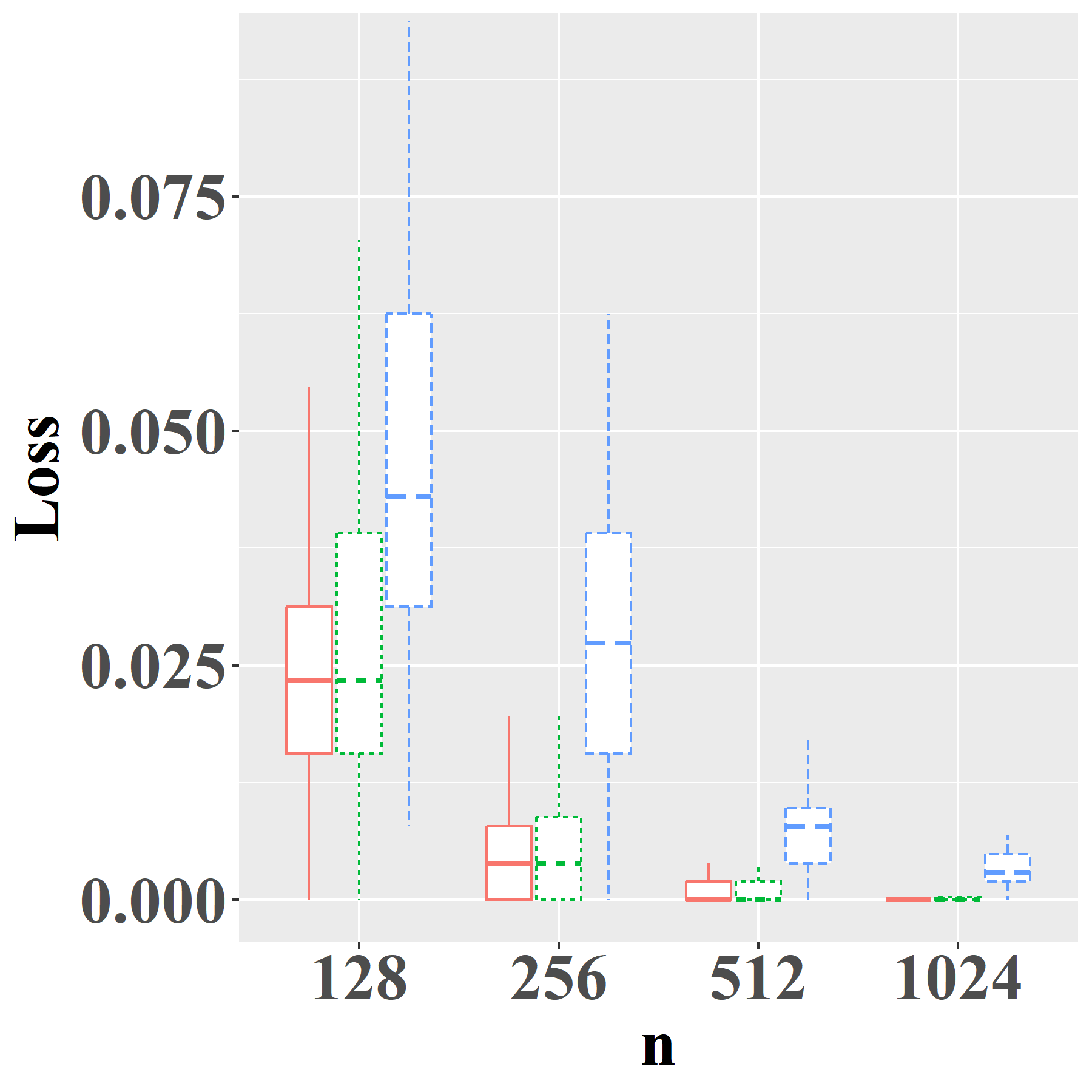}
  }
  \subfigure[$K=3$]{
    \label{cPABMfig2:subfig:b_multistep} 
    \includegraphics[width=1.98in]{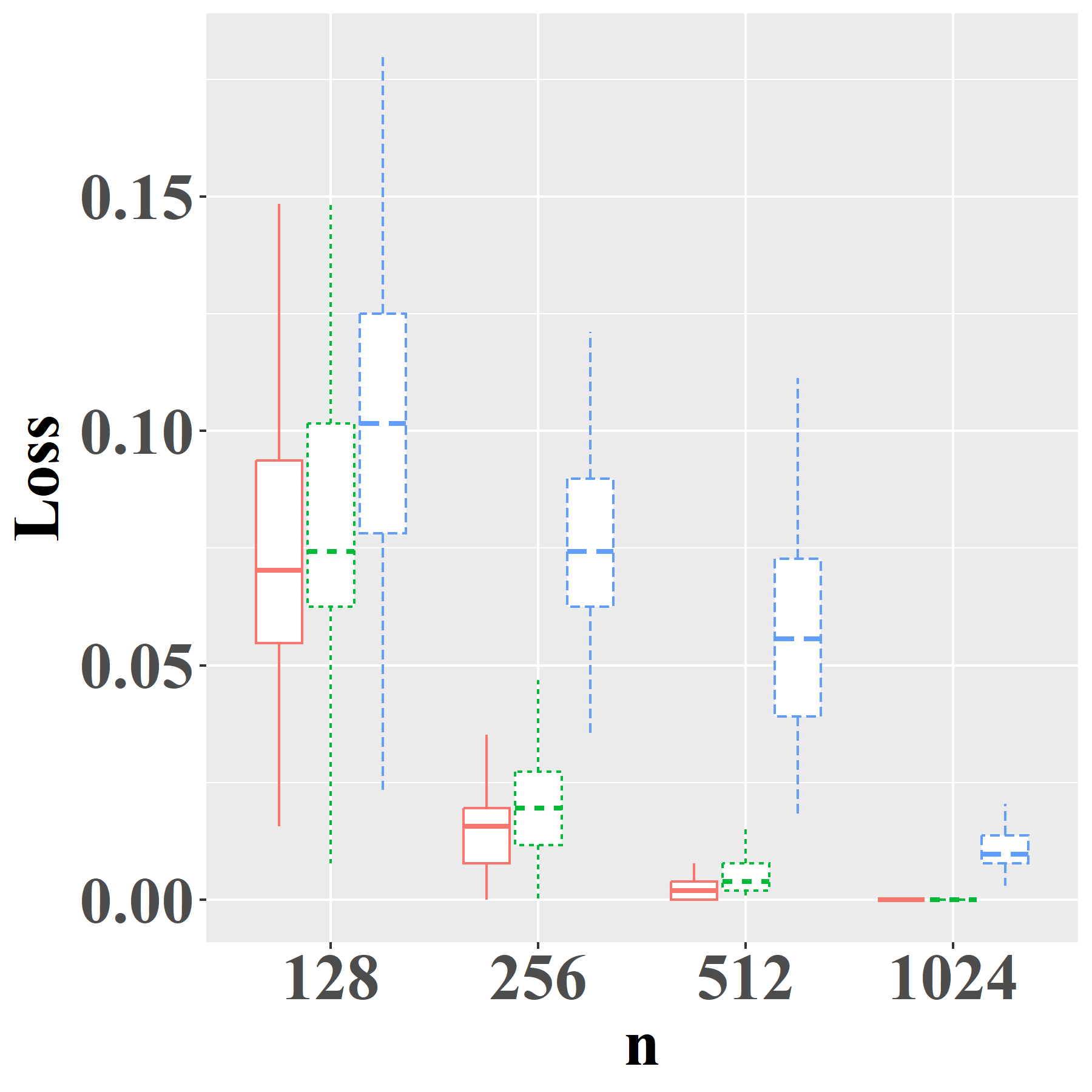}
  }
  \subfigure[$K=4$]{
    \label{cPABMfig2:subfig:c_multistep} 
    \includegraphics[width=1.98in]{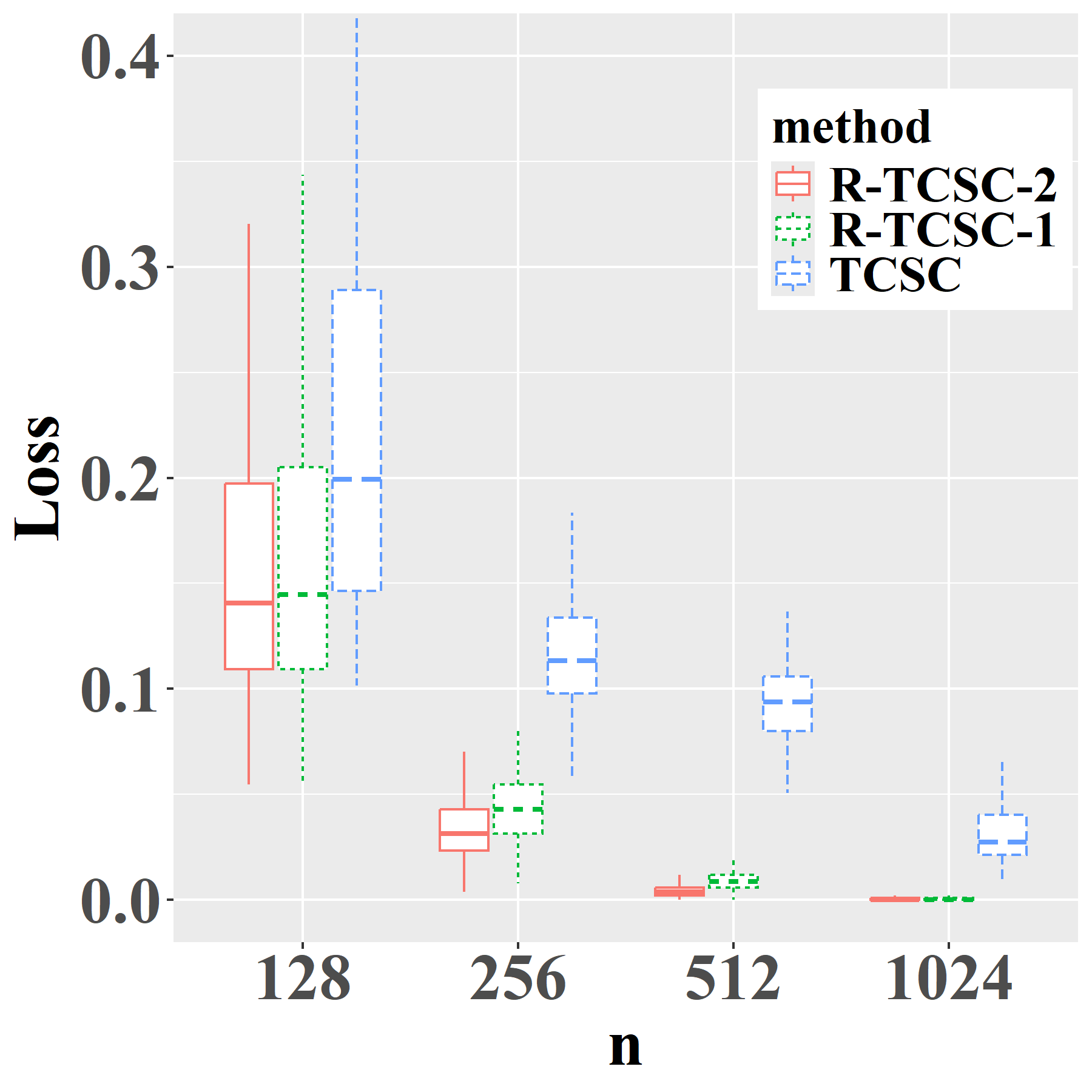}
  }
  \caption{\textsf{R-TCSC-1} vs \textsf{R-TCSC-2} in case of unbalanced communities.}
  \label{cPABMn_change_ab2112_unbalanced_multistep}
\end{figure}

\begin{figure}[t!]
  \centering
  \subfigure[$K=2$]{
    \label{cPABMfig3:subfig:a_multistep} 
    \includegraphics[width=1.98in]{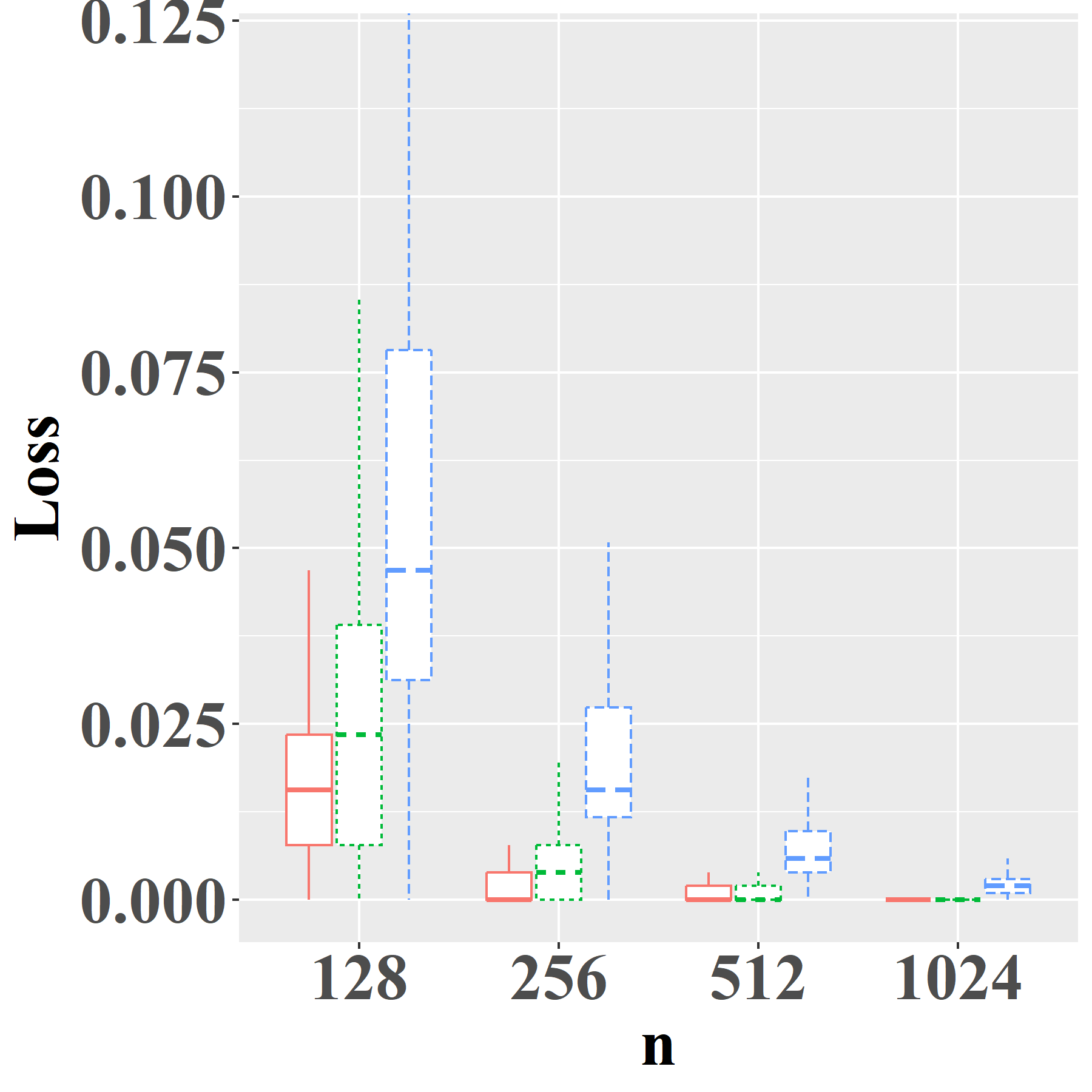}
  }
  \subfigure[$K=3$]{
    \label{cPABMfig3:subfig:b_multistep} 
    \includegraphics[width=1.98in]{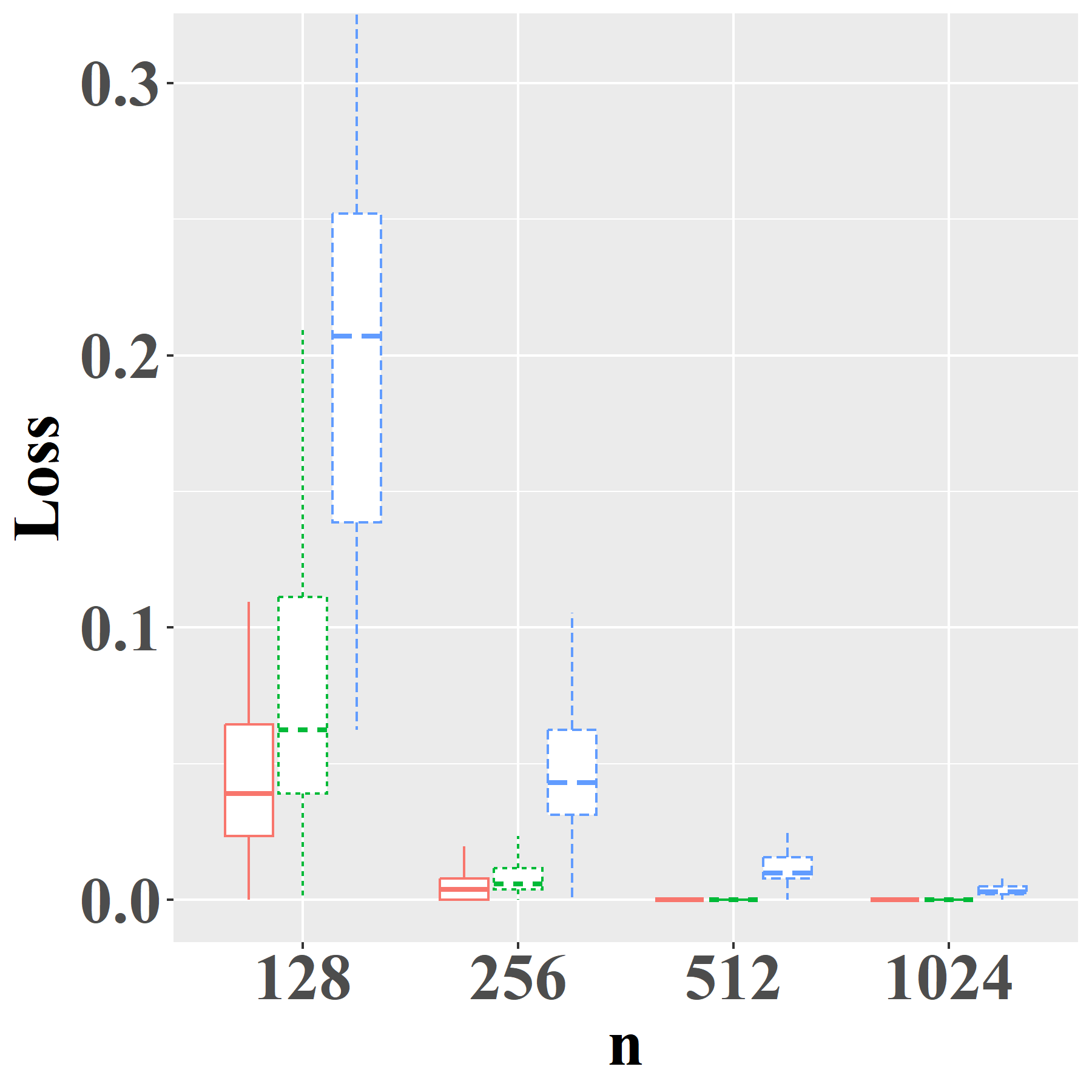}
  }
  \subfigure[$K=4$]{
    \label{cPABMfig3:subfig:c_multistep} 
    \includegraphics[width=1.98in]{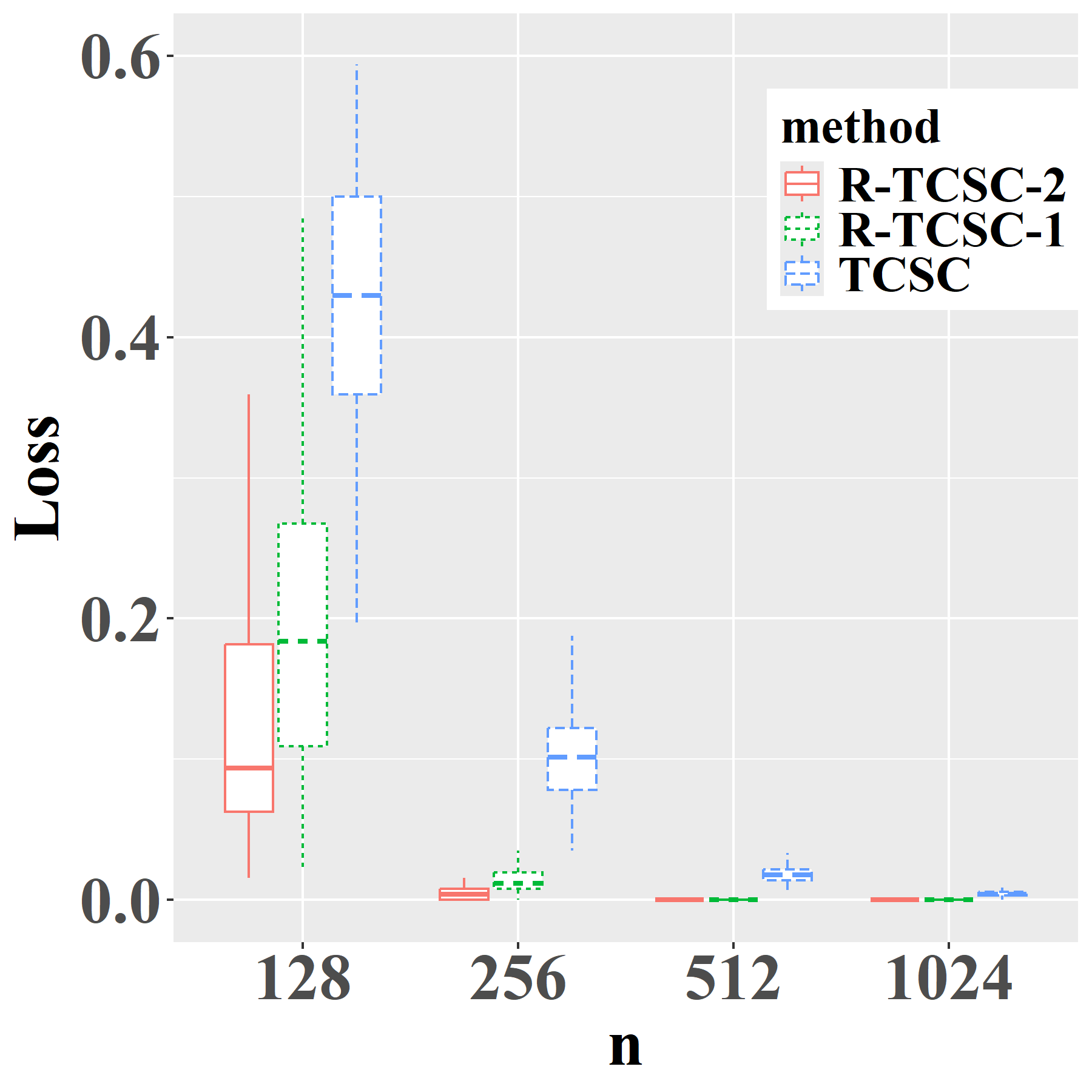}
  }
  \caption{\textsf{R-TCSC-1} vs \textsf{R-TCSC-2} in case of disassortative models.}
  \label{cPABMn_change_ab1221_multistep}
\end{figure}

\subsection{Heterogeneity within-community}
\label{S_simu_heterogeneity}
\begin{figure}[ht!]
  \centering
  \subfigure[$K=2$]{
    \label{cPABMfigiid:subfig:a} 
    \includegraphics[width=1.98in]{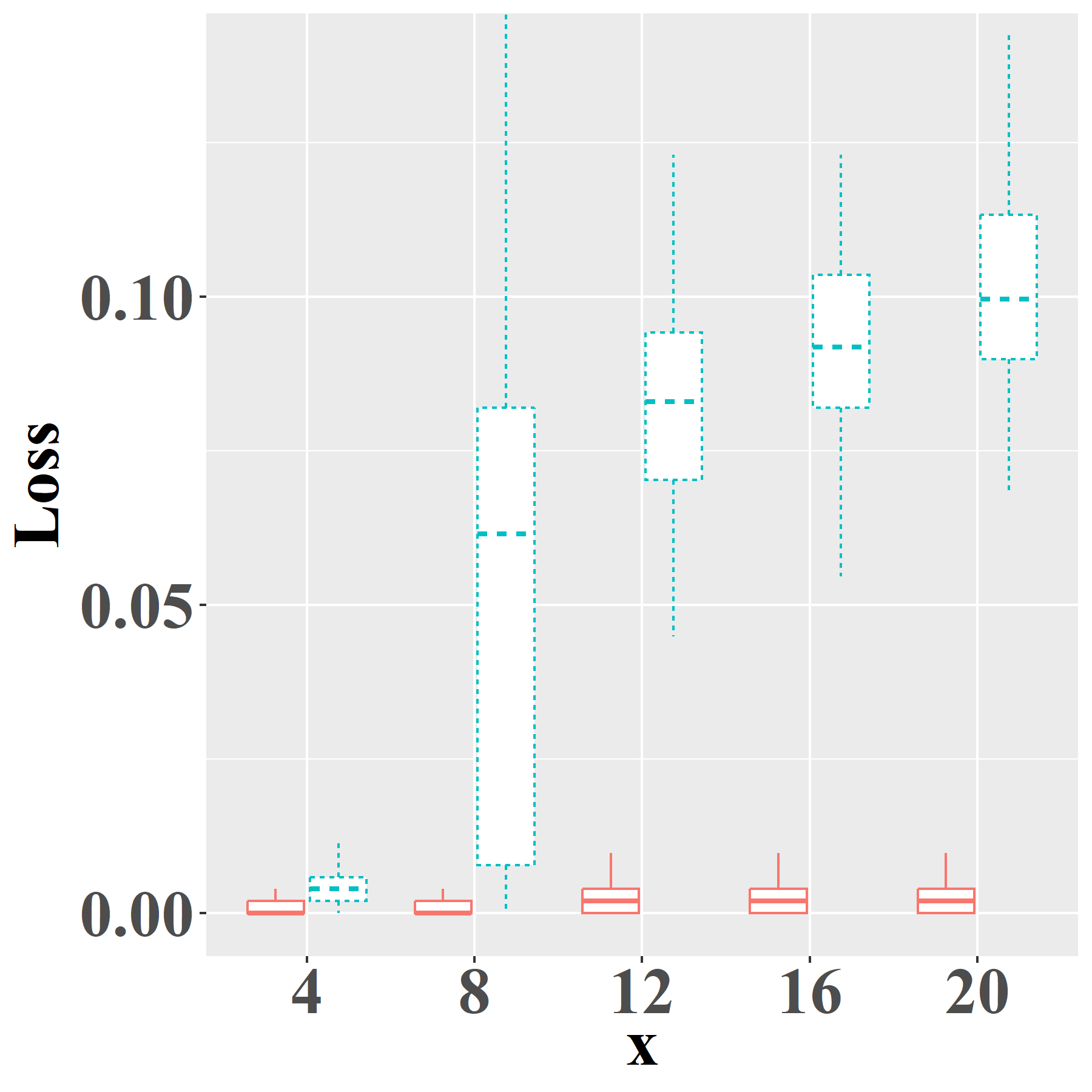}
  }
  \subfigure[$K=3$]{
    \label{cPABMfigiid:subfig:b} 
    \includegraphics[width=1.98in]{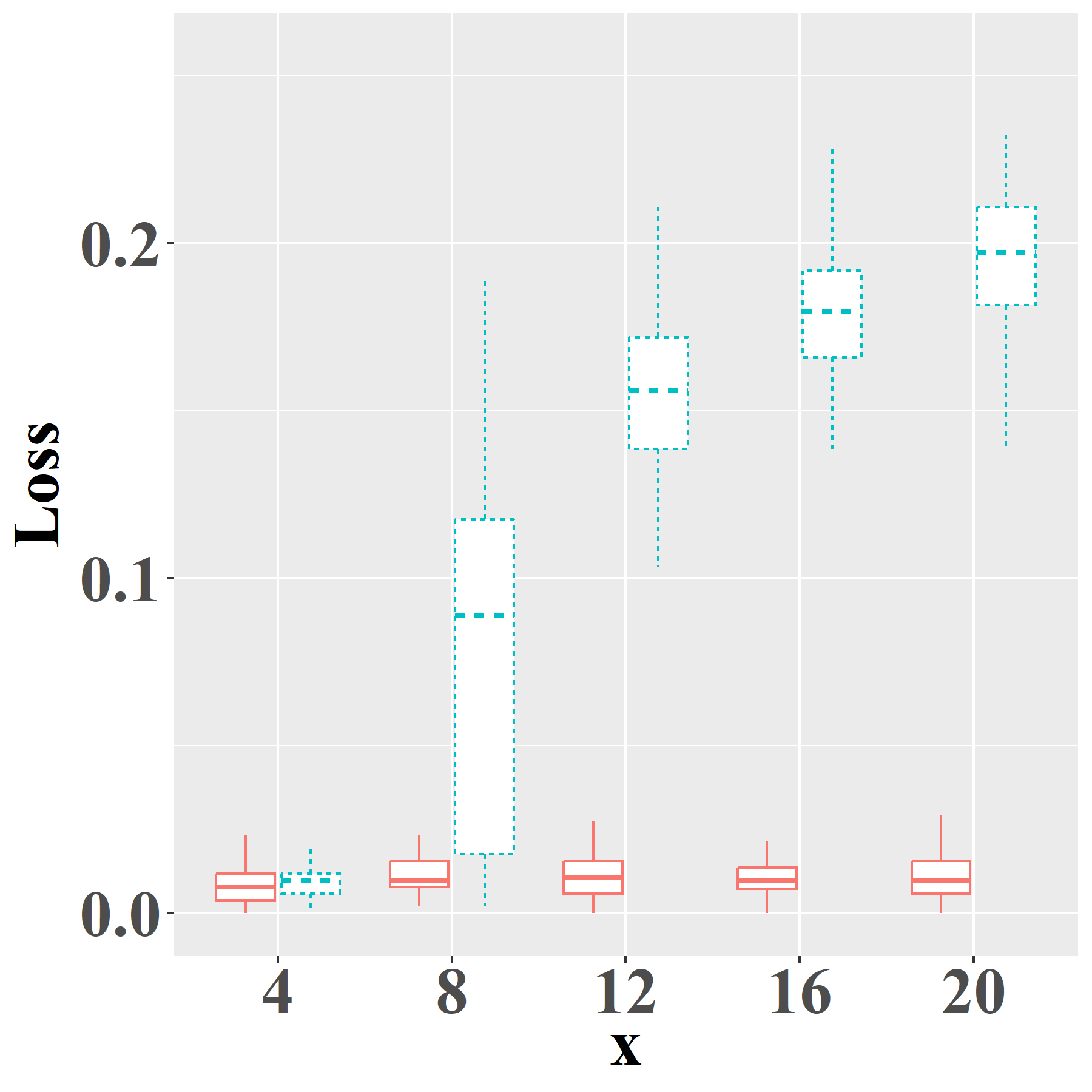}
  }
  \subfigure[$K=4$]{
    \label{cPABMfigiid:subfig:c} 
    \includegraphics[width=1.98in]{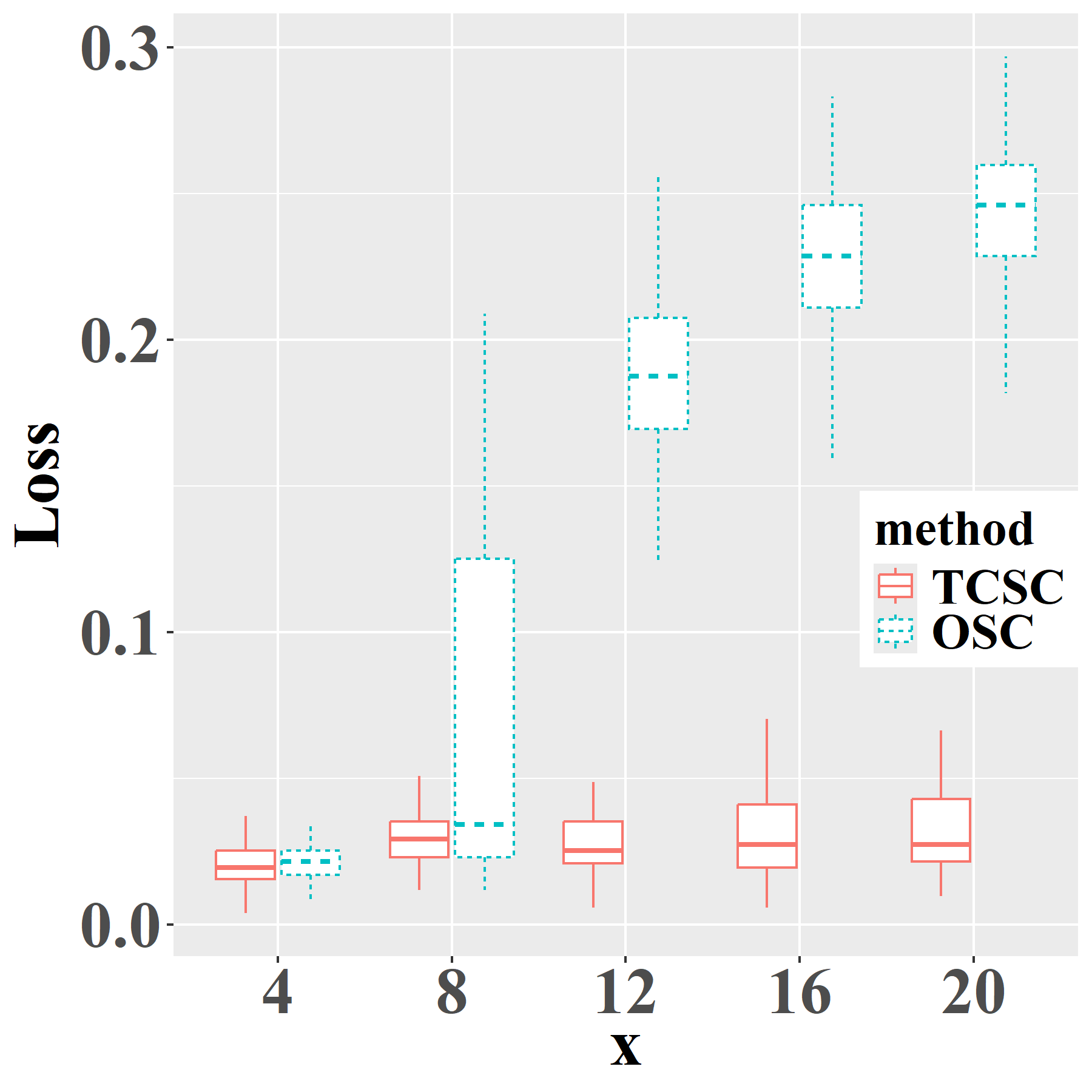}
  }
  \caption{The community detection performance when $(\lambda_{i 1},\cdots,\lambda_{i K})_{i: \bm c^{\ast}(i)=k}$ for nodes in the same community are drawn independently from heterogeneous distributions.}
  \label{cPABMiid_differ_plot}
\end{figure}

In this section, we consider the case of within-community heterogeneity. In previous settings, $(\lambda_{i 1},\cdots,\lambda_{i K})_{i, \bm c^{\ast}(i)=k}$ were assumed to be \emph{i.i.d.}, a key assumption underlying the theoretical guarantees of \textsf{OSC}. Here, we relax this to independent but non-identically distributed (non-\emph{i.i.d.}) samples and show that under this setting, the performance of \textsf{OSC} deteriorates significantly. To introduce heterogeneity, for each $k, l\in[K]$, we generate the first half of the entries in $\bm \lambda^{(k, l)}$ \emph{i.i.d.} from $\operatorname{Beta}(x,2)$ with $x=4,8,12,16,20$, and the second half \emph{i.i.d.} drawn from $\operatorname{Beta}(1,1)$. Figure~\ref{cPABMiid_differ_plot} presents the community detection results with $n=512$. As $x$ increases, the Hellinger distance between $\operatorname{Beta}(x,2)$ and $\operatorname{Beta}(1,1)$, used for the first and second halves of $\bm \lambda^{(k, l)}$, grows larger: $0.35,0.52,0.61,0.66,0.70$. Correspondingly, the performance of the \textsf{OSC} algorithm deteriorates noticeably with increasing heterogeneity.

\section{Selection of the Number of Communities}\label{cPABMsec:SelectionK}
Until now, we have assumed that the number of communities 
 $K$ is known. However, in practice, this is typically unknown. It is an important research topic to choose the number of communities in network data, such as those for SBM \citep{saldana2017many,Chen2018Network,li2020network}. To the best of our knowledge, the only method for estimating $K$ under the PABM is proposed by \citet{Noroozi2021Estimation}, which selects $K$ by minimizing the squared loss with a penalty term, denoted as \textsf{LP}. We propose a new approach to estimate $ K $ by {leveraging} the structure of $ \bm \Theta $ in \eqref{cPABMTheta:rank1}, which implies that $ \text{rank}(\bm \Theta^{(k, k)}({\bm c}^{\ast})) = 1 $ for any $ k \in [K] $. 
Define
\begin{equation}
\label{cPABMc.Ktilde}
 \bm c_{\tilde{K}} \in \underset{\substack{
 \tilde{\bm c} \in [\tilde{K}]^n, \forall k \in [\tilde{K}], \\ n_k(\tilde{\bm c}) \in \big[\frac{n}{C\tilde{K}},\frac{Cn}{\tilde{K}}\big]}}{\arg \min} f(\tilde{\bm c}, \bm \Theta),
\end{equation}
for $\tilde{K}\in[K_{\max}]$, where $f(\tilde{\bm c}, \bm \Theta)={\max}_{k \in [\tilde{K}]}  \sigma_{2}(\bm \Theta^{(k, k)}(\tilde{\bm c}))$ and $\bm \Theta^{(k, k)}(\tilde{\bm c})=(\theta_{ij})_{i,j:\tilde{\bm c}(i)=\tilde{\bm c}(j)=k}$.

The following proposition shows that $ f(\bm{c}_{\tilde{K}}, \bm{\Theta}) $ attains its minimum at the true number of communities $ K $.  Furthermore, replacing $ \bm{\Theta} $ with the observed adjacency matrix $ \bm{A} $, the empirical version $ f(\bm{c}_{\tilde{K}}, \bm{A}) $ continues to identify $ K $ accurately with high probability.

\begin{proposition}
\label{cPABMtheorem:chooseK}
Suppose that, for each $\tilde{K}<K$, $f(\bm c_{\tilde{K}}, \bm \Theta) \geq C n\rho_{n}^{2}/(\tilde{K}\sqrt{\log n})$ for a constant $C>0$. Then, we have the following results:
\begin{enumerate}[label=(\Roman*)]
  \item \label{cPABMlemma:chooseK_0}
If $\tilde{K} \geq K$, $f(\bm c_{\tilde{K}}, \bm \Theta)\equiv 0$;
  \item \label{cPABMlemma:chooseK_A}
When $K_{\max}\leq n^{1-C}$ for a constant $C>0$ and $\rho_{n}^{2}{{n}/{K_{\max}}} \gg (\log n)^{2} 
$, then,
\begin{eqnarray}
\label{cPABMlemma:Lemmaeq:chooseK_A1}
\begin{cases}
f(\bm c_{\tilde{K}}, \bm A)
\lesssim
\rho_{n}\sqrt{\frac{n}{\tilde{K}}},
\quad &K \leq \tilde{K} \leq K_{\max}; \\
\frac{n\rho_{n}^2}{\tilde{K}\sqrt{\log n}}
\lesssim
f(\bm c_{\tilde{K}}, \bm A)
\lesssim
\frac{n\rho_{n}^2}{\tilde{K}} ,
\quad &\tilde{K} < K,
\end{cases}
\end{eqnarray}
holds with probability at least $1-n^{-C'}$ for a constant $C'>0$.
\end{enumerate}
\end{proposition}

\begin{remark}
The assumption of Proposition \ref{cPABMtheorem:chooseK} ensures the identification of the community number $ K $ by providing the necessary conditions for the correct estimation of $ K $. This assumption is not restrictive. Taking $K=2$ as an example, it holds under the conditions of Theorem \ref{cPABMtheorem:Initial}. Specifically, for $\tilde{K} < K$, $f(\bm c_{\tilde{K}}, \bm \Theta)=\sigma_{2}(\bm \Theta) \geq C\sigma_{1}(\bm \Theta)/\sqrt{\log n} \geq C\|\bm \Theta\|_{\textrm{F}}/(2\sqrt{\log n})\gtrsim n \rho_{n}^{2}/\sqrt{\log n}$.
\end{remark}

Assume $2 \leq K \leq K_{\max}$ for some integer $ K_{\max} > 1 $.
We estimate $K$ by
\begin{equation}
\label{cPABMchooseK:eq1:0}
  \hat{K}(\bm c_{\tilde{K}}) = \underset{\tilde{K} \in \{2, \cdots, K_{\max}\}}{\arg \max} \frac{f(\bm c_{\tilde{K}-1}, \bm A)}{f(\bm c_{\tilde{K}}, \bm A)+\log n}.
\end{equation}
The addition of the $\log n$ penalty follows the same reasoning as in \cite{YU2022Projected} for selecting the number of factors in the factor model. Specifically, it prevents overestimation of $K$ by offsetting the rapid decay of $f(\bm c_{\tilde{K}}, \bm A)$ when $\tilde{K} >K$.

By Proposition \ref{cPABMtheorem:chooseK}, we directly obtain the following corollary.
\begin{corollary}
\label{cPABMcorollary:chooseK}
Under the assumptions of Proposition \ref{cPABMtheorem:chooseK}, we have  
$
\hat{K}(\bm c_{\tilde{K}})=K
$
holds with probability at least $1-n^{-C}$ for a constant $C>0$.
\end{corollary}

Building on Corollary \ref{cPABMcorollary:chooseK}, 
we transform the problem \eqref{cPABMchooseK:eq1:0} into the following form:
\begin{equation}
\label{cPABMchooseK:eq1}
  \hat{K} = \underset{\tilde{K} \in \{2, \cdots, K_{\max}\}}{\arg \max} \frac{f(\bm c_{\tilde{K}-1}, \bm A)}{\frac{1}{d}\sum_{K'=\tilde{K}}^{\tilde{K}+d-1}f(\bm c_{K'}, \bm A)+\log n},
\end{equation}
where $d$ is the window width parameter. While $ d=1$ suffices theoretically, we set $d = 2$ throughout for improved robustness in practice. We use \textsf{TCSC} to estimate $\bm c_{\tilde{K}}$ in \eqref{cPABMchooseK:eq1}, since the K-means step in \textsf{TCSC} tends to favor more balanced community sizes, which aligns better with the structure assumed in the optimization of \eqref{cPABMc.Ktilde}. As a result, we refer to the proposed method as \textsf{SVCP} (Singular Value Change Point).

\begin{table}[ht]
\centering
\caption{Accuracy of \textsf{SVCP} and \textsf{LP} in selecting the number of communities $K$.}
\label{cPABMTabChooseK_PABM}
\begin{tabular}{c | c c c c c c c}
  \hline	
Method	&	$K=2$	&	$3$	&	$4$	&	$5$&	$6$&	$7$&	$8$	\\
  \hline	
$\textsf{LP}$\citep{Noroozi2021Estimation}	&	0.92 	&	1.00 	&	 1.00	&	0.99  &	0.63&	0.76  &	 0.04 	\\
$\textsf{SVCP}$	(our method) &	0.92 	&	1.00 	&	1.00 	&	0.99  &	0.87 &	1.00  &	0.93  	\\
	\hline										
\end{tabular}
\end{table}

To examine the numerical performance of \textsf{SVCP} against \textsf{LP}, we consider following simulation setup:
number of nodes $n=512$; number of communities, $K=2,3,4,5,6,7,8$; community size parameters, $\pi_k=1 / K$ for $k=1, \ldots, K$; intra-community parameters $\bm \lambda^{(k, k)} \stackrel{\text { iid }}{\sim} \operatorname{Beta}(2,1)$ and inter-community parameters $\bm \lambda^{(k, l)} \stackrel{\text { iid }}{\sim} \operatorname{Beta}(1,2)$ for $k \neq l \in [K]$. Both $\textsf{LP}$ and \textsf{SVCP} achieve comparable performance when $K$ is small, but \textsf{SVCP} shows a clear advantage as $K$ increases, as shown in Table \ref{cPABMTabChooseK_PABM}. Since $n = 512$ is fixed, increasing $K$ raises model complexity and the difficulty of estimating $K$. Table~\ref{cPABMTabChooseK_PABM} indicates that \textsf{LP} starts to fail when $K=6$, $7$ or $8$, whereas \textsf{SVCP} remains robust.

\section{Real Applications}\label{cPABMRealdata}
In this section, we analyze two real-world networks with known community labels. We first estimate the number of communities using the method from Subsection~\ref{cPABMsec:SelectionK} and then compare the performance of our proposed \textsf{TCSC} and \textsf{R-TCSC}  against baseline methods \textsf{EP}, \textsf{SSC-A}, \textsf{OSC} and \textsf{SSC-ASE}, using the same implementation settings as in Section~\ref{cPABMsec:simulation}.

To assess both stability and performance, we repeatedly subsample each community in the original network $\bm A$ by independently selecting nodes with probability $p_{sample} \in \{0.5, 0.6, 0.7, 0.8, 0.9\}$. After removing isolated nodes (those with degree zero), we obtain a pruned network $\bm A_{sample}$. This process is repeated $100$ times, allowing us to evaluate the consistency of each community detection method across different subsampled networks.

\begin{figure}[ht!]
  \centering
    \subfigure[DBLP Network]{
    \label{cPABMRealdata:subfig:b} 
    \includegraphics[width=3.05in]{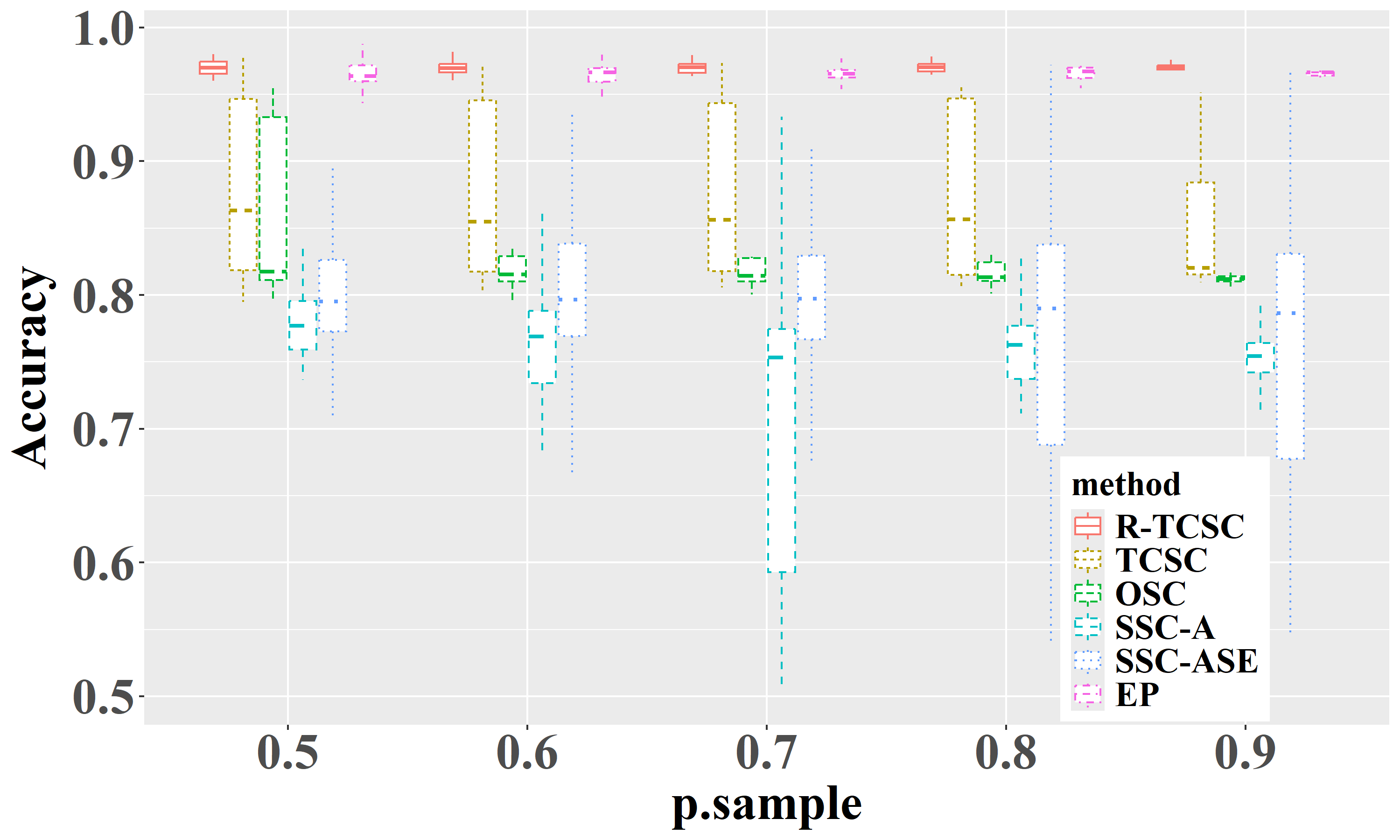}
  }
  \subfigure[Butterfly Network]{
    \label{cPABMRealdata:subfig:a} 
    \includegraphics[width=3.05in]{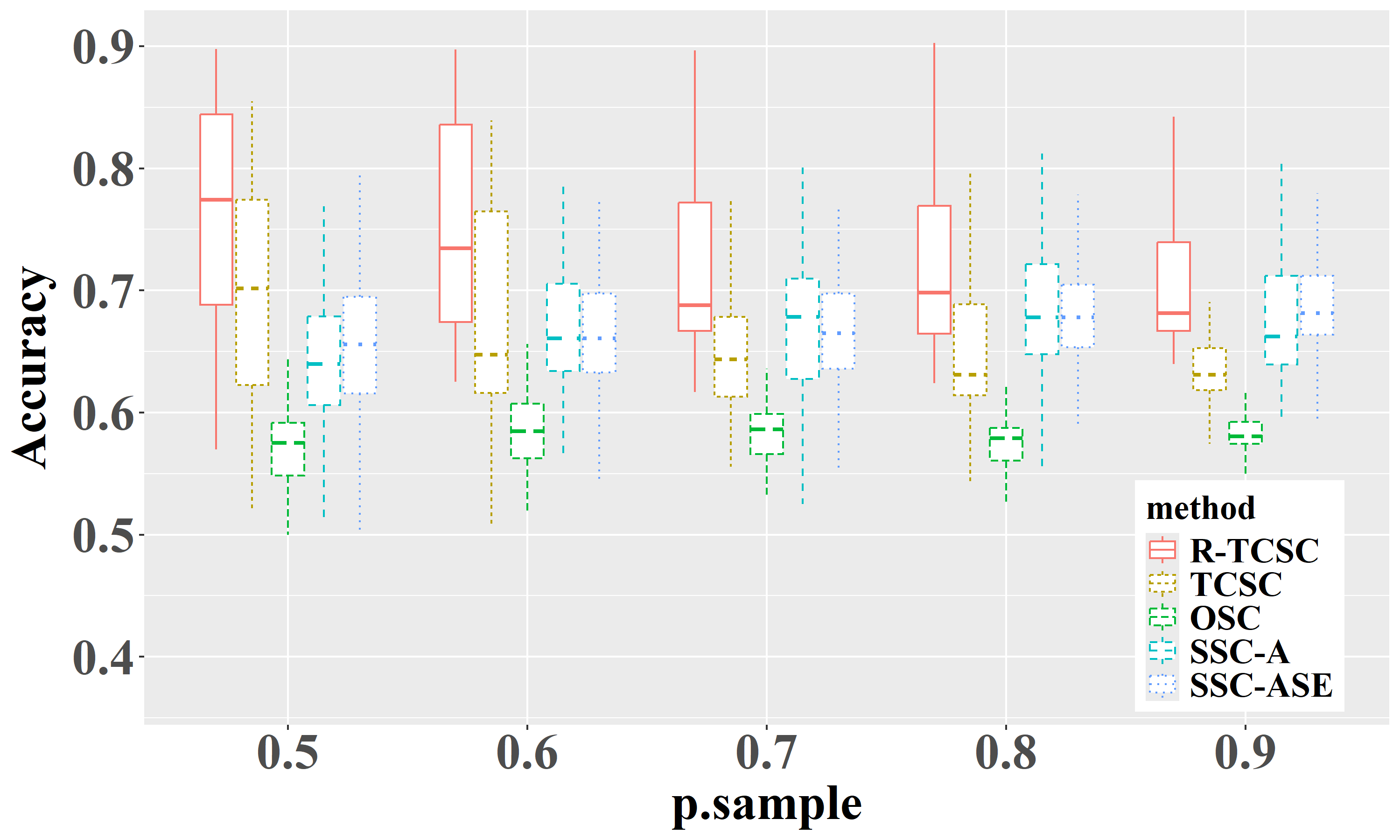}
  }
  \caption{Community detection performance on two real-world networks}
  \label{cPABMRealdata_plot}
\end{figure}

\textbf{Digital Bibliography \& Library Project (DBLP) Network.} The DBLP network extracted by \citet{Gao2009Graph} and \citet{Ming2010Graph}, represents authors as nodes, with an edge connecting any pair who attended the same conference. Following \citet{Sengupta2017aBlock}, we focus on authors in two research areas: databases and information retrieval.  The resulting network contains $2,203$ nodes and $1,148,044$ edges. We apply the \textsf{SVCP} to estimate the number of communities, obtaining $\hat{K} = 2$, which matches the ground truth. Figure~\ref{cPABMRealdata_plot} reports clustering accuracy (i.e., $1-\ell(\bm c^{\ast},  \hat{\bm c} )$), and Figure~\ref{cPABMRealdata:subfig:b} shows that \textsf{R-TCSC} {and \textsf{EP} consistently outperform others} in both accuracy and robustness.

\textbf{Butterfly Network.} This network, based on the Leeds butterfly dataset \citep{Wang2018Network}, represents species as nodes and draws edges between visually similar pairs. Following \citet{Noroozi2021Estimation} and \citet{John2023Popularity}, we analyze a subnetwork with $373$ nodes and $20,566$ edges, comprising the four largest species categories. The PABM is well-suited for this data, as species may resemble others outside their category. Using our proposed \textsf{SVCP} method, we estimated $\hat{K} = 4$, correctly recovering the true number of communities. As shown in Figure~\ref{cPABMRealdata:subfig:a}, \textsf{R-TCSC} outperforms all competing methods.

\section{Conclusion}\label{conclusion}
{
The Popularity Adjusted Block Model (PABM) captures heterogeneous node popularity across communities but introduces significant estimation challenges due to its complexity. To address these challenges in this paper, we first propose an effective spectral clustering algorithm, \textsf{TCSC}, and establish its weak consistency under the PABM. We then show that the proposed one-step refinement algorithm, \textsf{R-TCSC}, achieves strong consistency, and further provide a theoretical guarantee that the two-step refinement accelerates the convergence rate of the clustering error in finite samples. We also introduce a data-driven method for selecting the number of communities,  which outperforms existing approaches. Beyond these results, our framework is broadly applicable. It relies only on first-order moment information, avoiding likelihood-based assumptions, and is well suited to general network settings. While our refinement procedure uses spectral clustering for initialization, it can readily incorporate other fast and reliable methods for a wide range of applications.
}


\bibliography{Refers}

\begin{thebibliography}{}

\bibitem[Abbe and Sandon, 2015]{Abbe2015Community}
Abbe, E. and Sandon, C. (2015).
\newblock Community detection in general stochastic block models: Fundamental limits and efficient algorithms for recovery.
\newblock In {\em 2015 IEEE 56th Annual Symposium on Foundations of Computer Science}, page 670–688, Berkeley. IEEE Computer Society.

\bibitem[Agarwal and Xue, 2020]{agarwal2020model}
Agarwal, A. and Xue, L. (2020).
\newblock Model-based clustering of nonparametric weighted networks with application to water pollution analysis.
\newblock {\em Technometrics}, 62(2):161--172.

\bibitem[Amini et~al., 2013]{amini2013pseudo}
Amini, A.~A., Chen, A., Bickel, P.~J., and Levina, E. (2013).
\newblock Pseudo-likelihood methods for community detection in large sparse networks.
\newblock {\em The Annals of Statistics}, 41(4):2097--2122.

\bibitem[Ariu et~al., 2023]{ariu2023instanceoptimal}
Ariu, K., Proutiere, A., and Yun, S.-Y. (2023).
\newblock Instance-optimal cluster recovery in the labeled stochastic block model.
\newblock {\em arXiv}, page 2306.12968.

\bibitem[Bickel, 1975]{bickel1975one}
Bickel, P.~J. (1975).
\newblock One-step huber estimates in the linear model.
\newblock {\em Journal of the American Statistical Association}, 70(350):428--434.

\bibitem[Chen and Lei, 2018]{Chen2018Network}
Chen, K. and Lei, J. (2018).
\newblock Network cross-validation for determining the number of communities in network data.
\newblock {\em Journal of the American Statistical Association}, 113(521):241--251.

\bibitem[Chen et~al., 2024]{chen2024model}
Chen, Q., Agarwal, A., Fong, D.~K., DeSarbo, W.~S., and Xue, L. (2024).
\newblock Model-based co-clustering in customer targeting utilizing large-scale online product rating networks.
\newblock {\em Journal of Business \& Economic Statistics}, pages 1--13.

\bibitem[Erd\"os and R\'enyi, 1959]{Erdos1959}
Erd\"os, P. and R\'enyi, A. (1959).
\newblock On random graphs \textit{I}.
\newblock {\em Publicationes Mathematicae Debrecen}, 6:290--297.

\bibitem[Fan et~al., 2014]{fan2014strong}
Fan, J., Xue, L., and Zou, H. (2014).
\newblock Strong oracle optimality of folded concave penalized estimation.
\newblock {\em The Annals of Statistics}, 42(3):819--849.

\bibitem[Fortunato, 2010]{fortunato2010community}
Fortunato, S. (2010).
\newblock Community detection in graphs.
\newblock {\em Physics Reports}, 486(3):75--174.

\bibitem[Fortunato and Hric, 2016]{fortunato2016community}
Fortunato, S. and Hric, D. (2016).
\newblock Community detection in networks: A user guide.
\newblock {\em Physics Reports}, 659:1--44.

\bibitem[Gao et~al., 2017]{Gao2015}
Gao, C., Ma, Z., Zhang, A., and Zhou, H. (2017).
\newblock Achieving optimal misclassification proportion in stochastic block model.
\newblock {\em Journal of Machine Learning Research}, 18:1--45.

\bibitem[Gao et~al., 2018]{Gao2018}
Gao, C., Ma, Z., Zhang, A.~Y., and Zhou, H.~H. (2018).
\newblock {Community detection in degree-corrected block models}.
\newblock {\em The Annals of Statistics}, 46(5):2153--2185.

\bibitem[Gao et~al., 2009]{Gao2009Graph}
Gao, J., Liang, F., Fan, W., Sun, Y., and Han, J. (2009).
\newblock Graph-based consensus maximization among multiple supervised and unsupervised models.
\newblock In {\em Advances in Neural Information Processing Systems}, page 585–593, New York. Curran Associates Inc.

\bibitem[Goldenberg et~al., 2010]{Anna2010}
Goldenberg, A., Zheng, A.~X., Fienberg, S.~E., and Airoldi, E.~M. (2010).
\newblock A survey of statistical network models.
\newblock {\em Foundations and Trends in Machine Learning}, 2(2):129--233.

\bibitem[Holland et~al., 1983]{Holland1983stochastic}
Holland, P.~W., Laskey, K.~B., and Leinhardt, S. (1983).
\newblock Stochastic blockmodels: First steps.
\newblock {\em Social Networks}, 5(2):109--137.

\bibitem[Hu et~al., 2021]{Hu2021Using}
Hu, J., Zhang, J., Qin, H., Yan, T., and Zhu, J. (2021).
\newblock Using maximum entry-wise deviation to test the goodness of fit for stochastic block models.
\newblock {\em Journal of the American Statistical Association}, 116(535):1373--1382.

\bibitem[Huang et~al., 2024]{Sihan2023PCABM}
Huang, S., Sun, J., and Feng, Y. (2024).
\newblock Pcabm: Pairwise covariates-adjusted block model for community detection.
\newblock {\em Journal of the American Statistical Association}, 119(547):2092--2104.

\bibitem[Ji et~al., 2010]{Ming2010Graph}
Ji, M., Sun, Y., Danilevsky, M., Han, J., and Gao, J. (2010).
\newblock Graph regularized transductive classification on heterogeneous information networks.
\newblock In {\em Machine Learning and Knowledge Discovery in Databases}, pages 570--586, Berlin. Springer Berlin Heidelberg.

\bibitem[Jin, 2015]{Jin2015Fast}
Jin, J. (2015).
\newblock {Fast community detection by SCORE}.
\newblock {\em The Annals of Statistics}, 43(1):57--89.

\bibitem[Jin et~al., 2024]{Jin2024Mixed}
Jin, J., Ke, Z.~T., and Luo, S. (2024).
\newblock Mixed membership estimation for social networks.
\newblock {\em Journal of Econometrics}, 239(2):105369.

\bibitem[Jin et~al., 2023]{Jin2023Optimal}
Jin, J., Ke, Z.~T., Luo, S., and Wang, M. (2023).
\newblock Optimal estimation of the number of network communities.
\newblock {\em Journal of the American Statistical Association}, 118(543):2101--2116.

\bibitem[Karrer and Newman, 2011]{Karrer2011Stochastic}
Karrer, B. and Newman, M. E.~J. (2011).
\newblock Stochastic blockmodels and community structure in networks.
\newblock {\em Physical Review E}, 83:016107.

\bibitem[Kim et~al., 2018]{kim2018review}
Kim, B., Lee, K.~H., Xue, L., and Niu, X. (2018).
\newblock A review of dynamic network models with latent variables.
\newblock {\em Statistics Surveys}, 12:105--135.

\bibitem[Koo et~al., 2023]{John2023Popularity}
Koo, J., Tang, M., and Trosset, M.~W. (2023).
\newblock Popularity adjusted block models are generalized random dot product graphs.
\newblock {\em Journal of Computational and Graphical Statistics}, 32(1):131--144.

\bibitem[Kumar et~al., 2004]{Kumar2004}
Kumar, A., Sabharwal, Y., and Sen, S. (2004).
\newblock A simple linear time (1+$\xi$)-approximation algorithm for k-means clustering in any dimensions.
\newblock In {\em 45th Annual IEEE Symposium on Foundations of Computer Science}, pages 454--462, Rome. IEEE Symposium on Foundations of Computer Science.

\bibitem[Lei and Rinaldo, 2015]{Jing2015Consistency}
Lei, J. and Rinaldo, A. (2015).
\newblock {Consistency of spectral clustering in stochastic block models}.
\newblock {\em The Annals of Statistics}, 43(1):215 -- 237.

\bibitem[Li et~al., 2020]{li2020network}
Li, T., Levina, E., and Zhu, J. (2020).
\newblock Network cross-validation by edge sampling.
\newblock {\em Biometrika}, 107(2):257--276.

\bibitem[Moody and White, 2003]{moody2003structural}
Moody, J. and White, D.~R. (2003).
\newblock Structural cohesion and embeddedness: A hierarchical concept of social groups.
\newblock {\em American Sociological Review}, 68:103--127.

\bibitem[Noroozi et~al., 2021a]{Majid2021Sparse}
Noroozi, M., Pensky, M., and Rimal, R. (2021a).
\newblock Sparse popularity adjusted stochastic block model.
\newblock {\em Journal of Machine Learning Research}, 22(193):1--36.

\bibitem[Noroozi et~al., 2021b]{Noroozi2021Estimation}
Noroozi, M., Rimal, R., and Pensky, M. (2021b).
\newblock Estimation and clustering in popularity adjusted block model.
\newblock {\em Journal of the Royal Statistical Society Series B: Statistical Methodology}, 83(2):293--317.

\bibitem[Rohe et~al., 2010]{Rohe2010SpectralCA}
Rohe, K., Chatterjee, S., and Yu, B. (2010).
\newblock Spectral clustering and the high-dimensional stochastic blockmodel.
\newblock {\em The Annals of Statistics}, 39:1878--1915.

\bibitem[Saldana et~al., 2017]{saldana2017many}
Saldana, D.~F., Yu, Y., and Feng, Y. (2017).
\newblock How many communities are there?
\newblock {\em Journal of Computational and Graphical Statistics}, 26(1):171--181.

\bibitem[Sengupta and Chen, 2017]{Sengupta2017aBlock}
Sengupta, S. and Chen, Y. (2017).
\newblock {A block model for node popularity in networks with community structure}.
\newblock {\em Journal of the Royal Statistical Society Series B: Statistical Methodology}, 80(2):365--386.

\bibitem[Spirin and Mirny, 2003]{spirin2003protein}
Spirin, V. and Mirny, L.~A. (2003).
\newblock Protein complexes and functional modules in molecular networks.
\newblock {\em Proceedings of the National Academy of Sciences}, 100(21):12123--12128.

\bibitem[Su et~al., 2020]{Su2020Strong}
Su, L., Wang, W., and Zhang, Y. (2020).
\newblock Strong consistency of spectral clustering for stochastic block models.
\newblock {\em IEEE Transactions on Information Theory}, 66(1):324--338.

\bibitem[Sussman et~al., 2011]{Sussman2011ACA}
Sussman, D.~L., Tang, M., Fishkind, D.~E., and Priebe, C.~E. (2011).
\newblock A consistent adjacency spectral embedding for stochastic blockmodel graphs.
\newblock {\em Journal of the American Statistical Association}, 107:1119 -- 1128.

\bibitem[Wang et~al., 2018]{Wang2018Network}
Wang, B., Pourshafeie, A., Zitnik, M., Zhu, J., Bustamante, C., Batzoglou, S., and Leskovec, J. (2018).
\newblock Network enhancement as a general method to denoise weighted biological networks.
\newblock {\em Nature Communications}, 9:3108.

\bibitem[Wang et~al., 2020]{Wang2020Fast}
Wang, J., Zhang, J., Liu, B., Zhu, J., and Guo, J. (2020).
\newblock Fast network community detection with profile-pseudo likelihood methods.
\newblock {\em Journal of the American Statistical Association}, 118:1359--1372.

\bibitem[Wasserman and Faust, 1994]{wasserman1994}
Wasserman, S. and Faust, K. (1994).
\newblock {\em Social Network Analysis: Methods and Applications}.
\newblock Cambridge University Press, Cambridge.

\bibitem[Xu et~al., 2020]{Min2020Optimal}
Xu, M., Jog, V., and Loh, P. (2020).
\newblock Optimal rates for community estimation in the weighted stochastic block model.
\newblock {\em The Annals of Statistics}, 48(1):183--204.

\bibitem[Yu et~al., 2022]{YU2022Projected}
Yu, L., He, Y., Kong, X., and Zhang, X. (2022).
\newblock Projected estimation for large-dimensional matrix factor models.
\newblock {\em Journal of Econometrics}, 229(1):201--217.

\bibitem[Yuan et~al., 2022]{yuan2022testing}
Yuan, M., Liu, R., Feng, Y., and Shang, Z. (2022).
\newblock Testing community structure for hypergraphs.
\newblock {\em The Annals of Statistics}, 50(1):147--169.

\bibitem[Yun and Proutiere, 2016]{NIPS2016Yun}
Yun, S.-Y. and Proutiere, A. (2016).
\newblock Optimal cluster recovery in the labeled stochastic block model.
\newblock In {\em Advances in Neural Information Processing Systems}, pages 973--981, New York. Curran Associates, Inc.

\bibitem[Zhang and Zhou, 2016]{Harrison2016}
Zhang, A.~Y. and Zhou, H.~H. (2016).
\newblock {Minimax rates of community detection in stochastic block models}.
\newblock {\em The Annals of Statistics}, 44(5):2252--2280.

\bibitem[Zhang and Cao, 2017]{Zhang2017Finding}
Zhang, J. and Cao, J. (2017).
\newblock Finding common modules in a time-varying network with application to the drosophila melanogaster gene regulation network.
\newblock {\em Journal of the American Statistical Association}, 112(519):994--1008.

\bibitem[Zhang and Chen, 2017]{Zhang2017AHT}
Zhang, J. and Chen, Y. (2017).
\newblock A hypothesis testing framework for modularity based network community detection.
\newblock {\em Statistica Sinica}, 27(1):437--456.

\bibitem[Zhao et~al., 2024]{Zhao2024Variational}
Zhao, Y., Hao, N., and Zhu, J. (2024).
\newblock Variational estimators of the degree-corrected latent block model for bipartite networks.
\newblock {\em Journal of Machine Learning Research}, 25(150):1--42.

\bibitem[Zhao et~al., 2011]{Zhao2011Consistency}
Zhao, Y., Levina, E., and Zhu, J. (2011).
\newblock Consistency of community detection in networks under degree-corrected stochastic block models.
\newblock {\em The Annals of Statistics}, 40:2266--2292.

\end{thebibliography}

\end{document}